\documentclass{article}
\usepackage{bm,latexsym,amsmath,amsfonts,amssymb,fancyhdr,color,graphicx,slashed,cite}
\usepackage[a4paper,bottom=2.5cm,top=3.cm,head=5mm,width=17cm,dvipdfm]{geometry}
\usepackage[usenames,dvipsnames,svgnames,table]{xcolor}
\usepackage[colorlinks=true,
            linkcolor=blue,
            urlcolor=blue,
            citecolor=green,
            bookmarks=true,
            bookmarksnumbered=true,
            breaklinks=true,
            pdfpagemode=Fullscreen,
            pdfstartview=FitBH]{hyperref}

\usepackage[dotinlabels]{titletoc}
 \usepackage[normalem]{ulem}
\usepackage{float}
\usepackage{extarrows}
\usepackage{xcolor}
\usepackage{cancel}
\usepackage[bbgreekl]{mathbbol}
\usepackage{bbm}
\usepackage{siunitx}
\usepackage{enumitem}
\usepackage{tikz,pgfplots}
\usetikzlibrary{positioning}
\usepackage[title]{appendix}

\usepackage{multirow}
\usepackage{feynmf}
\numberwithin{equation}{section}
\allowdisplaybreaks[4]

\newcommand{\be}{\begin{equation}}
	\newcommand{\ee}{\end{equation}}
\newcommand{\ba}{\begin{array}}
	\newcommand{\ea}{\end{array}}
\newcommand{\bea}{\begin{eqnarray}}
	\newcommand{\eea}{\end{eqnarray}}

\newcommand{\C}{ {\tt C} }
\newcommand{\cL}{ {\tt L} }
\newcommand{\cR}{ {\tt R} }

\newcommand{\calO}{ {\cal O} }

\newcommand{\hc}{\text{h.c.}}
\newcommand{\beqn}{\begin{eqnarray}}
	\newcommand{\eeqn}{\end{eqnarray}}

\usepackage[titles]{tocloft}

\begin{document}

\fontsize{10pt}{12pt}\selectfont

\title{\textbf{Dark Sector Effective Field Theory} }

\author{
{Jin-Han Liang~$^{a,b}$}\footnote{jinhanliang@m.scnu.edu.cn},\,
{Yi Liao~$^{a,b}$}\footnote{liaoy@m.scnu.edu.cn},\,
{Xiao-Dong Ma~$^{a,b}$}\footnote{maxid@scnu.edu.cn}\, and 
{Hao-Lin Wang~$^{a,b}$}\footnote{whaolin@m.scnu.edu.cn}
 \\[3mm]
{\small\it $^a$~Key Laboratory of Atomic and Subatomic Structure and Quantum Control (MOE),}\\
{\small\it Guangdong Basic Research Center of Excellence for Structure and Fundamental Interactions of Matter,}\\
{\small\it Institute of Quantum Matter, South China Normal University, Guangzhou 510006, China}\\[1mm]
{\small\it $^b$~Guangdong-Hong Kong Joint Laboratory of Quantum Matter,}\\
{\small\it Guangdong Provincial Key Laboratory of Nuclear Science, Southern Nuclear Science Computing Center,}\\
{\small\it South China Normal University, Guangzhou 510006, China}
}

\date{}
\maketitle

\vspace{-0.75cm}

\begin{abstract}
We introduce the effective field theory of two different light dark particles interacting with the standard model (SM) light states in a single vertex, termed dark sector effective field theory (DSEFT). We focus on the new light particles with spin up to 1 and being real in essence, namely, new real scalars $\phi$ and $S$, Majorana fermions $\chi$ and $\psi$, and real vectors $X_\mu$ and $V_\mu$. In the framework of low energy effective field theory with QED and QCD symmetry, the DSEFT can be classified into six categories, including the scalar-scalar-SM ($\phi S$-SM), fermion-fermion-SM ($\chi\psi$-SM), vector-vector-SM ($X V$-SM), scalar-fermion-SM ($\phi \chi$-SM), scalar-vector-SM ($\phi X$-SM), and fermion-vector-SM ($\chi X$-SM) cases. For each case, we construct the effective operator basis up to canonical dimension 7, which will cover most interesting phenomenology at low energy. As a phenomenological example, we investigate the longstanding neutron lifetime anomaly through the neutron dark decay modes $n \to \chi  \phi \text{ or } \chi X$ from the effective 
interactions in the fermion-scalar-SM or fermion-vector-SM case. When treating the light fermion as a dark matter candidate, we also explore the constraints from DM-neutron annihilation signal at Super-Kamiokande. 
We find the neutron dark decay in each scenario can accommodate the anomaly, at the same time, without contradicting with the Super-Kamiokande
limit. 
\end{abstract}

\newpage
\hypersetup{linkcolor=black}
\tableofcontents
\hypersetup{linkcolor=red}

\hspace{2cm}
\section{Introduction}

~~The nature of dark matter (DM) is one of the most significant unsolved mysteries in particle physics and cosmology. A plethora of compelling evidence supports the existence of DM, such as the galactic rotation curves \cite{Rubin:1970zza}, the gravitational lensing effect \cite{Refregier:2003ct}, the power spectra of cosmic microwave background (CMB) \cite{Planck:2019nip}, the formation of large scale structure \cite{Allen:2002eu}, and the multi-wavelength observations of the Bullet Cluster \cite{Clowe_2006}, etc (for reviews, see Refs.\,\cite{Bertone:2004pz,Young:2016ala,Arbey:2021gdg}). Confronted with these facts, various DM candidates have been proposed in the past several decades, including primordial black holes (PBHs) \cite{Carr:2016drx,Escriva:2022duf}, weakly interacting massive particles (WIMPs) \cite{Jungman:1995df}, feebly interacting DM \cite{Bernal:2017kxu}, self-interacting DM \cite{Spergel:1999mh}, asymmetric DM \cite{Petraki:2013wwa}, ultra-light DM \cite{Ferreira:2020fam}, to name but a few. The mass for these candidates spans a huge gap from ${\cal O}(10^{-30}\,\rm eV)$ for axion or dark photon to ${\cal O}(M_{\rm sun})$ like PBHs. Among them, the WIMPs have been extensively searched for from terrestrial direct detection experiments, but without any positive signal, which has put stringent constraints on WIMP scenarios \cite{Roszkowski:2017nbc,Bottaro:2021snn}. This null result has spurred investigations into alternative DM scenarios beyond WIMP, especially the light DM such as sterile neutrino, axion or axion-like particle, and dark photon \cite{Davis:2015vla,Drewes:2016upu,Boyarsky:2018tvu}. For such light DM particles, their interactions with SM particles at low energy can be well-described within the framework of effective field theory (EFT), that has already constituted a major tool for the search of DM in direct and indirect detection experiments for a long time \cite{Harnik:2008uu,Fan:2010gt,Goodman:2010ku,Ding:2012sm,Ding:2013nvx,Kumar:2013iva,Alanne:2017oqj,Barman:2021hhg}.

Roughly speaking, the EFT approach offers a model-independent way to study DM interactions with visible matter like the electron or nucleon in low energy environment. The EFT analysis begins with the traditional $Z_2$ protected stable DM case \cite{Bishara:2017pfq,Brod:2017bsw,Bhattacharya:2021edh,Catena:2019gfa}, mainly for DM-electron or $u,d,s$ quark interactions relevant to direct detection. However, in recent years, under the framework of standard model effective field theory (SMEFT), this DM EFT analysis has been systematized and extended to more general and higher dimensional cases. Under the SMEFT framework, the EFT construction of single component DM has been extended up to dimension 8 (dim 8), which encompasses the scalar, fermion and vector cases \cite{Rajaraman:2012fu,DelNobile:2011uf,Criado:2021trs,Song:2023lxf,Song:2023jqm,Grojean:2023tsd}. On the other hand, since DM contributes four times more energy density in the universe than the ordinary visible matter and the latter has rich particle spectra, there is no reason to assume that DM is composed of a single type of particle.

In this sense, extending the single component DM scenario to a multi-component dark sector is a more viable option. The DM candidates can be provided by several of them while the rest acts as dark partners or mediators. This picture is perfectly illustrated by the SM itself---almost all abundance of visible matter in the universe consists of the proton, neutron and the electron, while the gluon and photon serve as force mediators. In between the dark and visible sectors there are ``portals'' to connect them, including the well known vector portal \cite{Holdom:1985ag}, axion portal \cite{Nomura:2008ru}, higgs portal \cite{Baek:2014jga}, and neutrino portal \cite{Falkowski:2009yz}. Thus far, the majority of studies have regarded portals as the interactions between a single component of dark sector and the SM fields, including the EFT analysis \cite{Arina:2021nqi,Contino:2020tix}. It is proposed in~\cite{Kaneta:2016wvf}, the new “dark-axion” portal connecting the dark photon and axion physics emerges when a dark photon and an axion coexist. Regarding EFT studies for the multi-component dark sector, only few works have appeared and focused on a very limited type of interactions \cite{GonzalezMacias:2015rxl,Aebischer:2022wnl}.

Hence the connection between the dark and visible sectors could be very diverse in the presence of multi-component dark particles, which involves new types of interactions. It is thus necessary to make a more general consideration on the interactions between the two sectors. The minimal choice of such a dark sector is to assume there are two light dark particles that intertwine with each other for the responsibility of the observed universe today. And in the existence of two distinct dark particles, we investigate their connections with the visible sector. 

In this work, we will systematically classify the complete and independent operator bases involving two distinct light dark sector particles and SM particles up to dim 7, we name it as dark sector effective field theory (DSEFT). We focus on the interactions in which the dark sector particles couple to at least one light standard model (SM) particle in the broken phase with the heavy weak gauge bosons, Higgs boson and top quark integrated out. In the DSEFT, the operators are required to be invariant under the gauge group $G\equiv \rm SU(3)_c\times U(1)_{em}$, and are organized in the form of DS1-DS2-SM, where DS1 and DS2 are new dark sector particles. Working within the low energy effective field theory (LEFT) framework, we can cover scenarios that contain both light dark particles and new weak scale mediators needed to be integrated out. The DSEFT can be viewed as an extension of the LEFT by those additional new interactions involving dark sector particles, which are singlets under the gauge group $G$.

The DSEFT offers a phenomenologically rich landscape to explore, from DM direct and indirect detections to astrophysical objects and cosmological observables. It also could be used to explain experimental anomalies. One example is the neutron lifetime anomaly from two different measuring methods, the beam-type \cite{Byrne:1996zz,Yue:2013qrc} and bottle-type \cite{Mampe:1993an,Serebrov:2004zf,Pichlmaier:2010zz,Steyerl:2012zz,Arzumanov:2015tea} experiments. 
This anomaly can be accounted for by additional neutron dark decay modes, which can be easily realized in the DSEFT framework. We perform a detailed analysis for this possibility by taking into account the constraints from matter instability induced by dark matter. 

The paper is organized as follows. We first construct in Sec.\,\ref{opebas} the operator bases in DSEFT up to dim 7 describing the local interactions between light dark particles and visible sector, which is followed by a brief discussion on the phenomenological implications at the beginning of Sec.\,\ref{neudark}. As a concrete application, we explore the phenomenology of baryon number violating interactions in neutron dark decay and DM-neutron annihilation in Sec.\,\ref{neudark}. Our concluding remarks are presented in Sec.\,\ref{sec:conc}. Finally, in Appendices\,\ref{app1} and \ref{app2}, we make some detailed analyses on the operator construction for two special cases. 

\section{Operator basis in DSEFT}
\label{opebas}

~~In this section, we construct the effective operators involving a pair of dark sector particles up to dim 7 within the LEFT framework. LEFT works below the electroweak scale $\Lambda_{\rm EW}$ with the symmetry group $G=\rm SU(3)_{c}\times U(1)_{em}$. It contains all SM light fields without the weak gauge bosons $W^\pm$, $Z$, the Higgs boson $h$, and the top quark $t$, which have been integrated out as heavy degrees of freedom \cite{Jenkins:2017jig,Liao:2020zyx}. We focus on interactions with the form of DS1-DS2-SM that would be induced in multi-component dark sector theory. These interactions have received limited investigation in previous research. We explore the light dark sector particles with spin up to $1$, i.e., light scalars ($\phi$ and $S$), light fermions ($\chi$ and $\psi$) and light vectors ($X_\mu$ and $V_\mu$). These dark sector particles possess masses below $\Lambda_{\rm EW}$ and are characterized as singlets under the gauge group $G$. For simplicity, we assume those new states are “real", i.e., real scalars, Majorana fermions, and real vectors. A straightforward generalization can be applied to encompass the cases of Dirac fermions and complex bosons, without introducing any distinct phenomenological features.
As an extension of the LEFT, the DSEFT contains all dynamical degrees of freedom existing within the LEFT of SM light fields. The fermionic constituents of the visible sector include up-type quarks, down-type quarks, charged leptons and left-handed neutrinos, denoted as $\Psi_p\in \{u_p, d_p, \ell_p, \nu_{L,p}\}$
\footnote{In the forthcoming discussions, the symbol $\Psi$ is occasionally employed to represent the light SM fermionic fields excluding the $\nu_{L}$ field.}, where $p$ denotes the generation index.  Specifically, there are $n_u=2$ up-type quarks $(u,c)$, $n_d=3$ down-type quarks $(d,s,b)$, $n_\ell=3$ charged leptons $(e,\mu,\tau)$, and $n_\nu=3$ neutrinos $(\nu_{e L},\nu_{\mu   L},\nu_{\tau L})$. The quarks and charged leptons are in the mass basis, represented by four-component Dirac spinors, while the neutrinos are in the chiral flavor basis. The gauge bosons include the photon $(A_\mu)$ and the gluons $(G^A_\mu)$ whose field strength tensors are $F^{\mu\nu}$ and $G^{A,\mu\nu}$.

We use all the dynamical degrees of freedom including dark sector and light SM particles to construct the complete and independent operators invariant under the gauge group $G$. We have exhausted all the following possibilities that involve two different dark particles interacting with at least one SM particle: (1) scalar-scalar-SM ($\phi S$-SM); (2) fermion-fermion-SM ($\chi\psi$-SM); (3) vector-vector-SM ($X V$-SM); (4) scalar-fermion-SM ($\phi \chi$-SM); (5) scalar-vector-SM ($\phi X$-SM); (6) fermion-vector-SM ($\chi X$-SM). 
The cases $\phi \phi$-SM, $\chi\chi$-SM, $XX$-SM with a pair of identical dark particles have been extensively considered and we omit them here.
To avoid ambiguity, we use distinct symbols for fields with the same spin, ensuring that they are different from each other. For the massive vector field, a consistent approach for constructing its Lagrangian is usually through either the Higgs mechanism \cite{Higgs:1964pj,Englert:1964et} or the Stueckelberg mechanism \cite{Stueckelberg:1938hvi,Ruegg:2003ps}. 
Here, we focus on the unitary gauge, which is suitable for tree-level calculations. Moreover, our set of operators can also be applied to loop-level processes by incorporating the corresponding would-be Goldstone boson $\pi$ of the massive vector $X_\mu$, achieved through the Stueckelberg mechanism with the replacement $X_\mu \to X_\mu-\partial_\mu\pi/m_X$, and similar replacement for $V_\mu$.

We have systematically formulated an extensive set of operators, spanning up to dim 7, as listed sequentially in Tables\ \ref{tab:phiS} to \ref{tab:Xchi}, in order to cover all of the aforementioned scenarios. In these tables, we classify the operators based on their dimensions and field contents. The number of operators are counted by considering $n_u$ up-type quarks, $n_d$ down-type quarks, $n_\nu$ neutrinos, and $n_\ell$ charged leptons. For non-hermitian operators, we use $+\hc$ to include their hermitian conjugates. Our result has been cross-checked by using the package {\tt Basisgen} \cite{Criado:2019ugp} with a modified representation of dark vectors, which was first realized in \cite{He:2022ljo}.
Together with the EFT operators with a single new field enumerated in \cite{Brod:2017bsw,Criado:2021trs,Song:2023lxf,DelNobile:2011uf,Song:2023jqm,Grojean:2023tsd,Goodman:2010ku}, the complete DSEFT operators will cover most interesting physics related to light dark particles at low energy.

Now we briefly discuss the procedures we have taken for constructing the operator basis. As mentioned before, we primarily used Dirac fermionic fields, except for the left-handed neutrino fields. 
The Lorentz structure $\Gamma$ between two fermionic spinors was chosen from the standard basis for the Clifford algebra. This includes  $\{1_4,~\gamma_5,~\gamma^\mu,~\gamma^{\mu}\gamma_5,~\sigma^{\mu\nu},~\sigma^{\mu\nu}\gamma^5\}$, where $\sigma^{\mu\nu}\gamma_5=\frac{i}{2}\epsilon^{\mu\nu\rho\sigma}\sigma_{\rho\sigma}$. 
However, when building explicit forms of operators, the possible Lorentz structures $\Gamma$ are further constrained by Lorentz invariance and the manner in which they are contracted. It's important to note that $\gamma_5$-related structures are reducible when dealing with the neutrino field due to its left-handed nature. To systematically enumerate all possibilities, we had to eliminate redundant operators using various techniques. These techniques include: integration by parts (IBP), equations of motion (EoM), generalized Fierz identities (FI), Schouten identity (SI), various Dirac gamma matrix identities (GI), and the Bianchi identity (BI). We employed the SI for the Lorentz group, given by:
\begin{eqnarray}
	\label{eq:SI}
	g_{\mu \nu} \epsilon_{\rho \sigma \tau \eta}+g_{\mu \rho} \epsilon_{\sigma \tau \eta \nu}+g_{\mu \sigma} \epsilon_{\tau \eta \nu \rho}+g_{\mu \tau} \epsilon_{\eta \nu \rho \sigma}+g_{\mu \eta} \epsilon_{\nu \rho \sigma \tau}=0.
\end{eqnarray}
This identity can help us simplify and reduce operators involving pure vectors. We also used a set of identities for Dirac gamma matrices during operator reduction:
\begin{subequations}
    \label{eq:GI}
\begin{eqnarray}
	\label{eq:GI:1}
	\sigma^{\mu\nu}&=&\frac{i}{2}\left[\gamma^\mu,\gamma^\nu\right]=i\gamma^\mu\gamma^\nu-i g^{\mu\nu}=i g^{\mu\nu}-i\gamma^\nu\gamma^\mu,
	\\
	\label{eq:GI:2}
	\gamma^\mu\gamma^\alpha\gamma^\nu&=&g^{\mu\alpha}\gamma^\nu+g^{\alpha\nu}\gamma^\mu-g^{\mu\nu}\gamma^\alpha+i\epsilon^{\mu\alpha\nu\beta}\gamma_\beta\gamma_5,
	\\
	\label{eq:GI:3}
	\left[\sigma^{\mu \nu}, \gamma^\rho\right]&=& 2 i\left(g^{\nu \rho} \gamma^\mu-g^{\mu \rho} \gamma^\nu\right),
	\\
	\label{eq:GI:4}
	\{\sigma^{\mu \nu}, \gamma^\rho\} &=& -2 \epsilon^{\mu\nu\rho\sigma}\gamma_\sigma\gamma_5.
\end{eqnarray}
\end{subequations}
Additionally, we utilized the BI for the field strength tensors of gauge fields:
\begin{eqnarray}
	\label{eq:BI}
	D_\mu {\bm X}_{\nu\rho}+D_\nu {\bm X}_{\rho\mu}+D_\rho {\bm X}_{\mu\nu}=0,~\text{or}~D_\nu \tilde{{\bm X}}^{\mu\nu}=0,
\end{eqnarray}
where $\Tilde{\bm X}^{\mu\nu}=\frac{1}{2}\epsilon^{\mu\nu\alpha\beta} {\bm X}_{\alpha\beta}$. Here $\bm X_{\mu\nu}$ can be SM photon and gluon field strength tensors $F_{\mu\nu}$ and $G_{\mu\nu}^A$, 
or dark vector field strength tensors $X_{\mu\nu}$ and $V_{\mu\nu}$ with $D_\rho \to \partial_\rho$ in the above identity.
The covariant derivative $D_\mu=\partial_\mu-ieQA_\mu-ig_s T^A G_\mu^A$ includes the electric charge $Q$ and color generators $T^A$ along with gauge couplings $e$ and $g_s$. In the following, each scenario is considered in turn.

\subsection{Scalar-scalar-SM case}

We first consider the interactions between two distinct real scalar fields from dark sector, denoted as $\phi$ and $S$, with the visible sector. At dim 5, these dark scalars couple to the scalar or pseudo-scalar components of fermionic bilinears within the visible sector. The resulting expression takes the form $(\overline{\Psi} \Gamma \Psi)\phi S$ with $\Gamma=1_4 (\gamma_5)$. By incorporating additional derivatives onto the dim-5 operators, we can extend them into dim-6 and dim-7 operators. However, this process introduces some intricacies when reducing operators. When constructing dim-6 operators $(\overline\Psi\Gamma\Psi)\phi S D$ by introducing an extra derivative, the only feasible choice is to place the derivative on the dark scalars themselves, since it would otherwise result in the EoMs of fermionic field. The operators with a derivative acting on $S$ can always be reshuffled on $\phi$ through the use of IBP. 
Thus, we choose them to be in the form, $\phi\overleftrightarrow{\partial_\mu} S \equiv \phi \partial_\mu S - (\partial_\mu \phi)S$.
Consequently, for a SM fermion denoted as $\Psi_p\in \{u_p, d_p, \ell_p\}$, only two independent operators exist for the $(\overline\Psi \Psi)\phi S D$-type. These operators correspond to the vector and the axial-vector bilinear, respectively. For the left-handed neutrino field, it is sufficient to keep its vector bilinear. Subsequently, we make a more detailed discussion regarding the reduction of the dim-7 $(\overline\Psi\Gamma\Psi)\phi S D^2$ operators. In order to maintain Lorentz invariance, the Dirac matrix $\Gamma$ must belong to the set of possibilities:  $\{1_4,\gamma_5,\sigma^{\mu\nu},\sigma^{\mu\nu}\gamma_5\}$. The two covariant derivatives contracted with each other or with $\sigma^{\mu\nu}$ cannot act on one field simultaneously, which leads to field strength tensors  and (or) trivial mass terms. Through the use of IBP, we are able to convert operators that contain two derivatives, one acting on $\overline{\Psi}$ and the other on $\Psi$, into operators with one derivative acting on a fermion field and the other on a dark scalar, respectively. We parameterize these operators as follows,
\begin{table}[!h]
	\center
	\resizebox{\linewidth}{!}{
	\renewcommand\arraystretch{1.12}
	\begin{tabular}{|l | l |  l|  l| l| l|}
		\hline
		Operator & Specific form & \# & Operator & Specific form & \#
		\\
		\hline
		\hline
		\multicolumn{6}{|c|}{\cellcolor{magenta!25}dim-5: $(\overline{\Psi} \Psi)\phi S $}
		\\
		\hline
		$\calO_{\nu \phi S}^{\tt S}+\hc$ & 
		$(\overline{ \nu_{L,p}^\C}\nu_{L,r}) \phi S$ & $n_\nu (n_\nu+1)$ & & &
		\\
		$\calO_{\ell \phi S}^{\tt S}$ & $(\overline{\ell_p}\ell_r)\phi S$ & $n_\ell^2$ &
		$\calO_{\ell \phi S}^{\tt P}$ & $(\overline{\ell_p} i\gamma_5 \ell_r)\phi S$ & $n_\ell^2$
		\\
		$\calO_{u\phi S}^{\tt S}$ & $(\overline{u_p} u_r)\phi S$ & $n_u^2$ &
		$\calO_{u\phi S}^{\tt P}$ & $(\overline{u_p} i\gamma_5 u_r)\phi S$ & $n_u^2$
		\\
		$\calO_{d\phi S}^{\tt S}$ & $(\overline{d_p} d_r)\phi S$ & $n_d^2$ &
		$\calO_{d\phi S}^{\tt P}$ & $(\overline{d_p} i\gamma_5 d_r)\phi S$ & $n_d^2$
		\\
		\hline
		\multicolumn{6}{|c|}{\cellcolor{magenta!25}dim-6: $(\overline\Psi \Psi)\phi S D$}
		\\
		\hline
		$\calO_{\nu \phi S D}^{\tt V}$  & $(\overline{ \nu_{L,p} }\gamma^{\mu}\nu_{L,r})\phi\overleftrightarrow{\partial_\mu} S$ & $n_\nu^2$ & & &
		\\
		$\calO_{\ell \phi SD}^{\tt V}$  & $(\overline{ \ell_{p} }\gamma^{\mu}\ell_r)\phi\overleftrightarrow{\partial_\mu} S$ & $n_\ell^2$ &
		$\calO_{\ell \phi SD}^{\tt A}$  & $(\overline{ \ell_{p} }\gamma^{\mu}\gamma_5\ell_r)\phi\overleftrightarrow{\partial_\mu} S$ & $n_\ell^2$
		\\
		$\calO_{u\phi SD}^{\tt V}$  & $(\overline{ u_{p} }\gamma^{\mu} u_r) \phi\overleftrightarrow{\partial_\mu} S$ & $n_u^2$ &
		$\calO_{u\phi SD}^{\tt A}$  & $(\overline{ u_{p} }\gamma^{\mu}\gamma_5 u_r) \phi\overleftrightarrow{\partial_\mu} S$ & $n_u^2$
		\\
		$\calO_{d\phi SD}^{\tt V}$  & $(\overline{ d_{p} }\gamma^{\mu} d_r) \phi\overleftrightarrow{\partial_\mu} S$ & $n_d^2$ &
		$\calO_{d\phi SD}^{\tt A}$  & $(\overline{ d_{p} }\gamma^{\mu}\gamma_5 d_r) \phi\overleftrightarrow{\partial_\mu} S$ & $n_d^2$
		\\
		\multicolumn{6}{|c|}{\cellcolor{magenta!25}dim-6: $\phi S F^2,\, \phi S G^2$}
		\\
		\hline
		$\calO_{\phi S F^2 1}$  & $\phi S F^{\mu\nu} F_{\mu\nu}$ & 1 &
		$\calO_{\phi S F^2 2}$  & $\phi S F^{\mu\nu} \tilde{F}_{\mu\nu}$ & 1
		\\
		$\calO_{\phi S G^2 1}$  & $\phi S G^{A,\mu\nu} G_{\mu\nu}^A$ & 1 &
		$\calO_{\phi S G^2 2}$  & $\phi S G^{A,\mu\nu} \tilde{G}_{\mu\nu}^A$ & 1
		\\
		\multicolumn{6}{|c|}{\cellcolor{magenta!25}dim-7: $(\overline\Psi \Psi)\phi S D^2$}
		\\
		\hline
		$\calO_{\nu\phi S D^2}^{\tt S}+\hc$ & $(\overline{\nu_{L,p}^\C}\nu_{L,r})(\partial^\mu\phi) (\partial_\mu S)$ & $n_\nu (n_\nu+1)$ &
		$\calO_{ \nu\phi S D^2}^{\tt T}+\hc $ & $(\overline{\nu_{L,p}^\C}\sigma_{\mu\nu}\nu_{L,r})(\partial^\mu\phi) (\partial^\nu S)$ & $n_\nu (n_\nu-1)$
		\\        
		$\calO_{\ell \phi SD^2}^{\tt S}$ & $(\overline{ \ell_p }\ell_{r}) (\partial^\mu\phi) (\partial_\mu S)$ & $n_\ell^2$ &
		$\calO_{\ell \phi S D^2}^{\tt P}$ & $(\overline{ \ell_p }\gamma_5\ell_{r}) (\partial^\mu\phi) (\partial_\mu S)$ & $n_\ell^2$
		\\
		$\calO_{\ell \phi S D^2}^{\tt T1}$ & $(\overline{ \ell_p }\sigma_{\mu\nu}\ell_{r})(\partial^\mu\phi)(\partial^\nu  S)$ & $n_\ell^2$ &
		$\calO_{\ell \phi S D^2}^{\tt T2}$ & $(\overline{ \ell_p }\sigma_{\mu\nu}i\gamma_5\ell_{r})(\partial^\mu\phi) (\partial^\nu S)$ & $n_\ell^2$
		\\
		$\calO_{u\phi S D^2}^{\tt S}$ & $(\overline{ u_p } u_{r}) (\partial^\mu\phi) (\partial_\mu S)$ & $n_u^2$ &
		$\calO_{u\phi S D^2}^{\tt P}$ & $(\overline{ u_p }\gamma_5 u_{r}) (\partial^\mu\phi) (\partial_\mu S)$ & $n_u^2$
		\\
		$\calO_{u\phi S D^2}^{\tt T1}$ & $(\overline{ u_p }\sigma_{\mu\nu}u_{r}) (\partial^\mu\phi) (\partial^\nu S)$ & $n_u^2$ &
		$\calO_{u\phi S D^2}^{\tt T2}$ & $(\overline{ u_p }\sigma_{\mu\nu}i\gamma_5 u_{r}) (\partial^\mu\phi)(\partial^\nu S)$  & $n_u^2$
		\\
		$\calO_{d\phi S D^2}^{\tt S}$ & $(\overline{ d_p } d_{r})(\partial^\mu\phi) (\partial_\mu S)$ & $n_d^2$ &
		$\calO_{d\phi S D^2}^{\tt P}$ & $(\overline{ d_p }\gamma_5 d_{r}) (\partial^\mu\phi) (\partial_\mu S)$ & $n_d^2$
		\\
		$\calO_{d\phi S D^2}^{\tt T1}$ & $(\overline{ d_p }\sigma_{\mu\nu}d_{r}) (\partial^\mu\phi) (\partial^\nu S)$ & $n_d^2$ &
		$\calO_{d\phi S D^2}^{\tt T2}$ & $(\overline{ d_p }\sigma_{\mu\nu}i\gamma_5 d_{r}) (\partial^\mu\phi) (\partial^\nu S)$ & $n_d^2$
		\\
		\multicolumn{6}{|c|}{\cellcolor{magenta!25}dim-7: $(\overline\Psi \Psi)\phi S F,\, (\overline\Psi \Psi)\phi S G$}
		\\
		\hline
		$\calO_{\nu F\phi S}^{\tt T}+\hc$ & $(\overline{\nu_{L,p}^\C}\sigma_{\mu\nu}\nu_{L,r})F^{\mu\nu}\phi S$ & $n_\nu (n_\nu-1)$ & & &
		\\
		$\calO_{\ell F \phi S}^{\tt T1}$ & $(\overline{ \ell_p }\sigma_{\mu\nu}\ell_{r}) F^{\mu\nu} \phi S$  & $n_\ell^2$ &
		$\calO_{\ell F\phi S}^{\tt T2}$ & $(\overline{ \ell_p }i\sigma_{\mu\nu}\gamma_5\ell_{r}) F^{\mu\nu} \phi S$ & $n_\ell^2$
		\\
		$\calO_{u F\phi S }^{\tt T1}$ & $(\overline{ u_p }\sigma_{\mu\nu} u_{r}) F^{\mu\nu} \phi S$  & $n_u^2$ &
		$\calO_{u F\phi S }^{\tt T2}$ & $(\overline{ u_p }i\sigma_{\mu\nu}\gamma_5 u_{r}) F^{\mu\nu} \phi S$ & $n_u^2$
		\\
		$\calO_{d F\phi S }^{\tt T1}$ & $(\overline{ d_p }\sigma_{\mu\nu} d_{r}) F^{\mu\nu} \phi S$  & $n_d^2$ &
		$\calO_{d F\phi S }^{\tt T2}$ & $(\overline{ d_p }i\sigma_{\mu\nu}\gamma_5 d_{r}) F^{\mu\nu} \phi S$ & $n_d^2$
		\\
		$\calO_{u G\phi S}^{\tt T1}$ & $(\overline{ u_p }\sigma_{\mu\nu} T^A u_{r}) G^{A, \mu\nu} \phi S$  & $n_u^2$ &
		$\calO_{u G\phi S}^{\tt T2}$ & $(\overline{ u_p }i\sigma_{\mu\nu} \gamma_5 T^A u_{r}) G^{A, \mu\nu} \phi S$ & $n_u^2$
		\\
		$\calO_{d G \phi S}^{\tt T1}$ & $(\overline{ d_p }\sigma_{\mu\nu} T^A d_{r}) G^{A, \mu\nu} \phi S$  & $n_d^2$ &
		$\calO_{d G \phi S}^{\tt T2}$ & $(\overline{ d_p }i\sigma_{\mu\nu} \gamma_5 T^A d_{r}) G^{A, \mu\nu} \phi S$ & $n_d^2$
		\\
		\hline
	\end{tabular}}
	\caption{Operator basis up to dim 7 for the scalar-scalar-SM case. Here $\phi$ and $S$ are real scalars in the dark sector. And $\Psi^{\tt C}=C\overline{\Psi}^{T}$ with $C=i\gamma^2\gamma^0$ denotes the charge conjugation of the field $\Psi$. We have counted the number of operators with general $n_\nu$, $n_\ell$, $n_u$ and $n_d$. For non-hermitian operators, we have used  $+\hc$ to include their hermitian conjugations and also counted the $\hc$ parts. }
	\label{tab:phiS}
\end{table}
\begin{subequations}
	\begin{align}
		\label{phisd2:eq1}
		&(\overline{\Psi}i\overleftrightarrow{D_\mu}\Psi)\phi(\partial^\mu S),&\, (\overline{\Psi}i\overleftrightarrow{D_\mu}\Psi)(\partial^\mu\phi) S,
		\\
		\label{phisd2:eq2}
		&(\overline{\Psi}i\gamma_5 i\overleftrightarrow{D_\mu}\Psi)\phi(\partial^\mu S),&\, (\overline{\Psi}i\gamma_5 i\overleftrightarrow{D_\mu}\Psi)(\partial^\mu\phi) S,
		\\
		\label{phisd2:eq3}
		&(\overline{\Psi}\sigma^{\mu\nu}\overleftrightarrow{D_\mu}\Psi)\phi(\partial_\nu S),&\, (\overline{\Psi}\sigma^{\mu\nu}\overleftrightarrow{D_\mu}\Psi)(\partial_\nu \phi)S,
		\\
		\label{phisd2:eq4}
		&(\overline{\Psi}\sigma^{\mu\nu}\gamma_5\overleftrightarrow{D_\mu}\Psi)\phi(\partial_\nu S),&\, (\overline{\Psi}\sigma^{\mu\nu}\gamma_5\overleftrightarrow{D_\mu}\Psi)(\partial_\nu \phi)S,
	\end{align}
\end{subequations}
where $\overline A \overleftrightarrow{D_\mu} B=\overline{A} D_\mu B-(\overline{D_\mu A}) B$. For the scalar-type operators:
\begin{eqnarray}
	\nonumber
	(\overline{\Psi}i\overleftrightarrow{D_\mu}\Psi)(\partial^\mu\phi)S
	&\overset{\tt GI}{=}&\frac{1}{2}\left(\overline{\Psi}\{\gamma_{\mu},i\slashed{D}\}\Psi-\overline{\Psi}\{i\overleftarrow{\slashed{D}},\gamma_\mu\}\Psi\right)(\partial^\mu\phi)S
	\\
	\nonumber
	&\overset{\tt EoM}{=}&\frac{1}{2}(\overline{\Psi}i\slashed{D}\gamma_{\mu}\Psi-\overline{\Psi}\gamma_{\mu}i\overleftarrow{\slashed{D}}\Psi) (\partial^\mu\phi)S+\fbox{EoM}
	\\
	\nonumber
	&\overset{\tt GI}{\underset{\tt IBP}{=}}&\frac{1}{2}(\overline{\Psi}i\overleftrightarrow{D_\mu}\Psi)(\partial^\mu\phi)S+\frac{1}{2}(\overline{\Psi}\sigma_{\mu\nu}\Psi) (\partial^\mu\phi) (\partial^\nu S)+\fbox{EoM}+\fbox{T}.
\end{eqnarray}
We have used the GIs listed in Eq.\,\eqref{eq:GI} and the EoM for fermion fields in the second step, as well as IBP in the last step. The $\fbox{EoM}$ represents terms proportional to lower dimensional EoM operators, and $\fbox{T}$ accounts for total derivative terms that can be discarded. Then we arrive at
\begin{eqnarray}
	(\overline{\Psi}i\overleftrightarrow{D_\mu}\Psi)(\partial^\mu\phi)S &=&(\overline{\Psi}\sigma_{\mu\nu}\Psi)(\partial^\mu\phi) (\partial^\nu S)+\fbox{EoM}.
\end{eqnarray}
Similar manipulations apply to Eq.\,\eqref{phisd2:eq2}. For the tensor-type operators, we have
\begin{eqnarray}
	\nonumber
	(\overline{\Psi}\sigma^{\mu\nu}\overleftrightarrow{D_\mu}\Psi)(\partial_\nu\phi)S
	&\overset{\tt GI}{=}&\left(iD^\mu(\overline{\Psi}\Psi)-\overline{\Psi}i\gamma^\mu\slashed{D}\Psi-\overline{\Psi}i\overleftarrow{\slashed{D}}\gamma^\mu\Psi\right)(\partial_\mu\phi)S
	\\
	\nonumber
	&\overset{\tt EoM}{=}&iD^\mu(\overline{\Psi}\Psi)(\partial_\mu\phi)S+\fbox{EoM}
	\\
	\nonumber
	&\overset{\tt IBP}{\underset{\tt EoM}{=}}&-i(\overline{\Psi}\Psi)(\partial_\mu\phi)(\partial^\mu S)+\fbox{EoM}.
\end{eqnarray}
The same derivations as above apply to the operators in Eq.\,\eqref{phisd2:eq3} and Eq.\,\eqref{phisd2:eq4}. The conclusion is that operators which feature two derivatives with one on a SM fermion and the other on a dark scalar can always be converted to operators with each of the two dark scalars being acted by one derivative. As a result, we choose the operators with two derivatives on dark scalar fields to be the independent constituents of the $(\overline{\Psi} \Psi)\phi SD^2$ class, as shown in Table \ref{tab:phiS}.  There are also other potential operator configurations for both dim-6 and dim-7 scenarios that incorporate the field strength tensors $F^{\mu\nu}$ and $G^{A,\mu\nu}$, which can be formulated accordingly. The complete operator basis for the scalar-scalar-SM case can be found in Table \ref{tab:phiS}.

\subsection{Fermion-fermion-SM case}
 \begin{table}[!h]
 	\center
 	\resizebox{\linewidth}{!}{
 	\renewcommand\arraystretch{1.03}
 	\begin{tabular}{|l | l |  l|  l| l| l|}
 		\hline
 		Operator & Specific form & \# & Operator & Specific form & \#
 		\\
 		\hline
 		\hline
 		\multicolumn{6}{|c|}{\cellcolor{magenta!25}dim-5: $(\overline\chi \psi)F $}
 		\\
 		\hline
 		$\calO_{\chi\psi F}^{\tt T1}$ & $(\overline \chi i \sigma^{\mu\nu} \psi) F_{\mu\nu}$ & & $\calO_{\chi\psi F}^{\tt T2}$ & $(\overline \chi \sigma^{\mu\nu}\gamma_5 \psi) F_{\mu\nu}$ &
 		\\
 		\multicolumn{6}{|c|}{\cellcolor{magenta!25}dim-6: $(\overline\chi \psi)(\overline\Psi \Psi)$}
 		\\
 		\hline
 		$\calO_{\nu\chi\psi}^{\tt SS}+\hc$ & $(\overline{\nu^\C_{L,p}}\nu_{L,r}) (\overline \chi  \psi) $ & $n_\nu(n_\nu+1)$ & 
 		$\calO_{\nu\chi\psi}^{\tt PS}+\hc$ & $(\overline{\nu^\C_{L,p}}\nu_{L,r}) (\overline \chi i\gamma_5 \psi) $ & $n_\nu(n_\nu+1)$ 
 		\\
 		$\calO_{\nu\chi\psi}^{\tt VV}$ & $(\overline{\nu_{L,p}}\gamma_{\mu}\nu_{L,r}) (\overline \chi  i\gamma^\mu \psi) $ & $n_\nu^2$ & $\calO_{\nu\chi\psi}^{\tt VA}$ & $(\overline{\nu_{L,p}}\gamma_{\mu}\nu_{L,r}) (\overline \chi\gamma^\mu\gamma_5 \psi) $ & $n_\nu^2$
 		\\
 		$\calO_{\nu\chi\psi}^{\tt TT}+\hc$ & $(\overline{\nu^\C_{L,p}}\sigma_{\mu\nu}\nu_{L,r}) (\overline \chi  \sigma^{\mu\nu} \psi) $ & $n_\nu(n_\nu-1)$ & & &
 		\\
 		\hline
 		$\calO_{\ell\chi\psi}^{\tt SS}$ & $(\overline{\ell_p} \ell_{r}) (\overline \chi  \psi) $ & $n_\ell^2$ &
 		$\calO_{\ell\chi\psi}^{\tt SP}$ & $(\overline{\ell_p} \ell_{r}) (\overline \chi i\gamma_5  \psi) $ & $n_\ell^2$
 		\\
 		$\calO_{\ell\chi\psi}^{\tt PS}$ & $(\overline{\ell_p} i\gamma_5  \ell_{r}) (\overline \chi  \psi) $ & $n_\ell^2$ &
 		$\calO_{\ell\chi\psi}^{\tt PP}$ & $(\overline{\ell_p} i\gamma_5 \ell_{r}) (\overline \chi i \gamma_5  \psi)$ & $n_\ell^2$
 		\\
 		$\calO_{\ell\chi\psi}^{\tt VV}$ & $(\overline{\ell_p} \gamma_{\mu}  \ell_{r}) (\overline \chi  i\gamma^{\mu}  \psi)$ & $n_\ell^2$  & $\calO_{\ell \chi\psi}^{\tt VA}$ & $(\overline{\ell_p} \gamma_{\mu}  \ell_{r}) (\overline \chi\gamma^{\mu}\gamma_5  \psi)$ & $n_\ell^2$
 		\\
 		$\calO_{\ell\chi\psi}^{\tt AV}$ & $(\overline{\ell_p} \gamma_{\mu}\gamma_5  \ell_{r}) (\overline \chi  i\gamma^{\mu}  \psi)$
 		& $n_\ell^2$ & $\calO_{\ell \chi\psi}^{\tt AA}$ & $(\overline{\ell_p} \gamma_{\mu}\gamma_5  \ell_{r}) (\overline \chi\gamma^{\mu}\gamma_5  \psi)$& $n_\ell^2$
 		\\
 		$\calO_{\ell\chi\psi}^{\tt T1}$ & $ (\overline{\ell_p} \sigma_{\mu\nu}  \ell_{r}) (\overline \chi  i\sigma^{\mu\nu} \psi)$ & $n_\ell^2$ & $\calO_{\ell\chi\psi}^{\tt T2}$ & $(\overline{\ell_p} \sigma_{\mu\nu}  \ell_{r})(\overline \chi\sigma^{\mu\nu}\gamma_5 \psi)$ & $n_\ell^2$
 		\\
 		\hline
 		$\calO_{u\chi\psi}^{\tt SS}$ & $(\overline{u_p} u_{r}) (\overline \chi  \psi) $ & $n_u^2$ &
 		$\calO_{u\chi\psi}^{\tt SP}$ & $(\overline{u_p}   u_{r}) (\overline \chi i\gamma_5  \psi) $ & $n_u^2$
 		\\
 		$\calO_{u\chi\psi}^{\tt PS}$ & $(\overline{u_p} i\gamma_5  u_{r}) (\overline \chi  \psi) $ & $n_u^2$ &
 		$\calO_{u\chi\psi}^{\tt PP}$ & $(\overline{u_p}  i\gamma_5  u_{r}) (\overline \chi i\gamma_5  \psi) $ & $n_u^2$
 		\\
 		$\calO_{u\chi\psi}^{\tt VV}$ & $(\overline{u_p} \gamma_{\mu}  u_{r}) (\overline \chi  i\gamma^{\mu}  \psi) $ & $n_u^2$  &
 		$\calO_{u\chi\psi}^{\tt VA}$ & $(\overline{u_p} \gamma_{\mu}  u_{r}) (\overline \chi\gamma^{\mu}\gamma_5  \psi) $ & $n_u^2$
 		\\
 		$\calO_{\chi\psi u}^{\tt AV}$ & $(\overline{u_p} \gamma_{\mu}\gamma_5  u_{r}) (\overline \chi  i\gamma^{\mu}  \psi) $  &  $n_u^2$ & $\calO_{u\chi\psi}^{\tt AA}$ & $(\overline{u_p} \gamma_{\mu}\gamma_5  u_{r}) (\overline \chi\gamma^{\mu}\gamma_5  \psi) $ & $n_u^2$
 		\\
 		$\calO_{u\chi\psi}^{\tt T1}$ & $(\overline{u_p} \sigma_{\mu\nu} u_{r}) (\overline \chi  i\sigma^{\mu\nu} \psi) $ & $n_u^2$ &
 		$\calO_{u\chi\psi}^{\tt T2}$ & $(\overline{u_p} \sigma_{\mu\nu}  u_{r}) (\overline \chi\sigma^{\mu\nu}\gamma_5 \psi) $ & $n_u^2$
 		\\ 
 		\hline
 		$\calO_{d\chi\psi}^{\tt SS}$ & $(\overline{d_p} d_{r}) (\overline \chi  \psi)$ & $n_d^2$ &
 		$\calO_{d\chi\psi}^{\tt SP}$ & $(\overline{d_p}   d_{r}) (\overline \chi i\gamma_5  \psi)$ & $n_d^2$
 		\\
 		$\calO_{d\chi\psi}^{\tt PS}$ & $(\overline{d_p} i\gamma_5  d_{r}) (\overline \chi  \psi)$  & $n_d^2$ &
 		$\calO_{d\chi\psi}^{\tt PP}$ & $(\overline{d_p}  i\gamma_5  d_{r}) (\overline \chi i\gamma_5  \psi)$ & $n_d^2$
 		\\
 		$\calO_{d\chi\psi}^{\tt VV}$ & $(\overline{d_p} \gamma_{\mu}  d_{r}) (\overline \chi  i\gamma^{\mu}  \psi) $ & $n_d^2$ &
 		$\calO_{d\chi\psi}^{\tt VA}$ & $(\overline{d_p} \gamma_{\mu}  d_{r}) (\overline \chi\gamma^{\mu}\gamma_5  \psi) $ & $n_d^2$
 		\\
 		$\calO_{d\chi\psi}^{\tt AV}$ & $(\overline{d_p} \gamma_{\mu}\gamma_5  d_{r}) (\overline \chi  i\gamma^{\mu}  \psi) $ & $n_d^2$ &
 		$\calO_{d\chi\psi}^{\tt AA}$ & $(\overline{d_p} \gamma_{\mu}\gamma_5  d_{r}) (\overline \chi\gamma^{\mu}\gamma_5  \psi) $ & $n_d^2$
 		\\
 		$\calO_{d\chi\psi}^{\tt T1}$ & $(\overline{d_p} \sigma_{\mu\nu} d_{r}) (\overline \chi  i\sigma^{\mu\nu} \psi)$ & $n_d^2$ &
 		$\calO_{d\chi\psi}^{\tt T2}$ & $(\overline{d_p} \sigma_{\mu\nu}  d_{r}) (\overline \chi\sigma^{\mu\nu} \gamma_5 \psi)$ & $n_d^2$
 		\\
 		\multicolumn{6}{|c|}{\cellcolor{magenta!25}dim-7: $(\overline\chi \psi)F^2$,  $(\overline\chi \psi)G^2$}
 		\\
 		\hline
 		$\calO_{\chi\psi F^2 1}^{\tt S}$ & $(\overline \chi  \psi) F^{\mu\nu}F_{\mu\nu}$ & 1 &
 		$\calO_{\chi\psi F^2 2}^{\tt S}$ & $(\overline \chi  \psi) F^{\mu\nu}\tilde{F}_{\mu\nu}$ & 1
 		\\
 		$\calO_{\chi\psi F^2 1}^{\tt P}$ & $(\overline \chi i\gamma_5 \psi) F^{\mu\nu}F_{\mu\nu}$ & 1 &
 		$\calO_{\chi\psi F^2 2}^{\tt P}$ & $(\overline \chi i\gamma_5 \psi)  F^{\mu\nu}\tilde{F}_{\mu\nu}$ & 1
 		\\
 		$\calO_{\chi\psi G^2 1}^{\tt S}$ & $(\overline \chi  \psi) G^{A,\mu\nu}G^A_{\mu\nu}$ & 1 &
 		$\calO_{\chi\psi G^2 2}^{\tt S}$ & $(\overline \chi  \psi) G^{A,\mu\nu}\tilde{G}^A_{\mu\nu}$ & 1
 		\\
 		$\calO_{\chi\psi G^2 1}^{\tt P}$ & $(\overline \chi i\gamma_5 \psi) G^{A,\mu\nu}G^{A}_{\mu\nu}$ & 1 &
 		$\calO_{\chi\psi G^2 2}^{\tt P}$ & $(\overline \chi i\gamma_5 \psi)  G^{A,\mu\nu}\tilde{G}^A_{\mu\nu}$ & 1
 		\\
 		\multicolumn{6}{|c|}{\cellcolor{magenta!25}dim-7: $(\overline\chi  \psi)(\overline\Psi \Psi)D$}
 		\\
 		\hline
 		$\calO_{\nu\chi\psi D}^{\tt VS}$ & $(\overline {\nu_{L,p}} \gamma_{\mu}\nu_{L,r}) (\overline \chi \overleftrightarrow{\partial^\mu}  \psi)$ & $n_\nu^2$ &
 		$\calO_{\nu\chi\psi D}^{\tt VP}$ & $(\overline {\nu_{L,p}} \gamma_{\mu}\nu_{L,r}) (\overline\chi i\gamma_5 \overleftrightarrow{\partial^\mu}  \psi)$ & $n_\nu^2$
 		\\
 		$\calO_{\nu\chi\psi D}^{\tt TV}+\hc$ & $(\overline{ \nu_{L,p}^\C } \sigma_{\mu\nu}\nu_{L,r}) (\overline\chi\gamma^{[\mu} i\overleftrightarrow{\partial^{\nu]}}\psi)$ & $n_\nu(n_\nu-1)$ &
 		$\calO_{\nu\chi\psi D}^{\tt TA}+\hc$ & $(\overline{ \nu_{L,p}^\C } \sigma_{\mu\nu}\nu_{L,r}) (\overline\chi \gamma^{[\mu} i\overleftrightarrow{\partial^{\nu]}}\gamma_5\psi)$  & $n_\nu(n_\nu-1)$ 
 		\\
 		\hline
 		$\calO_{\ell \chi\psi D}^{\tt VS}$ &  $(\overline{\ell_p} \gamma_{\mu}\ell_r) (\overline\chi \overleftrightarrow{\partial^\mu}  \psi)$ & $n_\ell^2$ & $\calO_{\ell\chi\psi D}^{\tt VP}$ &  $(\overline{\ell_p} \gamma_{\mu}\ell_r) (\overline\chi i\gamma_5 \overleftrightarrow{\partial^\mu}  \psi)$ & $n_\ell^2$
 		\\
 		$\calO_{\ell\chi\psi D}^{\tt AS}$ &  $(\overline{\ell_p} \gamma_{\mu}\gamma_5\ell_r) (\overline\chi \overleftrightarrow{\partial^\mu}  \psi)$ & $n_\ell^2$ &
 		$\calO_{\ell\chi\psi D}^{\tt AP}$ &  $(\overline{\ell_p} \gamma_{\mu}\gamma_5\ell_r)(\overline\chi i\gamma_5 \overleftrightarrow{\partial^\mu}  \psi)$ & $n_\ell^2$
 		\\
 		$\calO_{\ell\chi\psi D}^{\tt T1V}$ &  $ (\overline{\ell_p} \sigma_{\mu\nu}\ell_r)(\overline\chi \gamma^{[\mu} i\overleftrightarrow{\partial^{\nu]}}\psi)$  & $n_\ell^2$ & $\calO_{\ell\chi\psi D}^{\tt T1A}$ & $ (\overline{\ell_p} \sigma_{\mu\nu}\ell_r) (\overline\chi \gamma^{[\mu} \overleftrightarrow{\partial^{\nu]}}\gamma_5\psi)$ & $n_\ell^2$
 		\\ 
 		$\calO_{\ell\chi\psi D}^{\tt T2V}$ &  $(\overline{\ell_p} \sigma_{\mu\nu}i\gamma_5\ell_r) (\overline \chi \gamma^{[\mu} i\overleftrightarrow{\partial^{\nu]}}\psi) $  & $n_\ell^2$  &
 		$\calO_{\ell\chi\psi D}^{\tt T2A}$ & $(\overline{\ell_p} \sigma_{\mu\nu}i\gamma_5\ell_r) (\overline \chi \gamma^{[\mu} \overleftrightarrow{\partial^{\nu]}}\gamma_5\psi) $ & $n_\ell^2$
 		\\
 		\hline
 		$\calO_{u \chi\psi D}^{\tt VS}$ &  $(\overline{u_p} \gamma_{\mu}u_r) (\overline \chi \overleftrightarrow{\partial^\mu}  \psi) $ & $n_u^2$  &
 		$\calO_{u \chi\psi D}^{\tt VP}$ &  $(\overline{u_p} \gamma_{\mu}u_r) (\overline \chi i\gamma_5 \overleftrightarrow{\partial^\mu}  \psi) $  & $n_u^2$
 		\\
 		$\calO_{u \chi\psi D}^{\tt AS}$ &  $(\overline{u_p} \gamma_{\mu}\gamma_5 u_r) (\overline \chi \overleftrightarrow{\partial^\mu}  \psi) $ & $n_u^2$ &
 		$\calO_{u \chi\psi D}^{\tt AP}$ &  $(\overline{u_p} \gamma_{\mu}\gamma_5 u_r) (\overline \chi i\gamma_5 \overleftrightarrow{\partial^\mu}  \psi) $ & $n_u^2$
 		\\
 		$\calO_{u \chi\psi D}^{\tt T1V}$ &  $(\overline{u_p} \sigma_{\mu\nu} u_r) (\overline \chi \gamma^{[\mu} i\overleftrightarrow{\partial^{\nu]}}\psi) $  & $n_u^2$ &
 		$\calO_{u \chi\psi D}^{\tt T1A}$ &  $(\overline{u_p} \sigma_{\mu\nu} u_r) (\overline \chi \gamma^{[\mu} \overleftrightarrow{\partial^{\nu]}}\gamma_5\psi) $ & $n_u^2$
 		\\
 		$\calO_{u \chi\psi D}^{\tt T2V}$ &  $(\overline{u_p} \sigma_{\mu\nu}i\gamma_5 u_r) (\overline \chi \gamma^{[\mu} i\overleftrightarrow{\partial^{\nu]}}\psi) $  & $n_u^2$ &
 		$\calO_{u \chi\psi D}^{\tt T2A}$ & $(\overline{u_p} \sigma_{\mu\nu}i\gamma_5 u_r) (\overline \chi \gamma^{[\mu} \overleftrightarrow{\partial^{\nu]}}\gamma_5\psi) $ & $n_u^2$
 		\\
 		\hline
 		$\calO_{d \chi\psi D}^{\tt VS}$ &  $ (\overline{d_p} \gamma_{\mu}d_r) (\overline \chi \overleftrightarrow{\partial^\mu}  \psi)$ & $n_d^2$ &
 		$\calO_{d\chi\psi D}^{\tt VP}$ &  $ (\overline{d_p} \gamma_{\mu}d_r) (\overline \chi i\gamma_5\overleftrightarrow{\partial^\mu}  \psi)$ & $n_d^2$
 		\\
 		$\calO_{d\chi\psi D}^{\tt AS}$ &  $ (\overline{d_p} \gamma_{\mu}\gamma_5 d_r) (\overline \chi \overleftrightarrow{\partial^\mu}  \psi)$ & $n_d^2$ &
 		$\calO_{d\chi\psi D}^{\tt AP}$ &  $ (\overline{d_p} \gamma_{\mu}\gamma_5 d_r) (\overline \chi i\gamma_5 \overleftrightarrow{\partial^\mu}  
 		 \psi)$ & $n_d^2$
 		\\
 		$\calO_{d\chi\psi D}^{\tt T1V}$ &  $ (\overline{d_p} \sigma_{\mu\nu} d_r) (\overline \chi \gamma^{[\mu} i\overleftrightarrow{\partial^{\nu]}}\psi)$  & $n_d^2$ &
 		$\calO_{d\chi\psi D}^{\tt T1A}$ &  $  (\overline{d_p} \sigma_{\mu\nu} d_r) (\overline \chi \gamma^{[\mu} \overleftrightarrow{\partial^{\nu]}}\gamma_5\psi)$ & $n_d^2$
 		\\
 		$\calO_{d\chi\psi D}^{\tt T2V}$ &  $ (\overline{d_p} \sigma_{\mu\nu}i\gamma_5 d_r) (\overline \chi \gamma^{[\mu} i\overleftrightarrow{\partial^{\nu]}}\psi)$  & $n_d^2$  &
 		$\calO_{d\chi\psi  D}^{\tt T2A}$ & $(\overline{d_p} \sigma_{\mu\nu}i\gamma_5 d_r) (\overline \chi \gamma^{[\mu} \overleftrightarrow{\partial^{\nu]}}\gamma_5\psi) $ & $n_d^2$
 		\\
 		\hline
 	\end{tabular}}
 	\caption{Operator basis up to dim 7 for the fermion-fermion-SM case. Here $\chi$ and $\psi$ are Majorana fermions in the dark sector. We have counted the number of operators with general $n_\nu$, $n_\ell$, $n_u$ and $n_d$. For non-hermitian operators, we have used  $+\hc$ to include their hermitian conjugations and also counted the $\hc$ parts.}
 	\label{tab:chipsi}
 \end{table}
 
 Now we move on to examine the DSEFT operators that arise from two distinct dark fermions (represented by $\chi$ and $\psi$).  Without loss of generality, both of these dark fermions have been considered as Majorana fermions in this context. We utilized the techniques mentioned above to reduce the operator basis while ensuring its completeness. The final set of operators is displayed in Table \ref{tab:chipsi}. 
 These operators share similar structures as those containing a single component DM, which have been investigated in Refs.\,\cite{Goodman:2010ku,Brod:2017bsw,Li:2021phq}. For operators involving four fermion fields, we have arranged the fields by employing FIs to separate the dark fermion bilinear from the SM bilinear. For the dim-7 $(\overline\chi \psi)(\overline\Psi \Psi)D$-type operators involving a derivative, the derivative can always be arranged to act on the dark fermion bilinear \cite{Liao:2016qyd,Li:2021phq}. We thus select the eight independent dim-7 operators with derivatives in the dark particle current:
  \begin{subequations}
 \begin{align}
 	\calO_{\Psi \chi\psi D}^{{\tt VS},pr}&=(\overline{\Psi_p} \gamma_{\mu}\Psi_r) (\overline\chi \overleftrightarrow{\partial^\mu}  \psi),& \calO_{\Psi\chi\psi D}^{{\tt VP},pr}&=(\overline{\Psi_p} \gamma_{\mu}\Psi_r) (\overline\chi i\gamma_5 \overleftrightarrow{\partial^\mu}  \psi)
 	\\
 	\calO_{\Psi\chi\psi D}^{{\tt AS},pr}&=(\overline{\Psi_p} \gamma_{\mu}\gamma_5\Psi_r) (\overline\chi \overleftrightarrow{\partial^\mu}  \psi),&
 	\calO_{\Psi\chi\psi D}^{{\tt AP},pr}&=(\overline{\Psi_p} \gamma_{\mu}\gamma_5\Psi_r)(\overline\chi i\gamma_5 \overleftrightarrow{\partial^\mu}  \psi),
 	\\
 	\calO_{\Psi\chi\psi D}^{{\tt T1V},pr}&=(\overline{\Psi_p} \sigma_{\mu\nu}\Psi_r)(\overline\chi \gamma^{[\mu} i\overleftrightarrow{\partial}^{\nu]}\psi),& \calO_{\Psi\chi\psi D}^{{\tt T1A},pr}&=(\overline{\Psi_p} \sigma_{\mu\nu}\Psi_r) (\overline\chi \gamma^{[\mu} \overleftrightarrow{\partial}^{\nu]}\gamma_5\psi),
 	\\
 	\calO_{\Psi\chi\psi D}^{{\tt T2V},pr}&=(\overline{\Psi_p} \sigma_{\mu\nu}i\gamma_5\Psi_r) (\overline \chi \gamma^{[\mu} i\overleftrightarrow{\partial}^{\nu]}\psi),& 
 	\calO_{\Psi\chi\psi D}^{{\tt T2A},pr}&=(\overline{\Psi_p} \sigma_{\mu\nu}i\gamma_5\Psi_r) (\overline \chi \gamma^{[\mu} \overleftrightarrow{\partial}^{\nu]}\gamma_5\psi), 
 \end{align}
 \end{subequations}
where $\Psi$ denotes the fermionic SM fields $\{u, d, \ell\}$ with the flavor indices $p,r$. We use the superscript symbol $[...]$ to represent anti-symmetrization of indices in between, i.e., $A^{[\mu} B^{\nu]}= A^\mu B^\nu-A^\nu B^\mu$. For $\Psi=\nu_L$, it is not necessary to attach a $\gamma_5$ to its bilinear. Then structures with derivative in the SM current are redundant and can be represented as combinations of other operators \cite{Li:2021phq}. For instance,
 \begin{eqnarray}
 	(\overline{\Psi_p} i\overleftrightarrow{D^\mu}  \Psi_r)(\overline\chi i\gamma_\mu\psi)&=&-\frac{1}{2}\left[(m_\chi+m_\psi)\calO_{\Psi\chi\psi}^{{\tt T1},pr}-\calO_{\Psi\chi\psi D}^{{\tt T2A},pr}\right]+(m_{p}+m_{r}) \calO_{\Psi\chi\psi}^{{\tt VV},pr},
   	\nonumber
 	\\
 	(\overline{\Psi_p}\gamma_{[\mu}i\overleftrightarrow{D}_{\nu]}\Psi_r)(\overline\chi i\sigma^{\mu\nu}\psi) &=& 2\left[(m_\chi-m_\psi)\calO_{\Psi\chi\psi}^{\tt AA}-\calO_{\Psi \chi\psi D}^{\tt AP}\right]+i(m_p-m_r)\calO_{\Psi\chi\psi}^{{\tt T1},pr},
 \end{eqnarray}
and similar reductions can be made for the operators with fermion bilinears containing a $\gamma_5$.
All operators on the right-hand side of the above equations, both dim-6 and dim-7 ones, are part of our basis and can be found in Table \ref{tab:chipsi}.

\subsection{Vector-vector-SM case}

Now we explore the effective interactions induced by dark vectors $X_\mu$ and $V_\mu$. We start with the interactions between the SM fermions and the two dark vectors, whose leading terms arise at dim 5 in the form $(\overline{\Psi}\Gamma\Psi)XV$, where $\Gamma=1_4(\gamma_5)$ or $\sigma^{\mu\nu}(\gamma_5)$. To form higher dimensional operators, we attach derivatives and the field tensors of the photon and gluons. Attaching one derivative will lead to dim-6 $(\overline\Psi\Psi)XVD$-type operators. The construction of these operators is similar to the case involving only one complex dark vector, which has been established in Ref.\,\cite{He:2022ljo}. Hence we skip the detailed discussion here. At dim 7, there are three possible classes:
$$(\overline\Psi \Psi)FXV,\quad (\overline\Psi \Psi)GXV,\quad (\overline\Psi\Psi)XVD^2.$$
In the first two cases, we combine a field tensor $F$ or $G$ with two dark vectors. It is obvious that the fermion bilinear can only be that of quarks in the presence of the gluon tensor. Since the reduction of the third type of operators is complicated, we postpone its analysis in Appendix \ref{app1}.

\begin{table}
	\center
	\resizebox{\linewidth}{!}{
		\renewcommand\arraystretch{1.1}
		\begin{tabular}{| l | l | l| l | l| l|}
			\hline
			Operator & Specific form & \# & Operator & Specific form & \#
			\\
			\hline
			\hline
			\multicolumn{6}{|c|}{\cellcolor{magenta!25}dim-4: $FXV$}
			\\
			\hline
			$\calO_{F XV1}$ & $F^{\mu\nu} X_\mu V_\nu$ & &
			$\calO_{F XV2}$ & $\tilde F^{\mu\nu} X_\mu V_\nu$ &
			\\
			\hline
			\multicolumn{6}{|c|}{\cellcolor{magenta!25}dim-5: $(\overline{\Psi}\Psi)XV$}
			\\
			\hline
			$\calO_{\nu XV}^{\tt S} +\hc $ & $(\overline{\nu_{L,p}^\C} \nu_{L,r})X_\mu V^\mu$ & $n_\nu (n_\nu+1)$ &
			$\calO_{\nu XV}^{\tt T} +\hc$ &  ${1 \over 2} (\overline{\nu_{L,p}^\C}  \sigma^{\mu\nu} \nu_{L,r}) X_{[\mu}  V_{\nu]}$ & $n_\nu (n_\nu-1)$
			\\
			\hline
			$\calO_{\ell XV }^{\tt S} $ & $(\overline{\ell_p} \ell_r)X_\mu V^\mu$ & $n_\ell^2$ &
			$\calO_{\ell XV}^{\tt P} $ & $(\overline{\ell_p}i \gamma_5 \ell_r)X_\mu V^\mu$ & $n_\ell^2$
			\\
			$\calO_{\ell XV}^{\tt T1}$ & ${1 \over 2} (\overline{\ell_p}  \sigma^{\mu\nu} \ell_r) X_{[\mu} V_{\nu]}$ & $n_\ell^2$ &
			$\calO_{\ell XV}^{\tt T2} $ & ${1\over 2} (\overline{\ell_p}\sigma^{\mu\nu}i\gamma_5 \ell_r) X_{[\mu} V_{\nu]}$ & $n_\ell^2$
			\\
			\hline
			$\calO_{u XV}^{\tt S} $ & $(\overline{u_p} u_r)X_\mu V^\mu $ & $n_u^2$ &
			$\calO_{u XV}^{\tt P} $ & $(\overline{u_p}i \gamma_5 u_r)X_\mu V^\mu$ & $n_u^2$
			\\
			$\calO_{u XV}^{\tt T1}$ & ${1 \over 2} (\overline{u_p}  \sigma^{\mu\nu} u_r) X_{[\mu} V_{\nu]}$ & $n_u^2$ &
			$\calO_{u XV}^{\tt T2}$ & ${1\over 2} (\overline{u_p}\sigma^{\mu\nu}i\gamma_5 u_r) X_{[\mu} V_{\nu]}$ & $n_u^2$
			\\
			\hline
			$\calO_{d XV}^{\tt S} $ & $(\overline{d_p} d_r)(X_\mu V^\mu)$ & $n_d^2$ &
			$\calO_{d XV}^{\tt P} $ & $(\overline{d_p}i \gamma_5 d_r)X_\mu V^\mu$ & $n_d^2$
			\\
			$\calO_{d XV}^{\tt T1}$ & ${1 \over 2} (\overline{d_p}  \sigma^{\mu\nu} d_r) X_{[\mu} V_{\nu]}$ & $n_d^2$ &
			$\calO_{d XV}^{\tt T2}$ & ${1\over 2} (\overline{d_p}\sigma^{\mu\nu}i\gamma_5 d_r) X_{[\mu} V_{\nu]}$ & $n_d^2$
			\\
			\hline
			\multicolumn{6}{|c|}{\cellcolor{magenta!25}dim-6:  $FXVD^2,\,F^2XV,\, G^2XV$}
			\\
			\hline
			\cellcolor{green!35}$ \tilde\calO_{FXVD^2 1}$ & $F^\mu_{~\nu}X_{\mu\rho}V^{\nu\rho}$ & 1 &
			\cellcolor{green!35}$\tilde\calO_{FXVD^2 2}$ & $\tilde F^\mu_{~\nu}X_{\mu\rho}V^{\nu\rho}$ & 1
			\\
			$\calO_{F^2 XV 1}$ & $F_{\mu\nu} F^{\mu\nu} X_\rho V^\rho$ & 1 & $\calO_{F^2 XV 2}$ & $F_{\mu\nu} \tilde F^{\mu\nu} X_\rho V^\rho$ & 1
			\\
			$\calO_{F^2 XV 3}$ & $F^{\mu\rho} F_{\nu\rho} X_\mu V^\nu$ & 1 & &
			& 
			\\
			$\calO_{G^2 XV 1}$ & $G^A_{\mu\nu} G^{A,\mu\nu} X_\rho V^\rho$ & 1 &
			$\calO_{G^2 XV 2}$ & $G^A_{\mu\nu} \tilde G^{A,\mu\nu} X_\rho V^\rho$ & 1
			\\
			$\calO_{G^2 XV 3}$ & $G^{A,\mu\rho} G^{A}_{\nu\rho} X_\mu V^\nu$ & 1 & &
			& 
			\\
			\hline
			\multicolumn{6}{|c|}{\cellcolor{magenta!25}dim-6: $(\overline\Psi\Psi)XVD$}
			\\
			\hline
			$\calO_{\nu X V 1}^{\tt V}$ & ${1\over 2} [ \overline{\nu_{L,p}}\gamma_{(\mu} i \overleftrightarrow{D_{\nu)} } \nu_{L,r}] X^{(\mu} V^{\nu)}$ & $n_\nu^2$ &
			$\calO_{\nu X V 2}^{\tt V}$ &  $(\overline{\nu_{L,p}}\gamma_\mu \nu_{L,r})\partial_\nu (X^{(\mu} V^{\nu)})$ & $n_\nu^2$
			\\
			$\calO_{\nu X V 3}^{\tt V}$ & $(\overline{\nu_{L,p}}\gamma_\mu \nu_{L,r})( X_\rho \overleftrightarrow{\partial_\nu} V_\sigma )\epsilon^{\mu\nu\rho\sigma}$ & $n_\nu^2$  & $\calO_{\nu XV4}^{\tt V}$ & $(\overline{\nu_{L,p}}\gamma^\mu \nu_{L,r})(X_\nu \overleftrightarrow{\partial_\mu} V^\nu)$ & $n_\nu^2$
			\\
			$\calO_{\nu X V5}^{\tt V}$ & $(\overline{\nu_{L,p}}\gamma_\mu \nu_{L,r}) \partial_\nu (X^{[\mu } V^{\nu]})$ & $n_\nu^2$ & $\calO_{\nu X V6}^{\tt V}$ & $(\overline{\nu_{L,p}}\gamma_\mu \nu_{L,r})  \partial_\nu ( X_\rho V_\sigma )\epsilon^{\mu\nu\rho\sigma}$ & $n_\nu^2$
			\\
			\hline
			$\calO_{\ell XV 1}^{\tt V}$ & ${1\over 2} [ \overline{\ell_p}\gamma_{(\mu} i \overleftrightarrow{D_{\nu)} } \ell_r] X^{(\mu} V^{\nu)}$ & $n_\ell^2$ &
			$\calO_{\ell XV 2}^{\tt V}$ & $(\overline{\ell_p}\gamma_\mu \ell_r)\partial_\nu (X^{(\mu} V^{\nu)})$ & $n_\ell^2$
			\\
			$\calO_{\ell XV3}^{\tt V}$ & $(\overline{\ell_p}\gamma_\mu \ell_r)( X_\rho \overleftrightarrow{\partial_\nu} V_\sigma )\epsilon^{\mu\nu\rho\sigma}$ & $n_\ell^2$ &
			$\calO_{\ell XV 4}^{\tt V}$ & $(\overline{\ell_p}\gamma^\mu \ell_r)(X_\nu \overleftrightarrow{\partial_\mu} V^\nu)$ & $n_\ell^2$
			\\
			$\calO_{\ell XV 5}^{\tt V}$ & $(\overline{\ell_p}\gamma_\mu\ell_r) \partial_\nu (X^{[\mu } V^{\nu]})$ & $n_\ell^2$ &
			$\calO_{\ell XV 6}^{\tt V}$ & $(\overline{\ell_p}\gamma_\mu \ell_r) \partial_\nu ( X_\rho V_\sigma )\epsilon^{\mu\nu\rho\sigma}$ & $n_\ell^2$
			\\
			$\calO_{\ell XV1}^{\tt A}$ & ${1\over 2} [\overline{\ell_p}\gamma_{(\mu} \gamma_5 i \overleftrightarrow{D_{\nu)} }  \ell_r] X^{(\mu} V^{\nu)}$ & $n_\ell^2$ &
			$\calO_{\ell XV2}^{\tt A}$ & $(\overline{\ell_p}\gamma_\mu \gamma_5 \ell_r)\partial_\nu (X^{(\mu} V^{\nu)})$ & $n_\ell^2$
			\\ 
			$\calO_{\ell XV3}^{\tt A}$ & $(\overline{\ell_p}\gamma_\mu\gamma_5 \ell_r) 
			(X_\rho \overleftrightarrow{ \partial_\nu} V_\sigma )\epsilon^{\mu\nu\rho\sigma}$ & $n_\ell^2$ &
			$\calO_{\ell XV 4}^{\tt A}$ & $(\overline{\ell_p}\gamma^\mu\gamma_5 \ell_r)
			(X_\nu  \overleftrightarrow{\partial_\mu} V^\nu)$ & $n_\ell^2$
			\\
			$\calO_{\ell XV 5}^{\tt A}$ &  $(\overline{\ell_p}\gamma_\mu \gamma_5 \ell_r) \partial_\nu (X^{[\mu } V^{\nu]})$ & $n_\ell^2$ &
			$\calO_{\ell XV 6}^{\tt A}$ & $(\overline{\ell_p}\gamma_\mu\gamma_5 \ell_r) \partial_\nu (  X_\rho V_\sigma)\epsilon^{\mu\nu\rho\sigma}$ & $n_\ell^2$
			\\
			\hline
			$\calO_{uXV 1}^{\tt V}$ & ${1\over 2} [ \overline{u_p}\gamma_{(\mu} i \overleftrightarrow{D_{\nu)} } u_r] X^{(\mu} V^{\nu)}$ & $n_u^2$ &
			$\calO_{uXV 2}^{\tt V}$ & $(\overline{u_p}\gamma_\mu u_r)\partial_\nu (X^{(\mu} V^{\nu)})$ & $n_u^2$
			\\
			$\calO_{uXV3}^{\tt V}$ & $(\overline{u_p}\gamma_\mu u_r)( X_\rho \overleftrightarrow{\partial_\nu} V_\sigma )\epsilon^{\mu\nu\rho\sigma}$ & $n_u^2$ &
			$\calO_{uXV 4}^{\tt V}$ & $(\overline{u_p}\gamma^\mu u_r)(X_\nu \overleftrightarrow{\partial_\mu} V^\nu)$ & $n_u^2$
			\\
			$\calO_{uXV 5}^{\tt V}$ & $(\overline{u_p}\gamma_\mu u_r) \partial_\nu (X^{[\mu } V^{\nu]})$ & $n_u^2$ &
			$\calO_{uXV 6}^{\tt V}$ & $(\overline{u_p}\gamma_\mu  u_r) \partial_\nu ( X_\rho V_\sigma )\epsilon^{\mu\nu\rho\sigma}$ & $n_u^2$
			\\
			$\calO_{uXV1}^{\tt A}$ & ${1\over 2} [\overline{u_p}\gamma_{(\mu} \gamma_5 i \overleftrightarrow{D_{\nu)} }  u_r]X^{(\mu} V^{\nu)}$ & $n_u^2$ &
			$\calO_{uXV2}^{\tt A}$ & $(\overline{u_p}\gamma_\mu \gamma_5 u_r)\partial_\nu (X^{(\mu} V^{\nu)})$ & $n_u^2$
			\\ 
			$\calO_{uXV3}^{\tt A}$ & $(\overline{u_p}\gamma_\mu\gamma_5 u_r) 
			(X_\rho \overleftrightarrow{ \partial_\nu} V_\sigma )\epsilon^{\mu\nu\rho\sigma}$ & $n_u^2$ &
			$\calO_{uXV 4}^{\tt A}$ & $(\overline{u_p}\gamma^\mu\gamma_5 u_r)
			(X_\nu  \overleftrightarrow{\partial_\mu} V^\nu)$ & $n_u^2$
			\\
			$\calO_{uXV 5}^{\tt A}$ &  $(\overline{u_p}\gamma_\mu \gamma_5 u_r) \partial_\nu (X^{[\mu } V^{\nu]})$ & $n_u^2$ &
			$\calO_{uXV 6}^{\tt A}$ & $(\overline{u_p}\gamma_\mu\gamma_5 u_r)  \partial_\nu (  X_\rho V_\sigma)\epsilon^{\mu\nu\rho\sigma}$ & $n_u^2$
			\\
			\hline
			$\calO_{dXV 1}^{\tt V}$ & ${1\over 2} [ \overline{d_p}\gamma_{(\mu} i \overleftrightarrow{D_{\nu)} } d_r] X^{(\mu} V^{\nu)}$ & $n_d^2$ &
			$\calO_{dXV 2}^{\tt V}$ & $(\overline{d_p}\gamma_\mu d_r)\partial_\nu (X^{(\mu} V^{\nu)})$ & $n_d^2$
			\\
			$\calO_{d X V3}^{\tt V}$ & $(\overline{d_p}\gamma_\mu d_r)( X_\rho \overleftrightarrow{\partial_\nu} V_\sigma )\epsilon^{\mu\nu\rho\sigma}$ & $n_d^2$ &
			$\calO_{dXV 4}^{\tt V}$ & $(\overline{d_p}\gamma^\mu d_r)(X_\nu \overleftrightarrow{\partial_\mu} V^\nu)$ & $n_d^2$
			\\
			$\calO_{dXV 5}^{\tt V}$ & $(\overline{d_p}\gamma_\mu d_r) \partial_\nu (X^{[\mu } V^{\nu]})$ & $n_d^2$ &
			$\calO_{dXV 6}^{\tt V}$ & $(\overline{d_p}\gamma_\mu  d_r) \partial_\nu ( X_\rho V_\sigma )\epsilon^{\mu\nu\rho\sigma}$ & $n_d^2$
			\\
			$\calO_{dXV1}^{\tt A}$ & ${1\over 2} [\overline{d_p}\gamma_{(\mu} \gamma_5 i \overleftrightarrow{D_{\nu)} }  d_r] X^{(\mu} V^{\nu)}$ & $n_d^2$ &
			$\calO_{d X V2}^{\tt A}$ & $(\overline{d_p}\gamma_\mu \gamma_5 d_r)\partial_\nu (X^{(\mu} V^{\nu)})$ & $n_d^2$
			\\ 
			$\calO_{d X V3}^{\tt A}$ & $(\overline{d_p}\gamma_\mu\gamma_5 d_r) 
			(X_\rho \overleftrightarrow{ \partial_\nu} V_\sigma )\epsilon^{\mu\nu\rho\sigma}$ & $n_d^2$ &
			$\calO_{d X V 4}^{\tt A}$ & $(\overline{d_p}\gamma^\mu\gamma_5 d_r)
			(X_\nu  \overleftrightarrow{\partial_\mu} V^\nu)$ & $n_d^2$
			\\
			$\calO_{d X V 5}^{\tt A}$ &  $(\overline{d_p}\gamma_\mu \gamma_5 d_r) \partial_\nu (X^{[\mu } V^{\nu]})$ & $n_d^2$ &
			$\calO_{d X V 6}^{\tt A}$ & $(\overline{d_p}\gamma_\mu\gamma_5 d_r) \partial_\nu (  X_\rho V_\sigma)\epsilon^{\mu\nu\rho\sigma}$ & $n_d^2$
			\\
			\hline
	\end{tabular}}
	\caption{Operators for the vector-vector-SM case. Here $X_\mu$ and $V_\mu$ is real dark vectors. The symbols $(...)$ and $[...]$ represent the symmetrization and anti-symmetrization of indices in between, e.g., $A^{(\mu} B^{\nu)}=A^{\mu}B^{\nu}+A^{\nu}B^{\mu}$ and $A^{[\mu} B^{\nu]}=A^{\mu}B^{\nu}-A^{\nu}B^{\mu}$, respectively. For non-hermitian operators, we have used  $+\hc$ to include their hermitian conjugations and also counted the $\hc$ parts. The operators built purely with field tensors are highlighted in green.}
	\label{tab:XV:1}
\end{table}

\begin{table}
	\center
	\resizebox{\linewidth}{!}{
		\renewcommand\arraystretch{0.95}
		\begin{tabular}{| l| l | l| l | l| l|}
			\hline
			Operator & Specific form & \# & Operator & Specific form & \#
			\\
			\multicolumn{6}{|c|}{\cellcolor{magenta!25}dim-7: $(\overline\Psi \Psi)XVD^2$}
			\\
			\hline
			\cellcolor{green!25} $\tilde\calO_{\nu XV D^2 1}^{\tt S}+\hc$ & $(\overline{\nu_{L,p}^{\tt C} }\nu_{L,r} )
			X_{\mu\nu}  V^{\mu\nu}$ & $n_\nu(n_\nu+1)$ &
			\cellcolor{green!25} $\tilde\calO_{\nu XV D^2 2}^{\tt S}+\hc$ & $(\overline{\nu_{L,p}^{\tt C} }\nu_{L,r} )
			\tilde{X}_{\mu\nu}  V^{\mu\nu}$ & $n_\nu(n_\nu+1)$
			\\
			\cellcolor{green!25} $\tilde\calO_{\nu XV D^2 1}^{\tt T}+\hc$ & $(\overline{\nu_{L,p}^{\tt C} } \sigma^{\mu\nu} \nu_{L,r} ) X_{\mu\rho}  V_{\nu}^{~\rho}$ &  $n_\nu(n_\nu-1)$ & & &
			\\
			\cellcolor{green!25} $\tilde\calO_{\ell XV D^2 1}^{\tt S}$ & $(\overline{\ell_p }\ell_r )
			X_{\mu\nu}  V^{\mu\nu}$ & $n_\ell^2$ &
			\cellcolor{green!25} $\tilde\calO_{\ell XV D^2 2}^{\tt S}$ & $(\overline{\ell_p }\ell_r )
			\tilde X_{\mu\nu}  V^{\mu\nu}$ & $n_\ell^2$
			\\
			\cellcolor{green!25} $\tilde\calO_{\ell XV D^2 1}^{\tt P}$ & $(\overline{\ell_p } i\gamma_5 \ell_r ) X_{\mu\nu}  V^{\mu\nu}$ & $n_\ell^2$ &
			\cellcolor{green!25} $\tilde\calO_{\ell XV D^2 2}^{\tt P}$ & $(\overline{\ell_p } i\gamma_5 \ell_r ) \tilde X_{\mu\nu}  V^{\mu\nu}$ & $n_\ell^2$
			\\
			\cellcolor{green!25} $\tilde\calO_{\ell XV D^2}^{\tt T1}$ & $(\overline{\ell_p } \sigma^{\mu\nu} \ell_r ) X_{\mu\rho}  V_{\nu }^{~\rho}$ & $n_\ell^2$ &
			\cellcolor{green!25} $\tilde\calO_{\ell XV D^2}^{\tt T2}$ & $(\overline{\ell_p } \sigma^{\mu\nu}i \gamma_5\ell_r ) X_{\mu\rho}  V_{\nu}^{~\rho}$ & $n_\ell^2$
			\\
			\cellcolor{green!25} $\tilde\calO_{u XV D^2 1}^{\tt S}$ & $(\overline{u_p } u_r )
			X_{\mu\nu}  V^{\mu\nu}$ & $n_u^2$ &
			\cellcolor{green!25} $\tilde\calO_{u XV D^2 2}^{\tt S}$ & $(\overline{u_p } u_r )
			\tilde X_{\mu\nu}  V^{\mu\nu}$ & $n_u^2$
			\\
			\cellcolor{green!25}  $\tilde\calO_{u XV D^2 1}^{\tt P}$ & $(\overline{u_p } i\gamma_5 u_r )
			X_{\mu\nu}  V^{\mu\nu}$ & $n_u^2$ &
			\cellcolor{green!25} $\tilde\calO_{u XV D^2 2}^{\tt P}$ & $(\overline{u_p } i\gamma_5 u_r )
			\tilde X_{\mu\nu}  V^{\mu\nu}$ & $n_u^2$
			\\
			\cellcolor{green!25} $\tilde\calO_{u XV D^2}^{\tt T1}$ & $(\overline{u_p } \sigma^{\mu\nu} u_r )
			X_{\mu\rho}  V_{\nu}^{~\rho}$ & $n_u^2$  &
			\cellcolor{green!25} $\tilde\calO_{u XV D^2}^{\tt T2}$ & $(\overline{u_p } \sigma^{\mu\nu} i\gamma_5 u_r )
			X_{\mu\rho}  V_{\nu}^{~\rho}$ & $n_u^2$
			\\
			\cellcolor{green!25} $\tilde\calO_{d XV D^2 1}^{\tt S}$ & $(\overline{d_p } d_r )
			X_{\mu\nu}  V^{\mu\nu}$ & $n_d^2$ &
			\cellcolor{green!25}  $\tilde\calO_{d XV D^2 2}^{\tt S}$ & $(\overline{d_p } d_r )
			\tilde X_{\mu\nu}  V^{\mu\nu}$ & $n_d^2$
			\\
			\cellcolor{green!25} $\tilde\calO_{d XV D^2 1}^{\tt P}$ & $(\overline{d_p } i\gamma_5 d_r )
			X_{\mu\nu}  V^{\mu\nu}$ & $n_d^2$  &
			\cellcolor{green!25} $\tilde\calO_{d XV D^2 2}^{\tt P}$ & $(\overline{d_p} i\gamma_5 d_r )
			\tilde X_{\mu\nu}  V^{\mu\nu}$ & $n_d^2$
			\\
			\cellcolor{green!25} $\tilde\calO_{d XV D^2}^{\tt T1}$ & $(\overline{d_p } \sigma^{\mu\nu} d_r )
			X_{\mu\rho}  V_{\nu}^{~\rho}$ & $n_d^2$  &
			\cellcolor{green!25} $\tilde\calO_{d XV D^2}^{\tt T2}$ & $(\overline{d_p } \sigma^{\mu\nu} i\gamma_5 d_r ) X_{\mu\rho}  V_{\nu}^{~\rho}$ & $n_d^2$
			\\
			\hline
			$\calO_{\nu XVD^2 1}^{\tt S}+\hc$ & $(\overline{\nu_{L,p}^\C} i\overleftrightarrow{D_\mu} \nu_{L,r}) X_\nu V^{\mu\nu}$ & $n_\nu(n_\nu-1)$ &
			$\calO_{\nu XVD^2 2}^{\tt S}+\hc$ & $(\overline{\nu_{L,p}^\C} i\overleftrightarrow{D_\mu} \nu_{L,r}) X_\nu \tilde V^{\mu\nu} $ & $n_\nu(n_\nu-1)$
            \\
			$\calO_{\nu XVD^2 3}^{\tt S}+\hc$ & $(\overline{\nu_{L,p}^\C} i\overleftrightarrow{D_\mu} \nu_{L,r}) X^{\mu\nu} V_\nu $ & $n_\nu(n_\nu-1)$ &
			$\calO_{\nu XVD^2 4}^{\tt S}+\hc$ & $(\overline{\nu_{L,p}^\C} i\overleftrightarrow{D_\mu} \nu_{L,r}) \tilde X^{\mu\nu} V_\nu  $ & $n_\nu(n_\nu-1)$
			\\
			$\calO_{\nu XVD^2 1}^{\tt T}+\hc$ & $(\overline{\nu_{L,p}^\C} \sigma_{\mu\nu} i\overleftrightarrow{D_\rho} \nu_{L,r})X^\mu V^{\nu\rho} $ & $n_\nu(n_\nu+1)$ &
			$\calO_{\nu XVD^2 2}^{\tt T}+\hc$ & $(\overline{\nu_{L,p}^\C} \sigma_{\mu\nu} i\overleftrightarrow{D_\rho} \nu_{L,r}) X^{\mu\rho} V^\nu  $ & $n_\nu(n_\nu+1)$
			\\
			\hline
			$\calO_{\ell XVD^2 1}^{\tt S}$ & $(\overline{\ell_p} i \overleftrightarrow{D_\mu} \ell_r) X_\nu V^{\mu\nu}$ & $n_\ell^2$  &
			$\calO_{\ell XVD^2 2}^{\tt S}$ & $(\overline{\ell_p} i\overleftrightarrow{D_\mu} \ell_r) X_\nu \tilde V^{\mu\nu} $ & $n_\ell^2$
             \\
			$\calO_{\ell XVD^2 3}^{\tt S}$ & $(\overline{\ell_p} i\overleftrightarrow{D_\mu} \ell_r) X^{\mu\nu} V_\nu$ & $n_\ell^2$ &
			$\calO_{\ell XVD^2 4}^{\tt S}$ & $(\overline{\ell_p} i\overleftrightarrow{D_\mu} \ell_r)\tilde X^{\mu\nu} V_\nu  $ & $n_\ell^2$
			\\
			$\calO_{\ell XVD^2 1}^{\tt P}$ & $(\overline{\ell_p} \gamma_5 \overleftrightarrow{D_\mu} \ell_r) X_\nu V^{\mu\nu} $ & $n_\ell^2$ &
			$\calO_{\ell XVD^2 2}^{\tt P}$ & $(\overline{\ell_p} \gamma_5 \overleftrightarrow{D_\mu} \ell_r) X_\nu \tilde V^{\mu\nu} $ & $n_\ell^2$
            \\
			$\calO_{\ell XVD^2 3}^{\tt P}$ & $(\overline{\ell_p} \gamma_5 \overleftrightarrow{D_\mu} \ell_r) X^{\mu\nu} V_\nu $ & $n_\ell^2$ &
			$\calO_{\ell XVD^2 4}^{\tt P}$ & $(\overline{\ell_p} \gamma_5 \overleftrightarrow{D_\mu} \ell_r) \tilde X^{\mu\nu} V_\nu$ & $n_\ell^2$
			\\
			$\calO_{\ell XVD^2 1}^{\tt T1}$ & $(\overline{\ell_p} \sigma_{\mu\nu} i\overleftrightarrow{D_\rho} \ell_r)X^\mu V^{\nu\rho} $ & $n_\ell^2$ &
			$\calO_{\ell XVD^2 1}^{\tt T2}$ & $(\overline{\ell_p} \sigma_{\mu\nu}\gamma_5 \overleftrightarrow{D_\rho} \ell_r) X^\mu V^{\nu\rho} $  &  $n_\ell^2$
			\\
			$\calO_{\ell XVD^2 2}^{\tt T1}$ & $(\overline{\ell_p} \sigma_{\mu\nu} i\overleftrightarrow{D_\rho} \ell_r) X^{\mu\rho} V^\nu $ & $n_\ell^2$ &
			$\calO_{\ell XVD^2 2}^{\tt T2}$ & $(\overline{\ell_p} \sigma_{\mu\nu}\gamma_5 \overleftrightarrow{D_\rho} \ell_r) X^{\mu\rho} V^\nu $  & $n_\ell^2$
			\\
			\hline
			$\calO_{u XVD^2 1}^{\tt S}$ & $(\overline{u_p} i\overleftrightarrow{D_\mu} u_r) X_\nu V^{\mu\nu}$ & $n_u^2$ &
			$\calO_{u XVD^2 2}^{\tt S}$ & $(\overline{u_p} i\overleftrightarrow{D_\mu} u_r) X_\nu \tilde V^{\mu\nu} $ & $n_u^2$
			\\
			$\calO_{u XVD^2 3}^{\tt S}$ & $(\overline{u_p} i \overleftrightarrow{D_\mu} u_r)  X^{\mu\nu} V_\nu$ & $n_u^2$  &
			$\calO_{u XVD^2 4}^{\tt S}$ & $(\overline{u_p} i\overleftrightarrow{D_\mu} u_r) \tilde X^{\mu\nu} V_\nu $ & $n_u^2$
            \\
			$\calO_{u XVD^2 1}^{\tt P}$ & $(\overline{u_p} \gamma_5 \overleftrightarrow{D_\mu} u_r) X_\nu V^{\mu\nu} $ & $n_u^2$ &
			$\calO_{u XVD^2 2}^{\tt P}$ & $(\overline{u_p} \gamma_5 \overleftrightarrow{D_\mu} u_r) X_\nu \tilde V^{\mu\nu} $ & $n_u^2$
            \\
			$\calO_{u XVD^2 3}^{\tt P}$ & $(\overline{u_p} \gamma_5 \overleftrightarrow{D_\mu} u_r)X^{\mu\nu} V_\nu  $ & $n_u^2$ &
			$\calO_{u XVD^2 4}^{\tt P}$ & $(\overline{u_p} \gamma_5 \overleftrightarrow{D_\mu} u_r)\tilde X^{\mu\nu} V_\nu  $ & $n_u^2$
			\\
			$\calO_{u XVD^2 1}^{\tt T1}$ & $(\overline{u_p} \sigma_{\mu\nu} i\overleftrightarrow{D_\rho} u_r)X^\mu V^{\nu\rho} $ & $n_u^2$ & 
			$\calO_{u XVD^2 1}^{\tt T2}$ & $(\overline{u_p} \sigma_{\mu\nu}\gamma_5 \overleftrightarrow{D_\rho} u_r) X^\mu V^{\nu\rho} $  & $n_u^2$
			\\
			$\calO_{u XVD^2 2}^{\tt T1}$ & $(\overline{u_p} \sigma_{\mu\nu} i\overleftrightarrow{D_\rho} u_r) X^{\mu\rho} V^\nu  $ & $n_u^2$ &
			$\calO_{u XVD^2 2}^{\tt T2}$ & $(\overline{u_p} \sigma_{\mu\nu}\gamma_5 \overleftrightarrow{D_\rho} u_r)X^{\mu\rho}  V^\nu $ & $n_u^2$
			\\
			\hline 
			$\calO_{d XVD^2 1}^{\tt S}$ & $(\overline{d_p} i\overleftrightarrow{D_\mu} d_r) X_\nu V^{\mu\nu}$ & $n_d^2$ &
			$\calO_{d XVD^2 2}^{\tt S}$ & $(\overline{d_p} i\overleftrightarrow{D_\mu} d_r) X_\nu \tilde V^{\mu\nu} $ & $n_d^2$
            \\
			$\calO_{d XVD^2 3}^{\tt S}$ & $(\overline{d_p} i\overleftrightarrow{D_\mu} d_r) X^{\mu\nu} V_\nu $ & $n_d^2$ &
			$\calO_{d XVD^2 4}^{\tt S}$ & $(\overline{d_p} i\overleftrightarrow{D_\mu} d_r) \tilde X^{\mu\nu} V_\nu   $ & $n_d^2$
			\\
			$\calO_{d XVD^2 1}^{\tt P}$ & $(\overline{d_p} \gamma_5 \overleftrightarrow{D_\mu} d_r) X_\nu V^{\mu\nu} $ & $n_d^2$ &
			$\calO_{d XVD^2 2}^{\tt P}$ & $(\overline{d_p} \gamma_5 \overleftrightarrow{D_\mu} d_r) X_\nu \tilde V^{\mu\nu} $ & $n_d^2$
            \\
			$\calO_{d XVD^2 3}^{\tt P}$ & $(\overline{d_p} \gamma_5 \overleftrightarrow{D_\mu} d_r) X^{\mu\nu}  V_\nu $ & $n_d^2$ &
			$\calO_{d XVD^2 4}^{\tt P}$ & $(\overline{d_p} \gamma_5 \overleftrightarrow{D_\mu} d_r) \tilde X^{\mu\nu} V_\nu  $ & $n_d^2$
			\\
			$\calO_{d XVD^2 1}^{\tt T1}$ & $(\overline{d_p} \sigma_{\mu\nu} i\overleftrightarrow{D_\rho} d_r)X^\mu V^{\nu\rho} $ & $n_d^2$ &
			$\calO_{d XVD^2 1}^{\tt T2}$ & $(\overline{d_p} \sigma_{\mu\nu}\gamma_5 \overleftrightarrow{D_\rho} d_r) X^\mu V^{\nu\rho} $ & $n_d^2$
			\\
			$\calO_{d XVD^2 2}^{\tt T1}$ & $(\overline{d_p} \sigma_{\mu\nu} i\overleftrightarrow{D_\rho} d_r) X^{\mu\rho} V^\nu  $ & $n_d^2$ &
			$\calO_{d XVD^2 2}^{\tt T2}$ & $(\overline{d_p} \sigma_{\mu\nu}\gamma_5 \overleftrightarrow{D_\rho} d_r)  X^{\mu\rho} V^\nu  $ & $n_d^2$
			\\
			\hline
			$\calO_{\nu XVD^2 5}^{\tt S}+\hc$ & $(\overline{\nu_{L,p}^\C}\nu_{L,r}) \partial_{(\mu} X_{\nu)} \partial^{(\mu} V^{\nu)}$ & $n_\nu(n_\nu+1)$  &
			$\calO_{\nu XVD^2 6}^{\tt S}+\hc$ & $(\overline{\nu_{L,p}^\C} \overleftrightarrow{\partial_\mu} \nu_{L,r}) \partial^{(\mu} X^{\nu)} V_\nu$  & $n_\nu(n_\nu-1)$
			\\
			$\calO_{\nu XVD^2 7}^{\tt S}+\hc$ & $(\overline{\partial_{\mu} \nu_{L,p}^\C} \partial_{\nu} \nu_{L,r}) X^{(\mu} V^{\nu)}$ & $n_\nu(n_\nu+1)$ & & &
			\\
			$\calO_{\ell XVD^2 5}^{\tt S}$ & $(\overline{\ell_p}\ell_r) \partial_{(\mu} X_{\nu)} \partial^{(\mu} V^{\nu)}$ & $n_\ell^2$ &
			$\calO_{\ell XVD^2 5}^{\tt P}$ & $(\overline{\ell_p} i \gamma_5 \ell_r)\partial_{(\mu} X_{\nu)} \partial^{(\mu} V^{\nu)}$ & $n_\ell^2$
			\\
			$\calO_{\ell XVD^2 6}^{\tt S}$ & $(\overline{\ell_p} i\overleftrightarrow{D_\mu} \ell_r) \partial^{(\mu} X^{\nu)} V_\nu$ & $n_\ell^2$ &
			$\calO_{\ell XVD^2 6}^{\tt P}$ & $(\overline{\ell_p} \gamma_5 \overleftrightarrow{D_\mu} \ell_r)  \partial^{(\mu} X^{\nu)}  V_\nu$ & $n_\ell^2$
			\\
			$\calO_{\ell XVD^2 7}^{\tt S}$ & $(\overline{D_{\mu} \ell_p} D_{\nu} \ell_r) X^{(\mu} V^{\nu)}$ & $n_\ell^2$ & $\calO_{\ell XVD^2 7}^{\tt P}$ & $(\overline{D_{\mu} \ell_p}i\gamma_5 D_{\nu} \ell_r) X^{(\mu} V^{\nu)}$ & $n_\ell^2$
			\\
			$\calO_{uXVD^2 5}^{\tt S}$ & $(\overline{u_p}u_r) \partial_{(\mu} X_{\nu)} \partial^{(\mu} V^{\nu)}$ & $n_u^2$ &
			$\calO_{uXVD^2 5}^{\tt P}$ & $(\overline{u_p} i \gamma_5 u_r) \partial_{(\mu} X_{\nu)} \partial^{(\mu} V^{\nu)}$ & $n_u^2$
			\\
			$\calO_{uXVD^2 6}^{\tt S}$ & $(\overline{u_p} i\overleftrightarrow{D_\mu}  u_r) \partial^{(\mu} X^{\nu)} V_\nu$ & $n_u^2$ &
			$\calO_{uXVD^2 6}^{\tt P}$ & $(\overline{u_p} \gamma_5 \overleftrightarrow{D_\mu} u_r) \partial^{(\mu} X^{\nu)} V_\nu$ & $n_u^2$
			\\
			$\calO_{uXVD^2 7}^{\tt S}$ & $(\overline{D_{\mu} u_p} D_{\nu} u_r) X^{(\mu} V^{\nu)}$ & $n_u^2$ & $\calO_{uXVD^2 7}^{\tt P}$ & $(\overline{D_{\mu} u_p}i\gamma_5 D_{\nu} u_r) X^{(\mu} V^{\nu)}$ & $n_u^2$
			\\
			$\calO_{dXVD^2 5}^{\tt S}$ & $(\overline{d_p}d_r) \partial_{(\mu} X_{\nu)} \partial^{(\mu} V^{\nu)}$ &  $n_d^2$ &
			$\calO_{dXVD^2 5}^{\tt P}$ & $(\overline{d_p} i \gamma_5 d_r) \partial_{(\mu} X_{\nu)} \partial^{(\mu} V^{\nu)}$ & $n_d^2$
			\\
			$\calO_{dXVD^2 6}^{\tt S}$ & $(\overline{d_p} i\overleftrightarrow{D_\mu}  d_r) \partial^{(\mu} X^{\nu)} V_\nu$ & $n_d^2$ &
			$\calO_{dXVD^2 6}^{\tt P}$ & $(\overline{d_p} \gamma_5 \overleftrightarrow{D_\mu} d_r) \partial^{(\mu} X^{\nu)} V_\nu$ & $n_d^2$
			\\
			$\calO_{dXVD^2 7}^{\tt S}$ & $(\overline{D_{\mu} d_p} D_{\nu} d_r) X^{(\mu} V^{\nu)}$ & $n_d^2$ & $\calO_{dXVD^2 7}^{\tt P}$ & $(\overline{D_{\mu} d_p}i\gamma_5 D_{\nu} d_r) X^{(\mu} V^{\nu)}$ & $n_d^2$
			\\
            \hline
	\end{tabular}}
	\caption{Continuation of Table \ref{tab:XV:1}.}
	\label{tab:XV:2}
\end{table}

\begin{table}[!h]
	\center
	\resizebox{\linewidth}{!}{
		\renewcommand\arraystretch{1.2}
		\begin{tabular}{| l | l | l| l | l| l| }
			\hline
			Operator & Specific form & \# & Operator & Specific form & \#
            \\
			\hline
			\multicolumn{6}{|c|}{\cellcolor{magenta!25}dim-7: $ (\overline\Psi \Psi)FXV,\, (\overline\Psi \Psi)GXV$}  
			\\
			\hline
			$\calO_{\nu FXV}^{\tt S}+\hc$ & $(\overline{\nu_{L,p}^\C} \nu_{L,r})F^{\mu\nu}X_\mu V_\nu$ & $n_\nu(n_\nu+1)$ &
			$\tilde\calO_{\nu FXV}^{\tt S}+\hc$ & $(\overline{\nu_{L,p}^\C} \nu_{L,r})\tilde F^{\mu\nu}X_\mu V_\nu$ & $n_\nu(n_\nu+1)$ 
			\\
			$\calO_{\nu FXV1}^{\tt T}+\hc $ & $(\overline{\nu_{L,p}^\C} \sigma_{\mu\nu} \nu_{L,r})F^{\mu\nu}X_\rho V^\rho$ & $n_\nu(n_\nu-1)$  &
			$\calO_{\nu FXV2}^{\tt T}+\hc $ & $(\overline{\nu_{L,p}^\C} \sigma_{\mu\nu} \nu_{L,r})F^{\mu\rho}X_\rho V^\nu$ & $n_\nu(n_\nu-1)$ 
			\\
			$\calO_{\nu FXV3}^{\tt T}+\hc $ & $(\overline{\nu_{L,p}^\C} \sigma_{\mu\nu} \nu_{L,r})F^{\mu\rho}X^\nu V_\rho$ & $n_\nu(n_\nu-1)$  & & &     
			\\
			$\calO_{\ell FXV 1}^{\tt S} $ & $(\overline{\ell_p} \ell_r)F^{\mu\nu}X_\mu V_\nu$ & $n_\ell^2$ &
			$\calO_{\ell FXV 2}^{\tt S}$ & $(\overline{\ell_p} \ell_r)\tilde F^{\mu\nu}X_\mu V_\nu$ & $n_\ell^2$
			\\
			$\calO_{\ell FXV 1}^{\tt P} $ & $(\overline{\ell_p} i\gamma_5 \ell_r)F^{\mu\nu}X_\mu V_\nu$ & $n_\ell^2$ &
			$\calO_{\ell FXV 2}^{\tt P}$ & $(\overline{\ell_p} i\gamma_5 \ell_r)\tilde F^{\mu\nu}X_\mu V_\nu$ & $n_\ell^2$
			\\
			$\calO_{\ell FXV1}^{\tt T1} $ & $(\overline{\ell_p} \sigma_{\mu\nu} \ell_r)F^{\mu\nu}X_\rho V^\rho$ & $n_\ell^2$  &
			$\calO_{\ell FXV1}^{\tt T2}$ & $(\overline{\ell_p} \sigma_{\mu\nu}i\gamma_5 \ell_r) F^{\mu\nu}X_\rho V^\rho$ & $n_\ell^2$
			\\
			$\calO_{\ell FXV2}^{\tt T1} $ & $(\overline{\ell_p} \sigma_{\mu\nu} \ell_r)F^{\mu\rho}X_\rho V^\nu$ & $n_\ell^2$ &
			$\calO_{\ell FXV2}^{\tt T2}$ & $(\overline{\ell_p} \sigma_{\mu\nu}i\gamma_5 \ell_r) F^{\mu\rho}X_\rho V^\nu$ & $n_\ell^2$
			\\
			$\calO_{\ell FXV3}^{\tt T1} $ & $(\overline{\ell_p} \sigma_{\mu\nu} \ell_r)F^{\mu\rho}X^\nu V_\rho$ &$n_\ell^2$ &
			$\calO_{\ell FXV3}^{\tt T2}$ & $(\overline{\ell_p} \sigma_{\mu\nu} i\gamma_5 \ell_r) F^{\mu\rho}X^\nu V_\rho$ & $n_\ell^2$
            			\\
			$\calO_{uFXV 1}^{\tt S} $ & $(\overline{u_p} u_r)F^{\mu\nu}X_\mu V_\nu$ & $n_u^2$ &
			$\calO_{uFXV 2}^{\tt S}$ & $(\overline{u_p} u_r)\tilde F^{\mu\nu}X_\mu V_\nu$ & $n_u^2$
			\\
			$\calO_{uFXV 1}^{\tt P} $ & $(\overline{u_p} i\gamma_5 u_r)F^{\mu\nu}X_\mu V_\nu$ & $n_u^2$ &
			$\calO_{uFXV 2}^{\tt P}$ & $(\overline{u_p} i\gamma_5 u_r)\tilde F^{\mu\nu}X_\mu V_\nu$ & $n_u^2$
			\\
			$\calO_{uFXV1}^{\tt T1} $ & $(\overline{u_p} \sigma_{\mu\nu} u_r)F^{\mu\nu}X_\rho V^\rho$ & $n_u^2$ &
			$\calO_{uFXV1}^{\tt T2}$ & $(\overline{u_p} \sigma_{\mu\nu} i\gamma_5 u_r) F^{\mu\nu}X_\rho V^\rho$ & $n_u^2$
			\\
			$\calO_{uFXV2}^{\tt T1} $ & $(\overline{u_p} \sigma_{\mu\nu} u_r)F^{\mu\rho}X_\rho V^\nu$ & $n_u^2$ &
			$\calO_{uFXV2}^{\tt T2}$ & $(\overline{u_p} \sigma_{\mu\nu} i\gamma_5 u_r) F^{\mu\rho}X_\rho V^\nu$ & $n_u^2$
			\\
			$\calO_{uFXV3}^{\tt T1} $ & $(\overline{u_p} \sigma_{\mu\nu} u_r)F^{\mu\rho}X^\nu V_\rho$ & $n_u^2$ &
			$\calO_{uFXV3}^{\tt T2}$ & $(\overline{u_p} \sigma_{\mu\nu} i\gamma_5 u_r) F^{\mu\rho}X^\nu V_\rho$ & $n_u^2$
			\\
			$\calO_{dFXV 1}^{\tt S} $ & $(\overline{d_p} d_r)F^{\mu\nu}X_\mu V_\nu$ & $n_d^2$ &
			$\calO_{dFXV 2}^{\tt S}$ & $(\overline{d_p} d_r)\tilde F^{\mu\nu}X_\mu V_\nu$ & $n_d^2$
			\\
			$\calO_{dFXV 1}^{\tt P} $ & $(\overline{d_p} i\gamma_5 d_r)F^{\mu\nu}X_\mu V_\nu$ & $n_d^2$ &
			$\calO_{dFXV 2}^{\tt P}$ & $(\overline{d_p} i\gamma_5 d_r)\tilde F^{\mu\nu}X_\mu V_\nu$ & $n_d^2$
			\\
			$\calO_{dFXV1}^{\tt T1} $ & $(\overline{d_p} \sigma_{\mu\nu} d_r)F^{\mu\nu}X_\rho V^\rho$ & $n_d^2$ &
			$\calO_{dFXV1}^{\tt T2}$ & $(\overline{d_p} \sigma_{\mu\nu} i\gamma_5 d_r) F^{\mu\nu}X_\rho V^\rho$ & $n_d^2$
			\\
			$\calO_{dFXV2}^{\tt T1} $ & $(\overline{d_p} \sigma_{\mu\nu} d_r)F^{\mu\rho}X_\rho V^\nu$ & $n_d^2$ &
			$\calO_{dFXV2}^{\tt T2}$ & $(\overline{d_p} \sigma_{\mu\nu} i\gamma_5 d_r) F^{\mu\rho}X_\rho V^\nu$ & $n_d^2$
			\\
			$\calO_{dFXV3}^{\tt T1} $ & $(\overline{d_p} \sigma_{\mu\nu} d_r)F^{\mu\rho}X^\nu V_\rho$ & $n_d^2$ &
			$\calO_{dFXV3}^{\tt T2}$ & $(\overline{d_p} \sigma_{\mu\nu} i\gamma_5 d_r) F^{\mu\rho}X^\nu V_\rho$ &  $n_d^2$
            \\
            \hline
			$\calO_{uGXV1}^{\tt S} $ & $(\overline{u_p}T^A u_r) G^{A,\mu\nu}X_\mu V_\nu$ & $n_u^2$ &
			$\calO_{uGXV 2}^{\tt S}$ & $(\overline{u_p}T^A  u_r)\tilde G^{A,\mu\nu}X_\mu V_\nu$ & $n_u^2$
			\\
			$\calO_{uGXV1}^{\tt P} $ & $(\overline{u_p} i\gamma_5T^A  u_r) G^{A,\mu\nu}X_\mu V_\nu$ & $n_u^2$ &
			$\calO_{uGXV2}^{\tt P}$ & $(\overline{u_p} i\gamma_5T^A  u_r)\tilde G^{A,\mu\nu}X_\mu V_\nu$ & $n_u^2$
			\\
			$\calO_{uGXV1}^{\tt T1} $ & $(\overline{u_p} \sigma_{\mu\nu}T^A  u_r) G^{A,\mu\nu}X_\rho V^\rho$ & $n_u^2$ &
			$\calO_{uGXV1}^{\tt T2}$ & $(\overline{u_p} \sigma_{\mu\nu} i\gamma_5T^A  u_r) G^{A,\mu\nu}X_\rho V^\rho$ & $n_u^2$
			\\
			$\calO_{uGXV2}^{\tt T1} $ & $(\overline{u_p} \sigma_{\mu\nu}T^A  u_r) G^{A,\mu\rho}X_\rho V^\nu$ & $n_u^2$ &
			$\calO_{uGXV2}^{\tt T2}$ & $(\overline{u_p} \sigma_{\mu\nu} i\gamma_5T^A  u_r) G^{A,\mu\rho}X_\rho V^\nu$ & $n_u^2$
			\\
			$\calO_{uGXV3}^{\tt T1} $ & $(\overline{u_p} \sigma_{\mu\nu}T^A u_r) G^{A,\mu\rho}X^\nu V_\rho$ & $n_u^2$ &
			$\calO_{uGXV3}^{\tt T2}$ & $(\overline{u_p} \sigma_{\mu\nu} i\gamma_5  T^A u_r) G^{A,\mu\rho}X^\nu V_\rho$ & $n_u^2$
			\\
			$\calO_{dGXV 1}^{\tt S} $ & $(\overline{d_p} T^A  d_r)G^{A,\mu\nu}X_\mu V_\nu$ & $n_d^2$ &
			$\calO_{dGXV 2}^{\tt S}$ & $(\overline{d_p}T^A  d_r)\tilde G^{A,\mu\nu}X_\mu V_\nu$ & $n_d^2$
			\\
			$\calO_{dGXV 1}^{\tt P} $ & $(\overline{d_p} i\gamma_5 T^A  d_r) G^{A,\mu\nu}X_\mu V_\nu$ & $n_d^2$ &
			$\calO_{dGXV 2}^{\tt P}$ & $(\overline{d_p} i\gamma_5 T^A d_r)\tilde G^{A,\mu\nu}X_\mu V_\nu$ & $n_d^2$
			\\
			$\calO_{dGXV1}^{\tt T1} $ & $(\overline{d_p} \sigma_{\mu\nu}T^A d_r) G^{A,\mu\nu}X_\rho V^\rho$ & $n_d^2$ &
			$\calO_{dGXV1}^{\tt T2}$ & $(\overline{d_p} \sigma_{\mu\nu}i\gamma_5T^A d_r) G^{A,\mu\nu}X_\rho V^\rho$ & $n_d^2$
			\\
			$\calO_{dGXV2}^{\tt T1} $ & $(\overline{d_p} \sigma_{\mu\nu}T^A d_r) G^{A,\mu\rho}X_\rho V^\nu$ & $n_d^2$ &
			$\calO_{dGXV2}^{\tt T2}$ & $(\overline{d_p} \sigma_{\mu\nu}i\gamma_5T^A d_r) G^{A,\mu\rho}X_\rho V^\nu$ & $n_d^2$
			\\
			$\calO_{dGXV3}^{\tt T1} $ & $(\overline{d_p} \sigma_{\mu\nu}T^A d_r) G^{A,\mu\rho}X^\nu V_\rho$ & $n_d^2$ &
			$\calO_{dGXV3}^{\tt T2}$ & $(\overline{d_p} \sigma_{\mu\nu}i\gamma_5 T^A d_r) G^{A,\mu\rho}X^\nu V_\rho$  & $n_d^2$
			\\        
			\hline
	\end{tabular}}
	\caption{Continuation of Table \ref{tab:XV:2}.}
	\label{tab:XV:3}
\end{table}

The interactions of pure vectors are more intriguing in this case. At dim 4, the operators $\calO_{F XV1}=F^{\mu\nu} X_\mu V_\nu$ and 
$\calO_{F XV2}=\tilde F^{\mu\nu} X_\mu V_\nu$ emerge naturally. Constrained by Lorentz invariance, only an even number of derivatives can be attached to them. The operators of class $FXVD^2$ are at dim 6. We find that only two operators are of this type and can be represented by the field strength tensors of dark vectors:
\begin{eqnarray}
    \tilde\calO_{FXVD^2 1}=F^\mu_{~\nu}X_{\mu\rho}V^{\nu\rho}, \qquad\qquad
	\tilde\calO_{FXVD^2 2}=\tilde F^\mu_{~\nu}X_{\mu\rho}V^{\nu\rho}.
\end{eqnarray}
In addition, there exist operators composed of two strength tensors of the photon or gluons at dim 6, in the forms $F^2XV$ and $G^2XV$, each coming with three independent operators. For instance, for the $F^2XV$ case, they read:
\begin{eqnarray}
    \calO_{F^2 XV 1}=F_{\mu\nu} F^{\mu\nu} X_\rho V^\rho,\qquad \calO_{F^2 XV 2}=F_{\mu\nu} \tilde F^{\mu\nu} X_\rho V^\rho,\qquad \calO_{F^2 XV 3}=F^{\mu\rho} F_{\nu\rho} X_\mu V^\nu.
\end{eqnarray}
There is an operator that appears to be independent at first glance, but is actually redundant:
\begin{eqnarray}
	\nonumber
	\tilde{F}^{\mu\rho} F_{\nu\rho} X_{\mu} V^{\nu}
	&=& (\partial_\alpha A_\beta) (\partial_\nu A_\rho) X_\mu\left(\epsilon^{\mu\rho\alpha\beta} V^\nu-\epsilon^{\mu\nu\alpha\beta} V^\rho\right)
	\\
	\nonumber
	&\overset{\tt SI}{=}&(\partial_\alpha A_\beta) (\partial_\nu A_\rho) X_\mu (\epsilon^{\alpha\beta\nu\rho}V^\mu+\epsilon^{\mu\nu\rho\alpha}V^\beta-\epsilon^{\rho\beta\mu\nu}V^\alpha)
	\\
	&=&\frac{1}{2
	}F_{\mu\nu}\tilde{F}^{\mu\nu} X_\rho V^\rho -\tilde{F}^{\mu\rho} F_{\nu\rho} X_{\mu} V^{\nu},
\end{eqnarray}
so that
\begin{eqnarray}
	\tilde{F}^{\mu\rho} F_{\nu\rho} X_{\mu} V^{\nu}=\frac{1}{4}F_{\mu\nu}\tilde{F}^{\mu\nu} X_\rho V^\rho=\frac{1}{4}\calO_{F^2 XV 2}.
\end{eqnarray}
We have used the Schouten identity for Lorentz group in Eq.\,\eqref{eq:SI} in the above procedures.

The comprehensive operator basis for the vector-vector-SM scenario is shown in Table\,\ref{tab:XV:1}-Table\,\ref{tab:XV:3}. We found there are a plenty of operators in this scenario, especially at dim 7. And the majority of these operators is composed of dark vectors instead of dark tensors. We have highlighted the latter in the tables.

 \newpage
\subsection{Scalar-fermion-SM case}
\label{sec:sca-fer-SM}
 \begin{table}[!h]
 	\center
 	\resizebox{\linewidth}{!}{
 	\renewcommand\arraystretch{1.2}
 	\begin{tabular}{|l | l |  l|  l| l| l|}
 		\hline
 		Operator & Specific form & \# & Operator & Specific form & \#
 		\\
 		\hline
 		\hline
 		\multicolumn{6}{|c|}{\cellcolor{magenta!25}dim-4: $ (\overline{\nu_L}\chi)\phi+\hc$}
 		\\
 		\hline
 		$\calO_{\nu \chi\phi}^{\tt S}$ & $(\overline{\nu_{L,p}}\chi)\phi$ & $n_\nu$ & & &
 		\\
 		\hline
 		\multicolumn{6}{|c|}{\cellcolor{magenta!25}dim-6: $(\overline{\nu_L} \chi)\phi F+\hc$}
 		\\
 		\hline
 		$\calO_{\nu \chi \phi F}^{\tt T}$ & $(\overline{\nu_{L,p}}\sigma_{\mu\nu}\chi)\phi F^{\mu\nu} $ & $n_\nu$ & & &
 		\\
 		\hline
 		\multicolumn{6}{|c|}{\cellcolor{magenta!25}dim-7: $(\overline{\nu_L} \chi)\phi F D+\hc$}
 		\\
 		\hline
 		$\calO_{\nu\chi\phi FD1}^{\tt V}$ & $(\overline{\nu_{L,p}} \gamma_\mu \chi) \partial_\nu \phi F^{\mu\nu}$ & $n_\nu$ &
 		$\calO_{\chi\nu\phi FD2}^{\tt V}$ & $(\overline{\nu_{L,p}} \gamma_\mu \chi) \partial_\nu \phi \tilde  F^{\mu\nu}$ & $n_\nu$
 		\\
 		\multicolumn{6}{|c|}{\cellcolor{magenta!25}dim-7: $(\overline\Psi \Psi)(\overline{\nu_L}\chi)\phi+\hc$}
 		\\
 		\hline
 		$\calO_{\nu\nu\chi\phi}^{\tt SS}$ & $(\overline{ \nu_{L,p} }  \nu_{L,r}^{\C}) (\overline{ \nu_{L,s}}   \chi)\phi$ & $\frac{1}{3}n_\nu (n_\nu^2-1)$ &
        $\calO_{\nu\nu\chi\phi}^{\tt VV}$ & $(\overline{\nu_{L,p}} \gamma_\mu \nu_{L,r}) (\overline{ \nu_{L,s}}\gamma^\mu \chi)\phi$ & $\frac{1}{2} n_\nu^2 (n_\nu+1)$
 		\\
 		\hline
 		$\calO_{\ell\nu\chi\phi}^{\tt SS}$ & $(\overline{\ell_p} \ell_r) (\overline{ \nu_{L,s} } \chi)\phi$ & $n_\ell^2 n_\nu$ &
 		$\calO_{\ell\nu\chi\phi}^{\tt PS}$ & $(\overline{\ell_p} i\gamma_5 \ell_r) (\overline{ \nu_{L,s} }\chi)\phi$ & $n_\ell^2 n_\nu$
 		\\
 		$\calO_{\ell\nu\chi\phi}^{\tt VV}$ & $(\overline{\ell_p} \gamma_\mu \ell_r) (\overline{ \nu_{L,s} }\gamma^\mu \chi)\phi$ & $n_\ell^2 n_\nu$ &
 		$\calO_{\ell\nu\chi\phi}^{\tt AV}$ & $(\overline{\ell_p} \gamma_\mu\gamma_5 \ell_r) (\overline{ \nu_{L,s} } \gamma^\mu \chi)\phi$ & $n_\ell^2 n_\nu$
 		\\
 		$\calO_{\ell\nu\chi\phi}^{\tt TT}$ & $(\overline{\ell_p} \sigma_{\mu\nu} \ell_r) (\overline{ \nu_{L,s} } \sigma^{\mu\nu}  \chi)\phi$ & $n_\ell^2 n_\nu$ & & &
 		\\
 		\hline
 		$\calO_{u\nu\chi\phi}^{\tt SS}$ & $(\overline{u_p}  u_r) (\overline{ \nu_{L,s} } \chi)\phi$ & $n_u^2 n_\nu$ &
 		$\calO_{u\nu\chi\phi}^{\tt PS}$ & $(\overline{u_p} i\gamma_5 u_r) (\overline{ \nu_{L,s} }\chi)\phi$  & $n_u^2 n_\nu$
 		\\
 		$\calO_{u\nu\chi\phi}^{\tt VV}$ & $(\overline{u_p} \gamma_\mu u_r) (\overline{ \nu_{L,s} }\gamma^\mu \chi)\phi$ & $n_u^2 n_\nu$ &
 		$\calO_{u\nu\chi\phi}^{\tt AV}$ & $(\overline{u_p} \gamma_\mu\gamma_5 u_r) (\overline{ \nu_{L,s} } \gamma^\mu \chi)\phi$ & $n_u^2 n_\nu$
 		\\
 		$\calO_{u\nu\chi\phi}^{\tt TT}$ & $(\overline{u_p} \sigma_{\mu\nu} u_r) (\overline{ \nu_{L,s} } \sigma^{\mu\nu}  \chi)\phi$ & $n_u^2 n_\nu$ & & &
 		\\
 		\hline
 		$\calO_{d\nu\chi\phi}^{\tt SS}$ & $(\overline{d_p}  d_r) (\overline{ \nu_{L,s} } \chi)\phi$ & $n_d^2 n_\nu$ &
 		$\calO_{d\nu\chi\phi}^{\tt PS}$ & $(\overline{d_p} i\gamma_5 d_r) (\overline{ \nu_{L,s} }\chi)\phi$ & $n_d^2 n_\nu$
 		\\
 		$\calO_{d\nu\chi\phi}^{\tt VV}$ & $(\overline{d_p} \gamma_\mu d_r) (\overline{ \nu_{L,s} }\gamma^\mu \chi)\phi$ & $n_d^2 n_\nu$ &
 		$\calO_{d\nu\chi\phi}^{\tt AV}$ & $(\overline{d_p} \gamma_\mu\gamma_5 d_r) (\overline{ \nu_{L,s} } \gamma^\mu \chi)\phi$ & $n_d^2 n_\nu$
 		\\
 		$\calO_{d\nu\chi\phi}^{\tt TT}$ & $(\overline{d_p} \sigma_{\mu\nu} d_r) (\overline{ \nu_{L,s} } \sigma^{\mu\nu}  \chi)\phi$ & $n_d^2 n_\nu$ & & &
 		\\
 		\multicolumn{6}{|c|}{\cellcolor{magenta!25}dim-7: $(\overline d u)(\overline \chi \ell)\phi+\hc$}
 		\\
 		\hline
 		$\calO^{\tt SS}_{du\chi\ell\phi}$ & $(\overline{d_p} u_{r})(\overline{\chi} \ell_{s})\phi$ & $n_d n_u n_\ell$ &  $\calO^{\tt SP}_{du\chi\ell\phi}$ &$(\overline{d_p} u_{r})(\overline{\chi} i\gamma_5 \ell_{s})\phi$ & $n_d n_u n_\ell$
 		\\
 		$\calO^{\tt PS}_{du\chi\ell\phi}$ & $(\overline{d_p} i\gamma_5 u_{r})(\overline{\chi} \ell_{s})\phi$ & $n_d n_u n_\ell$ & $\calO^{\tt PP}_{du\chi\ell\phi}$& $(\overline{d_p} i\gamma_5 u_{r})(\overline{\chi} i\gamma_5 \ell_{s})\phi$ & $n_d n_u n_\ell$
 		\\
 		$\calO^{\tt VV}_{du\chi\ell\phi}$ & $(\overline{d_p}\gamma_\mu u_{r})(\overline{\chi} \gamma^\mu\ell_{s})\phi$ & $n_d n_u n_\ell$ & $\calO^{\tt VA}_{\chi\phi du\ell}$ & $(\overline{d_p} \gamma_\mu u_{r})(\overline{\chi} \gamma^\mu\gamma_5 \ell_{s})\phi$ & $n_d n_u n_\ell$
 		\\
 		$\calO^{\tt AV}_{du\chi\ell\phi}$ & $(\overline{d_p} \gamma_\mu\gamma_5 u_{r})(\overline{\chi} \gamma^\mu\ell_{s})\phi$ & $n_d n_u n_\ell$ &  $\calO^{\tt AA}_{du\chi\ell\phi}$ & $(\overline{d_p} \gamma_\mu\gamma_5 u_{r})(\overline{\chi} \gamma^\mu\gamma_5 \ell_{s})\phi$ & $n_d n_u n_\ell$
 		\\
 		$\calO^{\tt T1}_{du\chi\ell\phi}$ & $(\overline{d_p} \sigma_{\mu\nu} u_{r})(\overline{\chi} \sigma^{\mu\nu}\ell_{s})\phi$ & $n_d n_u n_\ell$ & $\calO^{\tt T2}_{du\chi\ell\phi}$ & $(\overline{d_p} \sigma_{\mu\nu} i\gamma_5 u_{r})(\overline{\chi} \sigma^{\mu\nu} \ell_{s})\phi$ & $n_d n_u n_\ell$
 		\\
 		\multicolumn{6}{|c|}{\cellcolor{magenta!25}dim-7: $(\overline\chi u)(d d)\phi+\hc\, (\slashed{B})$ }
 		\\
 		\hline
 		$\calO^{\tt SS}_{\phi\chi u dd}$ & 
 		$\epsilon^{ijk}(\overline{\chi} u_{i,p})(\overline{d^{\C}_{j,r} }d_{k,s})\phi $ & $\frac{1}{2} n_u  n_d(n_d-1)$ &
 		$\calO^{\tt SP}_{\phi\chi u dd}$ & 
 		$\epsilon^{ijk}(\overline{\chi} u_{i,p})(\overline{d^{\C}_{j,r} }i\gamma_5 d_{k,s})\phi $ & $\frac{1}{2} n_u  n_d(n_d-1)$
 		\\
 		$\calO^{\tt PS}_{\phi\chi u dd}$ & 
 		$\epsilon^{ijk}(\overline{\chi}i\gamma_5 u_{i,p})(\overline{d^{\C}_{j,r} }d_{k,s}) \phi$ & $\frac{1}{2} n_u  n_d(n_d-1)$ &
 		$\calO^{\tt PP}_{\phi\chi u dd}$ & 
 		$\epsilon^{ijk}(\overline{\chi} \gamma_5 u_{i,p})(\overline{d^{\C}_{j,r} }\gamma_5 d_{k,s})\phi $ & $\frac{1}{2} n_u  n_d(n_d-1)$
 		\\
 		$\calO^{\tt VV}_{\phi\chi u dd}$ & 
 		$\epsilon^{ijk}(\overline{\chi}\gamma_\mu u_{i,p})(\overline{d^{\C}_{j,r} }\gamma^\mu d_{k,s})\phi $ & $\frac{1}{2} n_u  n_d(n_d+1)$ &
 		$\calO^{\tt VA}_{\phi\chi u dd}$ & 
 		$\epsilon^{ijk}(\overline{\chi}\gamma_\mu u_{i,p})(\overline{d^{\C}_{j,r} }\gamma^\mu\gamma_5 d_{k,s})\phi $  & $\frac{1}{2} n_u  n_d(n_d-1)$
 		\\
 		$\calO^{\tt AV}_{\phi\chi u dd}$ & 
 		$\epsilon^{ijk}(\overline{\chi}\gamma_\mu \gamma_5 u_{i,p})(\overline{d^{\C}_{j,r} }\gamma^\mu d_{k,s})\phi $ & $\frac{1}{2} n_u  n_d(n_d+1)$ &
 		$\calO^{\tt AA}_{\phi\chi u dd}$ & 
 		$\epsilon^{ijk}(\overline{\chi}\gamma_\mu \gamma_5 u_{i,p})(\overline{d^{\C}_{j,r} }\gamma^\mu\gamma_5 d_{k,s})\phi $ & $\frac{1}{2} n_u  n_d(n_d-1)$
 		\\
 		$\calO^{\tt T1}_{\phi\chi u dd}$ & 
 		$\epsilon^{ijk}(\overline{\chi}\sigma_{\mu\nu} u_{i,p})(\overline{d^{\C}_{j,r} }\sigma^{\mu\nu} d_{k,s})\phi $ & $\frac{1}{2} n_u  n_d(n_d+1)$ &
 		$\calO^{\tt T2}_{\phi\chi u dd}$ & 
 		$\epsilon^{ijk}(\overline{\chi}\sigma_{\mu\nu}\gamma_5 u_{i,p})(\overline{d^{\C}_{j,r} }\sigma^{\mu\nu}  d_{k,s})\phi $ & $\frac{1}{2}n_u n_d(n_d+1)$
 		\\
 		\hline
 	\end{tabular}}
 	\caption{Operator basis up to dim 7 for scalar-fermion-SM. Here $\chi$ is Majorana fermion and $\phi$ is real scalar in dark sector. All operators listed are non-hermitian. The hermitian conjugate operators are indicated as $+\hc$ in the operator classifications, although they are not shown in explicit forms. The counts provided for the operators include only the displayed half of the operators. $\slashed{B}$ means the operators in the sector violate baryon number by one unit.}
 	\label{tab:chiphi}
 \end{table}

In this section, we consider the interactions involving one dark fermion $\chi$ and one dark scalar $\phi$. As both $\chi$ and $\phi$ are singlets under the gauge group $G$, the possibilities for $G$-neutral operators are quite limited. We note first that $\chi$ has to form a bilinear with a SM fermion. We start with the case $(\overline{{\nu}_{L}}\Gamma\chi)$. Attaching $\phi$ to it gives a unique dim-4 Yukawa type of interaction. To achieve a potential dim-5 operator, at most a derivative can be attached but this is reducible by EoMs. At dim 6 the only possibility is to introduce a photon field tensor, $(\overline{{\nu}_{L}}\sigma_{\mu\nu}\chi)\phi F^{\mu\nu}$. Finally, at dim 7, we can attach either a SM fermion bilinear or pure bosonic factors. The former case $(\overline{\Psi}\Gamma'\Psi)(\overline{{\nu}_{L}}\Gamma\chi)\phi$ is easy to exhaust. In the latter case, since the only available bosonic factor is the photon field strength tensor, we have to attach a derivative resulting in the form $(\overline{{\nu}_{L}}\Gamma\chi)\phi FD$. Lorentz invariance restricts the bilinear to the structure $(\overline{\nu_L}\gamma_{\mu}\chi)$. While operators with the derivative acting on $F^{\mu\nu}$ are reducible by the photon's EoM, the one with the derivative acting on $\tilde{F}^{\mu\nu}$ vanishes because of the BI in Eq.\,\eqref{eq:BI}. When the derivative acts on the bilinear, we proceed as follows:
\begin{eqnarray}
	\nonumber
	(\overline{{\nu}_{L,p}}\gamma_{[\mu} i\overleftrightarrow{\partial_{\nu]}}\chi)\phi F^{\mu\nu}
	&\overset{\tt GI}{=}&\left(\overline{\nu_{L,p}}\left[(\gamma_\mu\gamma_\nu i\slashed{\partial}-i\overleftarrow{\slashed{\partial}}\gamma_\nu\gamma_\mu)+(\gamma_\mu i\slashed{\partial}\gamma_\nu+\gamma_\mu i\overleftarrow{\slashed{\partial}}\gamma_\nu)\right]\chi\right)\phi F^{\mu\nu}
	\\
	\nonumber
	&\overset{\tt EoM}{=}& i\partial^\alpha(\overline{\nu_{L,p}}\gamma_\mu\gamma_\alpha\gamma_\nu\chi)\phi F^{\mu\nu}+\fbox{EoM}
	\\
	&\overset{\tt GI,BI}{\underset{\tt IBP}{=}}&-2(\overline{\nu_{L,p}}\gamma^\mu\chi)\partial^\nu\phi\tilde{F}_{\mu\nu}+\fbox{EoM}+\fbox{T}.
\end{eqnarray}
In a similar manner the operator $(\overline{{\nu}_{L,p}}\gamma_{[\mu} i\overleftrightarrow{\partial_{\nu]}}\chi)\phi \tilde{F}^{\mu\nu}$ can also be reduced. We thus choose the following two operators in our basis,
\begin{eqnarray}
    \calO_{\chi\nu\phi FD1}^{\tt V}=(\overline{\nu_{L,p}} \gamma_\mu \chi) F^{\mu\nu} \partial_\nu \phi,\qquad \calO_{\chi\nu\phi FD2}^{\tt V}=(\overline{\nu_{L,p}} \gamma_\mu \chi)  \tilde  F^{\mu\nu}\partial_\nu \phi.
\end{eqnarray}

Now we consider the case when $\chi$ forms a bilinear with a charged lepton or quark. The first one can only be multiplied by a $\overline{d}u$ combination due to charge conservation, i.e., $(\overline{d_p}\Gamma u_r)(\overline\chi\Gamma^\prime \ell_s)\phi$. These interactions induce semi-invisible charged lepton number violating processes, which are strongly constrained. When $\chi$ couples to quarks, complete gauge symmetry allows only the structure, $\epsilon^{ijk}(\overline{\chi}\Gamma u_{i,p})(\overline{d^\C_{j,r}}\Gamma^\prime d_{k,s})\phi$, where $i,j,k$ are color indices. These operators exhibit exact symmetry or antisymmetry in the down-type quark flavors, and thus antisymmetric operators vanish when restricted to a single generation. These operators can result in invisible baryon number violating (BNV) processes, for instance, the neutron dark decay, which could potentially solve the longstanding neutron lifetime discrepancy. We will present a detailed phenomenological analysis in the following section. The complete operator basis for the scalar-fermion-SM case is tabulated in Table\,\ref{tab:chiphi}.

\subsection{Scalar-vector-SM case}

 The interactions between light SM particles and the dark scalar $\phi$ and vector $X_\mu$ are explored in this section. We start by considering the interactions between dark particles with SM fermions. The lowest dimensional operators are of dim 5, i.e., $(\overline\Psi \Gamma \Psi) X\phi$, where $\Gamma=\gamma^\mu(\gamma_5)$. With the inclusion of derivatives, $(\overline\Psi \Psi)X\phi D$-type and $(\overline\Psi \Psi)X\phi D^2$-type operators emerge at dim 6 and 7, respectively. When dealing with dim-6 operators, with a scalar bilinear, the derivative can only be positioned on either the dark scalar field or the bilinear, because of the physical condition $\partial_\mu X^\mu=0$. With a tensor fermionic bilinear, a single independent operator exists, and we opt for the derivative on the dark vector to form the field tensor $X^{\mu\nu}$. In summary, the independent operators of the type $(\overline\Psi \Psi)X\phi  D$ are as follows:
\begin{subequations}
	\begin{align}
		\calO_{\Psi X \phi D 1}^{{\tt S},pr}&=(\overline{\Psi_p}\Psi_r)X^\mu (\partial_\mu \phi),& \calO_{\Psi X \phi D 1}^{{\tt P},pr}&=(\overline{\Psi_p} i\gamma_5 \Psi_r) X^\mu (\partial_\mu\phi),
		\\
		\calO_{\Psi  X\phi  D 2}^{{\tt S},pr}&=(\overline{\Psi_p}i\overleftrightarrow{\partial_\mu}\Psi_r)X^\mu \phi ,& \calO_{\Psi X  \phi D 2}^{{\tt P},pr}&=(\overline{\Psi_p} \gamma_5 \overleftrightarrow{\partial_\mu} \Psi_r) X^\mu\phi,
		\\
		\calO_{\Psi  X \phi D }^{{\tt T1},pr}&=(\overline{\Psi_p}\sigma_{\mu\nu}\Psi_r) X^{\mu\nu}\phi,& \calO_{\Psi X  \phi  D}^{{\tt T2},pr}&=(\overline{\Psi_p} \sigma_{\mu\nu}\gamma_5 \Psi_r) X^{\mu\nu} \phi.
	\end{align}
\end{subequations}
The dim-7 $(\overline{\Psi}\Psi)X\phi  D^2$-type operators are also divided into two classes. The operators involving a dark field strength tensor read:
\begin{subequations}
    \begin{align}
        \calO_{\Psi X\phi D^2 1}^{\tt V}&=(\overline{\Psi}\gamma_\mu\Psi) X^{\mu\nu} \partial_\nu\phi,& \calO_{\Psi X\phi D^2 2}^{\tt V}&=(\overline{\Psi}\gamma_\mu\Psi) \Tilde{X}^{\mu\nu} \partial_\nu\phi,
        \\
        \calO_{\Psi X\phi D^2 1}^{\tt A}&=(\overline{\Psi}\gamma_\mu\gamma_5\Psi) X^{\mu\nu} \partial_\nu\phi,& \calO_{\Psi X\phi D^2 2}^{\tt A}&=(\overline{\Psi}\gamma_\mu\gamma_5\Psi) \Tilde{X}^{\mu\nu} \partial_\nu\phi.
    \end{align}
\end{subequations}
The others are
\begin{subequations}
    \begin{align}
        \calO_{\Psi X\phi D^2 3}^{\tt V}&=(\overline{\Psi}\gamma_\mu\Psi)\partial_\nu\phi\partial^{(\mu} X^{\nu)},& \calO_{\Psi X\phi D^2 3}^{\tt A}&=(\overline{\Psi}\gamma_\mu\gamma_5\Psi)\partial_\nu\phi\partial^{(\mu} X^{\nu)},
        \\
		\calO_{\Psi X\phi D^2 4}^{\tt V}&=(\overline{\Psi_p}\gamma_\mu i\overleftrightarrow{\partial_\nu}\Psi)\phi \partial^{(\mu} X^{\nu)},&
		\calO_{\Psi X\phi D^2 4}^{\tt A}&=(\overline{\Psi}\gamma_\mu\gamma_5 i\overleftrightarrow{\partial_\nu}\Psi)\phi \partial^{(\mu} X^{\nu)},
    \end{align}
\end{subequations}
where the notation of symmetrization of indices is used, $A^{(\mu} B^{\nu)}= A^\mu B^\nu+A^\nu B^\mu$. Moreover, at dim 7, the two derivatives can be replaced by the field strength tensor $F^{\mu\nu}$ or $G^{A,\mu\nu}$ to form operators $(\overline\Psi\Gamma \Psi)FX\phi$ or $(\overline\Psi \Gamma \Psi)GX\phi$ as shown in Table \ref{tab:Xphi:2}.

\begin{table}[!h]
	\center
	\resizebox{\linewidth}{!}{
	\renewcommand\arraystretch{0.9}
	\begin{tabular}{| l | l | l| l | l| l| }
		\hline
		Operator & Specific form & \# & Operator & Specific form & \#
		\\
		\hline
		\multicolumn{6}{|c|}{\cellcolor{magenta!25}dim-5: $FX\phi D$}
		\\
		\hline
		\cellcolor{green!25} $\calO_{FX\phi D1}$ & $F^{\mu\nu}X_{\mu\nu}\phi$ & 1 &
		\cellcolor{green!25} $\calO_{FX\phi D2}$ & $\tilde F^{\mu\nu}X_{\mu\nu}\phi$ & 1
		\\
		\hline
		\multicolumn{6}{|c|}{\cellcolor{magenta!25}dim-5: $(\overline\Psi \Psi)X\phi$}
		\\
		\hline
		$\calO_{\nu X\phi}^{\tt V}$ & $(\overline{\nu_{L,p}}\gamma_\mu\nu_{L,r})X^\mu\phi$ & $n_\nu^2$ &  & &
		\\
		$\calO_{\ell X\phi}^{\tt V}$ & $(\overline{\ell_p}\gamma_\mu\ell_r)X^\mu\phi$ & $n_\ell^2$ & $\calO_{\ell X\phi}^{\tt A}$ & $(\overline{\ell_p}\gamma_\mu\gamma_5\ell_r)X^\mu\phi$ & $n_\ell^2$
		\\
		$\calO_{uX\phi}^{\tt V}$ & $(\overline{u_p}\gamma_\mu u_r)X^\mu\phi$ & $n_u^2$ &
		$\calO_{uX\phi}^{\tt A}$ & $(\overline{u_p}\gamma_\mu\gamma_5 u_r)X^\mu\phi$ & $n_u^2$
		\\
		$\calO_{dX\phi}^{\tt V}$ & $(\overline{d_p}\gamma_\mu d_r)X^\mu\phi$ & $n_d^2$ &
		$\calO_{dX\phi}^{\tt A}$ & $(\overline{d_p}\gamma_\mu\gamma_5 d_r)X^\mu\phi$ & $n_d^2$
		\\
		\hline
		\multicolumn{6}{|c|}{\cellcolor{magenta!25}dim-6: $(\overline\Psi \Psi)X \phi  D$}
		\\
		\hline
		{\cellcolor{green!25}$\calO_{\nu X\phi D}^{\tt T}+\hc$} & 
		$(\overline{\nu_{L,p}^\C} \sigma_{\mu\nu} \nu_{L,r})X^{\mu\nu} \phi $ & $n_\nu(n_\nu-1)$  & & &
        \\
		{\cellcolor{green!25}$\calO_{\ell X\phi D }^{\tt T1}$} & 
		$(\overline{\ell_p} \sigma_{\mu\nu} \ell_r) X^{\mu\nu} \phi$ & $n_\ell^2$ &
		{\cellcolor{green!25}$\calO_{\ell X\phi D}^{\tt T2}$} &
		$(\overline{\ell_p} \sigma_{\mu\nu} i\gamma_5 \ell_r)X^{\mu\nu} \phi $ & $n_\ell^2$
        \\
		{\cellcolor{green!25}$\calO_{uX\phi D}^{\tt T1}$} & 
		$(\overline{u_p} \sigma_{\mu\nu} u_r)X^{\mu\nu} \phi $ & $n_u^2$  &
		{\cellcolor{green!25}$\calO_{uX\phi D}^{\tt T2}$} &
		$(\overline{u_p} \sigma_{\mu\nu} i\gamma_5 u_r) X^{\mu\nu} \phi$ & $n_u^2$
        \\
		{\cellcolor{green!25}$\calO_{dX\phi D}^{\tt T1}$} & 
		$(\overline{d_p} \sigma_{\mu\nu} d_r)X^{\mu\nu} \phi $ & $n_d^2$ &
		{\cellcolor{green!25}$\calO_{dX\phi D}^{\tt T2}$} &
		$(\overline{d_p} \sigma_{\mu\nu}i \gamma_5 d_r) X^{\mu\nu} \phi$ & $n_d^2$
        \\
		$\calO_{\nu X\phi D 1}^{\tt S}+\hc$ & $(\overline{\nu_{L,p}^\C} \nu_{L,r})X^\mu \partial_\mu\phi $ & $n_\nu(n_\nu+1)$ &
		$\calO_{\nu X\phi D 2}^{\tt S}+\hc$ & $(\overline{\nu_{L,p}^\C} i\overleftrightarrow{\partial_\mu}\nu_{L,r} ) X^\mu \phi$ & $n_\nu(n_\nu-1)$ 
		\\
		$\calO_{\ell X\phi D 1}^{\tt S}$ & $(\overline{\ell_p} \ell_r)X^\mu \partial_\mu\phi $ & $n_\ell^2$ &
		$\calO_{\ell X\phi D1}^{\tt P}$ & $(\overline{\ell_p} i\gamma_5 \ell_r) X^\mu \partial_\mu\phi$ & $n_\ell^2$
		\\
		$\calO_{\ell X\phi D2}^{\tt S}$ &
		$(\overline{\ell_p} i\overleftrightarrow{\partial_\mu}\ell_r) X^\mu \phi$ &  $n_\ell^2$ &
		$\calO_{\ell X\phi D2}^{\tt P}$ & 
		$(\overline{\ell_p} \gamma_5 \overleftrightarrow{\partial_\mu}\ell_r)X^\mu \phi $ & $n_\ell^2$
		\\
		$\calO_{uX\phi D 1}^{\tt S}$ & $(\overline{u_p} u_r) X^\mu \partial_\mu\phi$ & $n_u^2$  &
		$\calO_{uX\phi D 1}^{\tt P}$ & $(\overline{u_p} i\gamma_5 u_r) X^\mu \partial_\mu\phi$ & $n_u^2$
		\\
		$\calO_{uX\phi D 2}^{\tt S}$ &
		$(\overline{u_p} i\overleftrightarrow{\partial_\mu}u_r)X^\mu \phi $ & $n_u^2$ &
		$\calO_{uX\phi D 2}^{\tt P}$ & 
		$(\overline{u_p} \gamma_5 \overleftrightarrow{\partial_\mu}u_r)X^\mu\phi$ & $n_u^2$
		\\
		$\calO_{dX\phi D 1}^{\tt S}$ & $(\overline{d_p} d_r)X^\mu \partial_\mu\phi $ & $n_d^2$ &
		$\calO_{dX\phi D 1}^{\tt P}$ & $(\overline{d_p} i\gamma_5 d_r) X^\mu \partial_\mu\phi$ & $n_d^2$
		\\
		$\calO_{ dX\phi D 2}^{\tt S}$ &
		$(\overline{d_p} i\overleftrightarrow{\partial_\mu}d_r) X^\mu \phi$ & $n_d^2$ &
		$\calO_{dX\phi D 2}^{\tt P}$ & 
		$(\overline{d_p} \gamma_5 \overleftrightarrow{\partial_\mu}d_r) X^\mu \phi$ & $n_d^2$
		\\
		\hline
		\multicolumn{6}{|c|}{\cellcolor{magenta!25}dim-7: $X\phi F^2 D,\, X\phi G^2 D$}
		\\
		\hline
		$\calO_{F^2X\phi D1}$ & $F_{\mu\nu} F^{\mu\nu} X_\rho \partial^\rho \phi$ &  &
		$\calO_{F^2X\phi D2}$ & $F_{\mu\nu} \tilde F^{\mu\nu} X_\rho \partial^\rho \phi$ &
		\\
		$\calO_{F^2X\phi D3}$ & $F^{\mu\nu} F_{\nu\rho} X_\mu \partial^\rho \phi$ & & & &
		\\
		$\calO_{G^2 X\phi D1}$ & $G^A_{\mu\nu} \tilde G^{A\mu\nu} X_\rho \partial^\rho \phi$ & &
		$\calO_{G^2X\phi D2}$ & $G^A_{\mu\nu} \tilde G^{A\mu\nu} X_\rho \partial^\rho \phi$ &
		\\
		$\calO_{G^2X\phi D3}$ & $G^{A,\mu\nu} G_{\nu\rho}^{A} X_\mu \partial^\rho \phi$ & & & &
		\\
		\hline
	\end{tabular}}
	\caption{Operator basis up to dim 7 for the scalar-vector-SM case. Here $\phi$ and $X_\mu$ are respectively a real scalar and vector in the dark sector. For non-hermitian operators, we have used $+\hc$ to include their hermitian conjugations and also counted the $\hc$ parts. We have highlighted the operators involving the dark field tensor $X^{\mu\nu}$.}
	\label{tab:Xphi:1}
\end{table}

\begin{table}[!h]
	\center
		\renewcommand\arraystretch{0.95}
		\begin{tabular}{| l | l | l| l | l| l| }
			\hline
			Operator & Specific form & \# & Operator & Specific form & \#
			\\
			\hline
			\multicolumn{6}{|c|}{\cellcolor{magenta!25}dim-7: $(\overline\Psi \Psi)FX\phi$, $(\overline\Psi \Psi)GX\phi$ }
			\\
			\hline
			$\calO_{\nu FX\phi 1}^{\tt V}$ & $(\overline{\nu_{L,p}}\gamma_\mu\nu_{L,r})F^{\mu\nu}X_\nu\phi$ & $n_\nu^2$ & $\calO_{\nu FX\phi 2}^{\tt V}$ & $(\overline{\nu_{L,p}}\gamma_\mu\nu_{L,r})\tilde F^{\mu\nu}X_\nu\phi$ & $n_\nu^2$
			\\
			$\calO_{\ell FX\phi 1}^{\tt V}$ & $(\overline{\ell_p}\gamma_\mu\ell_r)F^{\mu\nu}X_\nu\phi$ & $n_\ell^2$ & $\calO_{\ell FX\phi 2}^{\tt V}$ & $(\overline{\ell_p}\gamma_\mu\ell_r)\tilde F^{\mu\nu}X_\nu\phi$ & $n_\ell^2$
			\\
			$\calO_{\ell FX\phi 1}^{\tt A}$ & $(\overline{\ell_p}\gamma_\mu\gamma_5\ell_r)F^{\mu\nu}X_\nu\phi$ & $n_\ell^2$ &
			$\calO_{\ell FX\phi 2}^{\tt A}$ & $(\overline{\ell_p}\gamma_\mu\gamma_5\ell_r)\tilde F^{\mu\nu}X_\nu\phi$ & $n_\ell^2$
			\\
			$\calO_{uFX\phi 1}^{\tt V}$ & $(\overline{u_p}\gamma_\mu u_r)F^{\mu\nu}X_\nu\phi$ & $n_u^2$ &
			$\calO_{uFX\phi 2}^{\tt V}$ & $(\overline{u_p}\gamma_\mu u_r)\tilde F^{\mu\nu}X_\nu\phi$ & $n_u^2$ 
			\\
			$\calO_{uFX\phi 1}^{\tt A}$ & $(\overline{u_p}\gamma_\mu\gamma_5 u_r)F^{\mu\nu}X_\nu\phi$ & $n_u^2$  &  $\calO_{uFX\phi 2}^{\tt A}$ & $(\overline{u_p}\gamma_\mu\gamma_5 u_r)\tilde F^{\mu\nu}X_\nu\phi$ & $n_u^2$ 
			\\
			$\calO_{dFX\phi 1}^{\tt V}$ & $(\overline{d_p}\gamma_\mu d_r)F^{\mu\nu}X_\nu\phi$ & $n_d^2$  & 
			$\calO_{dFX\phi 2}^{\tt V}$ & $(\overline{d_p}\gamma_\mu d_r)\tilde F^{\mu\nu}X_\nu\phi$ & $n_d^2$
			\\
			$\calO_{dFX\phi 1}^{\tt A}$ & $(\overline{d_p}\gamma_\mu\gamma_5 d_r)F^{\mu\nu}X_\nu\phi$ & $n_d^2$ & $\calO_{dFX\phi 2}^{\tt A}$ & $(\overline{d_p}\gamma_\mu\gamma_5 d_r)\tilde F^{\mu\nu}X_\nu\phi$ & $n_d^2$
			\\
			$\calO_{uGX\phi 1}^{\tt V}$ & $(\overline{u_p}\gamma_\mu T^A u_r)G^{A,\mu\nu}X_\nu\phi$ & $n_u^2$ &   $\calO_{uGX\phi 2}^{\tt V}$ & $(\overline{u_p}\gamma_\mu T^A u_r)\tilde G^{A,\mu\nu}X_\nu\phi$ & $n_u^2$
			\\
			$\calO_{uGX\phi 1}^{\tt A}$ & $(\overline{u_p}\gamma_\mu\gamma_5 T^A u_r)G^{A,\mu\nu}X_\nu\phi$ & $n_u^2$ &  $\calO_{uGX\phi 2}^{\tt A}$ & $(\overline{u_p}\gamma_\mu\gamma_5 T^A u_r)\tilde G^{A,\mu\nu}X_\nu\phi$ & $n_u^2$
			\\
			$\calO_{dGX\phi 1}^{\tt V}$ & $(\overline{d_p}\gamma_\mu T^A d_r)G^{A,\mu\nu}X_\nu\phi$ & $n_d^2$ &  $\calO_{dGX\phi 2}^{\tt V}$ & $(\overline{d_p}\gamma_\mu T^A d_r)\tilde G^{A,\mu\nu}X_\nu\phi$ & $n_d^2$
			\\
			$\calO_{dGX\phi 1}^{\tt A}$ & $(\overline{d_p}\gamma_\mu\gamma_5  T^A d_r)G^{A,\mu\nu}X_\nu\phi$ & $n_d^2$ & $\calO_{dGX\phi 2}^{\tt A}$ & $(\overline{d_p}\gamma_\mu\gamma_5 T^A d_r)\tilde G^{A,\mu\nu}X_\nu\phi$ & $n_d^2$
			\\
			\hline
			\multicolumn{6}{|c|}{\cellcolor{magenta!25}dim-7: $(\overline\Psi \Psi)X\phi D^2$}
			\\
			\hline
            {\cellcolor{green!25}$\calO_{\nu X\phi D^2 1}^{\tt V}$} & $(\overline{\nu_{L,p}}\gamma_\mu\nu_{L,r}) X^{\mu\nu} \partial_\nu\phi$ & $n_\nu^2$ &
			{\cellcolor{green!25}$\calO_{\nu X\phi D^2 2}^{\tt V}$} & $(\overline{\nu_{L,p}}\gamma_\mu\nu_{L,r}) \Tilde{X}^{\mu\nu} \partial_\nu\phi$ & $n_\nu^2$
			\\
			{\cellcolor{green!25}$\calO_{\ell X\phi D^2 1}^{\tt V}$} & $(\overline{\ell_p}\gamma_\mu\ell_r) X^{\mu\nu} \partial_\nu\phi$ & $n_\ell^2$ &
			{\cellcolor{green!25}$\calO_{\ell X\phi D^2 2}^{\tt V}$}
			& $(\overline{\ell_p}\gamma_\mu\ell_r) \Tilde{X}^{\mu\nu} \partial_\nu\phi$ & $n_\ell^2$
			\\
			{\cellcolor{green!25}$\calO_{\ell X\phi D^2 1}^{\tt A}$} & 
			$(\overline{\ell_p}\gamma_\mu\gamma_5\ell_r) X^{\mu\nu} \partial_\nu\phi$ & $n_\ell^2$ &
			{\cellcolor{green!25}$\calO_{\ell X\phi D^2 2}^{\tt A}$} & $(\overline{\ell_p}\gamma_\mu\gamma_5\ell_r) \Tilde{X}^{\mu\nu} \partial_\nu\phi$ & $n_\ell^2$
			\\
			{\cellcolor{green!25}$\calO_{uX\phi D^2 1}^{\tt V}$} & $(\overline{u_p}\gamma_\mu u_r) X^{\mu\nu} \partial_\nu\phi$ & $n_u^2$ &
			{\cellcolor{green!25}$\calO_{uX\phi D^2 2}^{\tt V}$}
			& $(\overline{u_p}\gamma_\mu u_r) \Tilde{X}^{\mu\nu} \partial_\nu\phi$ & $n_u^2$
			\\
			{\cellcolor{green!25}$\calO_{uX\phi D^2 1}^{\tt A}$} & 
			$(\overline{u_p}\gamma_\mu\gamma_5 u_r) X^{\mu\nu} \partial_\nu\phi$ & $n_u^2$  &
			{\cellcolor{green!25}$\calO_{uX\phi D^2 2}^{\tt A}$} & $(\overline{u_p}\gamma_\mu\gamma_5 u_r) \Tilde{X}^{\mu\nu} \partial_\nu\phi$ & $n_u^2$
			\\
			{\cellcolor{green!25}$\calO_{dX\phi D^2 1}^{\tt V}$} & $(\overline{d_p}\gamma_\mu d_r) X^{\mu\nu} \partial_\nu\phi$ & $n_d^2$ &
			{\cellcolor{green!25}$\calO_{dX\phi D^2 2}^{\tt V}$}
			& $(\overline{d_p}\gamma_\mu d_r) \Tilde{X}^{\mu\nu} \partial_\nu\phi$ & $n_d^2$
			\\
			{\cellcolor{green!25}$\calO_{dX\phi D^2 1}^{\tt A}$} & 
			$(\overline{d_p}\gamma_\mu\gamma_5 d_r) X^{\mu\nu} \partial_\nu\phi$ & $n_d^2$ &
			{\cellcolor{green!25}$\calO_{dX\phi D^2 2}^{\tt A}$} & $(\overline{d_p}\gamma_\mu\gamma_5 d_r) \Tilde{X}^{\mu\nu} \partial_\nu\phi$ & $n_d^2$
			\\
			$\calO_{\nu X\phi D^2 3}^{\tt V}$ & $(\overline{\nu_{L,p}}\gamma_\mu\nu_{L,r})\partial_\nu\phi \partial^{(\mu} X^{\nu)}$ & $n_\nu^2$ &
			$\calO_{\nu X\phi D^2 4}^{\tt V}$ & $(\overline{\nu_{L,p}}\gamma_\mu i\overleftrightarrow{\partial_\nu}\nu_{L,r})\phi \partial^{(\mu} X^{\nu)}$  & $n_\nu^2$
			\\
			$\calO_{\ell X\phi D^2 3}^{\tt V}$ & $(\overline{\ell_p}\gamma_\mu\ell_r)\partial_\nu\phi\partial^{(\mu} X^{\nu)}$ & $n_\ell^2$ & $\calO_{\ell X\phi D^2 3}^{\tt A}$ &  $(\overline{\ell_p}\gamma_\mu\gamma_5\ell_r)\partial_\nu\phi\partial^{(\mu} X^{\nu)}$ & $n_\ell^2$ 
			\\
			$\calO_{\ell X\phi D^2 4}^{\tt V}$ & $(\overline{\ell_p}\gamma_\mu i\overleftrightarrow{\partial_\nu}\ell_r)\phi \partial^{(\mu} X^{\nu)}$ & $n_\ell^2$  &
			$\calO_{\ell X\phi D^2 4}^{\tt A}$ & $(\overline{\ell_p}\gamma_\mu\gamma_5 i\overleftrightarrow{\partial_\nu}\ell_r)\phi \partial^{(\mu} X^{\nu)}$ & $n_\ell^2$ 
			\\
			$\calO_{uX\phi D^2 3}^{\tt V}$ & $(\overline{u_p}\gamma_\mu u_r)\partial_\nu\phi \partial^{(\mu} X^{\nu)}$ & $n_u^2$ &
			$\calO_{uX\phi D^2 3}^{\tt A}$ & $(\overline{u_p}\gamma_\mu\gamma_5 u_r)\partial_\nu\phi \partial^{(\mu} X^{\nu)}$ & $n_u^2$ 
			\\
			$\calO_{uX\phi D^2 4}^{\tt V}$ & $(\overline{u_p}\gamma_\mu i\overleftrightarrow{\partial_\nu} u_r)\phi \partial^{(\mu} X^{\nu)}$ & $n_u^2$   &
			$\calO_{uX\phi D^2 4}^{\tt A}$ & $(\overline{u_p}\gamma_\mu\gamma_5 i \overleftrightarrow{\partial_\nu} u_r)\phi \partial^{(\mu} X^{\nu)}$ & $n_u^2$ 
			\\
			$\calO_{dX\phi D^2 3}^{\tt V}$ & $(\overline{d_p}\gamma_\mu d_r)\partial_\nu\phi \partial^{(\mu} X^{\nu)}$ & $n_d^2$ &
			$\calO_{dX\phi D^2 3}^{\tt A}$ & $(\overline{d_p}\gamma_\mu\gamma_5 d_r)\partial_\nu\phi \partial^{(\mu} X^{\nu)}$ & $n_d^2$
			\\
			$\calO_{dX\phi D^2 4}^{\tt V}$ & $(\overline{d_p}\gamma_\mu i\overleftrightarrow{\partial_\nu} d_r)\phi \partial^{(\mu} X^{\nu)}$ & $n_d^2$ &
			$\calO_{dX\phi D^2 4}^{\tt A}$ & $(\overline{d_p}\gamma_\mu\gamma_5 i\overleftrightarrow{\partial_\nu} d_r)\phi \partial^{(\mu} X^{\nu)}$ & $n_d^2$
			\\
			\hline
	\end{tabular}
	\caption{Continuation of Table \ref{tab:Xphi:1}.}
	\label{tab:Xphi:2}
\end{table}

We now direct our attention towards pure bosonic interactions between dark particles and gauge bosons. The lowest dimensional contributions appear at dim 5,
\begin{align}
	\calO_{FX\phi D1}=F^{\mu\nu}X_{\mu\nu}\phi, \qquad\qquad\calO_{FX\phi D2}= \tilde{F}^{\mu\nu}X_{\mu\nu}\phi.
\end{align}
Restricted by Lorentz invariance, there is no dim-6 bosonic operator with the form $FX\phi D^2$.  We find the dim-7 $FX\phi D^3$-type operators are reducible. Thus, two types of dim-7 operators survive: $X\phi FFD$ and $X\phi GGD$. As an illustrative example, we consider the class $X\phi FFD$ to confirm independence of our basis. First, Lorentz symmetry constrains such operators to the following possibilities,
\begin{eqnarray}
    \label{eq:FFphiXD}
	F^{\mu\nu}F_{\mu\nu}\phi X^\rho D_\rho,\qquad \tilde{F}^{\mu\nu}F_{\mu\nu}\phi X^\rho D_\rho,\qquad F^{\mu\nu}F_{\nu\rho}\phi X_\mu D^\rho,\qquad \tilde{F}^{\mu\nu}F_{\nu\rho}\phi X_\mu D^\rho.
\end{eqnarray}
To reduce to the above possibilities, we have taken the following relations into account:
\begin{eqnarray}
\epsilon_{\mu \nu \rho \sigma}\epsilon^{\rho\sigma\alpha\beta}
=-2 g_\mu^{[\alpha} g_\nu^{\beta]}, \quad\quad
\epsilon_{\mu \nu \rho \sigma} \epsilon^{\alpha \beta \gamma \sigma}= -g_\mu^{[\alpha} g_\nu^\beta g_\rho^{\gamma]}.
\label{eq:epsi}
\end{eqnarray}
For the first two cases, we can require both $F^{\mu\nu}$ and $\tilde{F}^{\mu\nu}$ to be derivative free upon using IBP, and in consideration of $\partial_\rho  X^\rho=0$, the derivative can only act on the dark scalar $\phi$. The resulting operators are
\begin{eqnarray}
	\calO_{FX\phi D1}=F_{\mu\nu} F^{\mu\nu} X_\rho \partial^\rho \phi,\qquad
	\calO_{FX\phi D2}=F_{\mu\nu} \tilde F^{\mu\nu} X_\rho \partial^\rho \phi.
\end{eqnarray}
For the last two cases in Eq.\,\eqref{eq:FFphiXD}, we require the first field tensor $F^{\mu\nu}$ or $\tilde F^{\mu\nu}$ to be derivative free by using IBP. And the operators with a derivative on $F_{\nu\rho}$ are reducible due to the EoM, $\partial^\rho F_{\nu\rho}$. Then there are four potential options for the last two cases with a derivative acting on dark scalar or vector. But actually we found there is only one independent operator, which is chosen as
\begin{eqnarray}
	\calO_{FX\phi D3}&=F^{\mu\nu} F_{\nu\rho} X_\mu \partial^\rho \phi.
\end{eqnarray}
The other possibilities are redundant and can be expressed in terms of the other basis operators:
\begin{subequations}
	\begin{eqnarray}
		F^{\mu\nu} F_{\nu\rho} (\partial^\rho X_\mu)\phi&=&-\frac{1}{4}\calO_{FX\phi D1}-\calO_{FX\phi D3}+\fbox{EoM},
		\\
		\tilde{F}^{\mu\nu} F_{\nu\rho} X_\mu \partial^\rho \phi&=&-\frac{1}{4}\calO_{FX\phi D2},
		\\
		\tilde{F}^{\mu\nu} F_{\nu\rho} (\partial^\rho X_\mu)\phi&=& \fbox{EoM}.
	\end{eqnarray}
\end{subequations}
To derive the above relations, we have used Eq.\,\eqref{eq:epsi}. The complete operator basis for the scalar-vector-SM scenario is presented in Table \ref{tab:Xphi:1} and Table \ref{tab:Xphi:2}.

\subsection{Fermion-vector-SM case}

\begin{table}[!h]
	\center
	\resizebox{\linewidth}{!}{
		\renewcommand\arraystretch{1.0}
		\begin{tabular}{| l | l | l| l | l| l|}
			\hline
			Operator & Specific form & \# & Operator & Specific form & \#
			\\
			\hline
			\hline
			\multicolumn{6}{|c|}{\cellcolor{magenta!25}dim-4: $(\overline{\nu_L}\chi)X+\hc$}
			\\
			\hline
			$\calO_{ \nu \chi X}^{\tt V} $ & $(\overline{\nu_{L,p}} \gamma^\mu\chi) X_\mu$ & $n_\nu$ & &
			& 
			\\
			\hline
			\multicolumn{6}{|c|}{\cellcolor{magenta!25}dim-5: $(\overline{\nu_L}\chi)XD+\hc$}
			\\
			\hline
			{\cellcolor{green!25}$\calO_{\nu \chi X D}^{\tt T} $} & $(\overline{\nu_{L,p}}\sigma^{\mu\nu}\chi) X_{\mu\nu}$ & $n_\nu$ & & & 
			\\
			\hline
			\multicolumn{6}{|c|}{\cellcolor{magenta!25}dim-6: $(\overline{\nu_L}\chi)XF+\hc$}
			\\
			\hline
			$\calO_{\nu \chi X F1}^{\tt V}$ & $(\overline{\nu_{L,p}} \gamma^\mu\chi) X^\nu F_{\mu\nu}$ & $n_\nu$ &
			$\calO_{\nu \chi X F2}^{\tt V}$ & $(\overline{\nu_{L,p}} \gamma^\mu\chi) X^\nu \Tilde{F}_{\mu\nu}$ & $n_\nu$
			\\
			\hline
			\multicolumn{6}{|c|}{\cellcolor{magenta!25}dim-7: $(\overline{\nu_L}\chi)XFD+\hc$}
			\\
			\hline
			\cellcolor{green!25}$\calO_{\nu \chi XFD1}^{\tt S}$ & $(\overline{\nu_{L,p}}\chi) X^{\mu\nu} F_{\mu\nu}$ & $n_\nu$  &
			\cellcolor{green!25}$\calO_{\nu \chi XFD2}^{\tt S}$ & $(\overline{\nu_{L,p}}\chi) X^{\mu\nu} \tilde F_{\mu\nu}$ & $n_\nu$
			\\
			\cellcolor{green!25}$\calO_{\nu \chi XFD}^{\tt T1}$ & $(\overline{\nu_{L,p}}\sigma^{\mu\nu}\chi) X_{\mu\alpha} F^{~\alpha}_{\nu}$ & $n_\nu$  & & &
			\\
			$\calO_{\nu \chi XFD3}^{\tt S}$ & $ (\overline{\nu_{L,p}} i\overleftrightarrow{\partial_\nu}\chi) X_\mu F^{\mu\nu}$ & $n_\nu$  &
			$\calO_{\nu \chi XFD4}^{\tt S}$ & $(\overline{\nu_{L,p}} i\overleftrightarrow{\partial_\nu}\chi) X_\mu \tilde{F}^{\mu\nu}$ & $n_\nu$
			\\
			$\calO_{\nu \chi XFD}^{\tt T2}$ & $(\overline{\nu_{L,p}}\sigma_{\mu\nu} i \overleftrightarrow{\partial_\rho}\chi)X^\mu F^{\nu\rho}$ & $n_\nu$  & & &
			\\
			\hline
			\multicolumn{6}{|c|}{\cellcolor{magenta!25}dim-7: $(\overline\Psi\Psi)(\overline{\nu_L}\chi)X+\hc$}
			\\
			\hline
			$\calO_{\nu\nu \chi X}^{\tt VS}$ & $(\overline{\nu_{L,p} } \gamma_\mu \nu_{L,r}) (\overline{\nu_{L,s}}\chi) X^\mu$ & $n_\nu^3$ &
			$\calO_{\nu\nu\chi X}^{\tt SV}$ & $(\overline{\nu_{L,p}^\C } \nu_{L,r}) (\overline{\nu_{L,s}^\C}\gamma_\mu\chi) X^\mu$ & $\frac{1}{3}n_\nu(n_\nu^2-1)$
			\\        
			$\calO_{\ell\nu\chi X}^{\tt VS}$ & $(\overline{\ell_p} \gamma_\mu \ell_r) (\overline{\nu_{L,s}}\chi)X^\mu$ & $n_\ell^2 n_\nu$  &
			$\calO_{\ell\nu\chi X}^{\tt AS}$ & $(\overline{\ell_p} \gamma_\mu\gamma_5 \ell_r) (\overline{\nu_{L,s}}\chi)X^\mu$ & $n_\ell^2 n_\nu$
			\\
			$\calO_{\ell\nu\chi X}^{\tt SV}$ & $(\overline{\ell_p} \ell_r) (\overline{\nu_{L,s}}\gamma_\mu\chi)X^\mu$ & $n_\ell^2 n_\nu$ &
			$\calO_{\ell\nu\chi X}^{\tt PV}$ & $(\overline{\ell_p} i\gamma_5 \ell_r) (\overline{\nu_{L,s}}\gamma_\mu\chi)X^\mu$ & $n_\ell^2 n_\nu$
			\\
			$\calO_{\ell\nu\chi X}^{\tt T1V}$ & $(\overline{\ell_p} \sigma_{\mu\nu}\ell_r) (\overline{\nu_{L,s}}\gamma^\mu\chi)X^\nu$ & $n_\ell^2 n_\nu$ &
			$\calO_{\ell\nu\chi X}^{\tt T2V}$ & $(\overline{\ell_p} \sigma_{\mu\nu} i\gamma_5\ell_r) (\overline{\nu_{L,s}}\gamma^\mu\chi)X^\nu$ & $n_\ell^2 n_\nu$
			\\
			$\calO_{\ell\nu\chi X}^{\tt VT1}$ & $(\overline{\ell_p} \gamma_\mu\ell_r) (\overline{\nu_{L,s}}\sigma^{\mu\nu}\chi)X_\nu$ & $n_\ell^2 n_\nu$ &
			$\calO_{\ell\nu\chi X}^{\tt AT1}$ & $(\overline{\ell_p} \gamma_\mu\gamma_5\ell_r) (\overline{\nu_{L,s}}\sigma^{\mu\nu}\chi)X_\nu$ & $n_\ell^2 n_\nu$
			\\
			$\calO_{u\nu\chi X}^{\tt VS}$ & $(\overline{u_p} \gamma_\mu u_r) (\overline{\nu_{L,s}}\chi)X^\mu$ & $n_u^2 n_\nu$ &
			$\calO_{u\nu\chi X}^{\tt AS}$ & $(\overline{u_p} \gamma_\mu\gamma_5 u_r) (\overline{\nu_{L,s}}\chi)X^\mu$ & $n_u^2 n_\nu$
			\\
			$\calO_{u\nu\chi X}^{\tt SV}$ & $(\overline{u_p} u_r) (\overline{\nu_{L,s}}\gamma_\mu\chi)X^\mu$ & $n_u^2 n_\nu$  &
			$\calO_{u\nu\chi X}^{\tt PV}$ & $(\overline{u_p} i\gamma_5 u_r) (\overline{\nu_{L,s}}\gamma_\mu\chi)X^\mu$ & $n_u^2 n_\nu$
			\\
			$\calO_{u\nu\chi X}^{\tt T1V}$ & $(\overline{u_p} \sigma_{\mu\nu}u_r) (\overline{\nu_{L,s}}\gamma^\mu\chi)X^\nu$ & $n_u^2 n_\nu$ &
			$\calO_{u\nu\chi X}^{\tt T2V}$ & $(\overline{u_p} \sigma_{\mu\nu} i\gamma_5 u_r) (\overline{\nu_{L,s}}\gamma^\mu\chi)X^\nu$ & $n_u^2 n_\nu$
			\\
			$\calO_{u\nu\chi X}^{\tt VT1}$ & $(\overline{u_p} \gamma_\mu u_r) (\overline{\nu_{L,s}}\sigma^{\mu\nu}\chi)X_\nu$ & $n_u^2 n_\nu$ &
			$\calO_{u\nu\chi X}^{\tt AT1}$ & $(\overline{u_p} \gamma_\mu\gamma_5 u_r) (\overline{\nu_{L,s}}\sigma^{\mu\nu}\chi)X_\nu$ & $n_u^2 n_\nu$
			\\
			$\calO_{d\nu\chi X}^{\tt VS}$ & $(\overline{d_p} \gamma_\mu d_r) (\overline{\nu_{L,s}}\chi)X^\mu$ & $n_d^2 n_\nu$ &
			$\calO_{d\nu\chi X}^{\tt AS}$ &
			$(\overline{d_p} \gamma_\mu\gamma_5 d_r) (\overline{\nu_{L,s}}\chi)X^\mu$ & $n_d^2 n_\nu$
			\\
			$\calO_{d\nu\chi X}^{\tt SV}$ & $(\overline{d_p} d_r) (\overline{\nu_{L,s}}\gamma_\mu\chi)X^\mu$ & $n_d^2 n_\nu$ &
			$\calO_{d\nu\chi X}^{\tt PV}$ & $(\overline{d_p} i\gamma_5 d_r) (\overline{\nu_{L,s}}\gamma_\mu\chi)X^\mu$ & $n_d^2 n_\nu$
			\\
			$\calO_{d\nu\chi X}^{\tt T1V}$ & $(\overline{d_p} \sigma_{\mu\nu}d_r) (\overline{\nu_{L,s}}\gamma^\mu\chi)X^\nu$ & $n_d^2 n_\nu$ &
			$\calO_{d\nu\chi X}^{\tt T2V}$ & $(\overline{d_p} \sigma_{\mu\nu}i\gamma_5 d_r) (\overline{\nu_{L,s}}\gamma^\mu\chi)X^\nu$ & $n_d^2 n_\nu$
			\\
			$\calO_{d\nu\chi X}^{\tt VT1}$ & $(\overline{d_p} \gamma_\mu d_r) (\overline{\nu_{L,s}}\sigma^{\mu\nu}\chi)X_\nu$ & $n_d^2 n_\nu$ & $\calO_{d\nu\chi X}^{\tt AT1}$ & $(\overline{d_p} \gamma_\mu\gamma_5 d_r) (\overline{\nu_{L,s}}\sigma^{\mu\nu}\chi)X_\nu$ & $n_d^2 n_\nu$
			\\
			\hline
			\multicolumn{6}{|c|}{\cellcolor{magenta!25}dim-7: $(\overline\chi\ell)(\overline du) X+\hc$ }
			\\
			\hline
			$\calO^{\tt SV}_{\chi\ell duX}$ & $(\overline{\chi} \ell_p) (\overline{d_r} \gamma_\mu u_s) X^\mu$ & $n_\ell n_d n_u$ &
			$\calO^{\tt VS}_{\chi\ell duX}$ & $(\overline{\chi}\gamma_\mu \ell_p ) (\overline{d_r} u_s) X^\mu $ & $n_\ell n_d n_u$
			\\
			$\calO^{\tt SA}_{\chi\ell duX}$ &  $(\overline{\chi} \ell_p) (\overline{d_r} \gamma_\mu\gamma_5 u_s) X^\mu$ & $n_\ell n_d n_u$ &
			$\calO^{\tt AS}_{\chi\ell duX}$ & $(\overline{\chi}\gamma_\mu\gamma_5 \ell_p ) (\overline{d_r} u_s) X^\mu $ & $n_\ell n_d n_u$
			\\
			$\calO^{\tt PV}_{\chi\ell duX}$ & $(\overline{\chi} i\gamma_5 \ell_p) (\overline{d_r} \gamma_\mu u_s) X^\mu$ & $n_\ell n_d n_u$ &
			$\calO^{\tt VP}_{\chi\ell duX}$ & $(\overline{\chi}\gamma_\mu \ell_p ) (\overline{d_r} i\gamma_5 u_s) X^\mu $ & $n_\ell n_d n_u$
			\\
			$\calO^{\tt PA}_{\chi\ell duX}$ & $(\overline{\chi} i\gamma_5 \ell_p) (\overline{d_r} \gamma_\mu\gamma_5 u_s) X^\mu$ & $n_\ell n_d n_u$ &
			$\calO^{\tt AP}_{\chi\ell duX}$ &  $(\overline{\chi}\gamma_\mu\gamma_5 \ell_p ) (\overline{d_r} i\gamma_5 u_s) X^\mu $ & $n_\ell n_d n_u$
			\\
			$\calO^{\tt T1V}_{\chi\ell duX}$ & $(\overline{\chi}\sigma_{\mu\nu}\ell_p) (\overline{d_r} \gamma^\mu u_s) X^\nu$  & $n_\ell n_d n_u$ &
			$\calO^{\tt VT1}_{\chi\ell duX}$ & $(\overline{\chi}\gamma^\mu \ell_p) (\overline{d_r} \sigma_{\mu\nu} u_s) X^\nu$ & $n_\ell n_d n_u$
			\\
			$\calO^{\tt T1A}_{\chi\ell duX}$ & $(\overline{\chi}\sigma_{\mu\nu}\ell_p)(\overline{d_r} \gamma^\mu\gamma_5 u_s)  X^\nu$ & $n_\ell n_d n_u$ &
			$\calO^{\tt AT1}_{\chi\ell duX}$ & $(\overline{\chi}\gamma^\mu\gamma_5 \ell_p) (\overline{d_r} \sigma_{\mu\nu} u_s)X^\nu$ & $n_\ell n_d n_u$
			\\
			$\calO^{\tt T2V}_{\chi\ell duX}$ & $(\overline{\chi}\sigma_{\mu\nu} i\gamma_5\ell_p) (\overline{d_r} \gamma^\mu u_s) X^\nu$  & $n_\ell n_d n_u$ &
			$\calO^{\tt VT2}_{\chi\ell duX}$ & $(\overline{\chi}\gamma^\mu \ell_p ) (\overline{d_r} \sigma_{\mu\nu} i\gamma_5 u_s) X^\nu$ & $n_\ell n_d n_u$
			\\
			$\calO^{\tt T2A}_{\chi\ell duX}$ & $(\overline{\chi}\sigma_{\mu\nu} i\gamma_5\ell_p)
			(\overline{d_r} \gamma^\mu\gamma_5 u_s) X^\nu$  & $n_\ell n_d n_u$ &
			$\calO^{\tt AT2}_{\chi\ell duX}$ & $(\overline{\chi}\gamma^\mu\gamma_5 \ell_p) (\overline{d_r} \sigma_{\mu\nu} i\gamma_5 u_s) X^\nu$ & $n_\ell n_d n_u$
			\\
			\hline
			\multicolumn{6}{|c|}{\cellcolor{magenta!25}dim-7: $(\overline\chi u)(dd)X+\hc$ ($\slashed{B}$)}
			\\
			\hline
			$\calO^{\tt SV}_{\chi u dd X}$ & 
			$\epsilon^{ijk}(\overline{\chi} u_{i,p})(\overline{d^{\C}_{j,r} }\gamma_\mu d_{k,s})X^\mu $ & $\frac{1}{2}n_un_d(n_d-1)$ &
			$\calO^{\tt VS}_{\chi u dd X}$ & 
			$\epsilon^{ijk}(\overline{\chi}\gamma_\mu u_{i,p})(\overline{d^{\C}_{j,r} }d_{k,s})X^\mu $ & $\frac{1}{2}n_un_d(n_d+1)$
			\\
			$\calO^{\tt SA}_{\chi u dd X}$ & 
			$\epsilon^{ijk}(\overline{\chi} u_{i,p})(\overline{d^{\C}_{j,r} }\gamma_\mu\gamma_5 d_{k,s})X^\mu $ & $\frac{1}{2}n_un_d(n_d+1)$ &
			$\calO^{\tt AS}_{\chi u dd X}$ & 
			$\epsilon^{ijk}(\overline{\chi}\gamma_\mu\gamma_5 u_{i,p})(\overline{d^{\C}_{j,r} }d_{k,s})X^\mu $ & $\frac{1}{2}n_un_d(n_d+1)$
			\\
			$\calO^{\tt PV}_{ \chi u dd X}$ & 
			$\epsilon^{ijk}(\overline{\chi}i\gamma_5 u_{i,p})(\overline{d^{\C}_{j,r} }\gamma_\mu d_{k,s})X^\mu $ & $\frac{1}{2}n_un_d(n_d-1)$ &
			$\calO^{\tt VP}_{\chi u dd X}$ & 
			$\epsilon^{ijk}(\overline{\chi}\gamma_\mu u_{i,p})(\overline{d^{\C}_{j,r} }i\gamma_5 d_{k,s}) X^\mu $ & $\frac{1}{2}n_un_d(n_d+1)$
			\\
			$\calO^{\tt PA}_{\chi u dd X}$ & 
			$\epsilon^{ijk}(\overline{\chi}i\gamma_5 u_{i,p})(\overline{d^{\C}_{j,r} }\gamma_\mu \gamma_5 d_{k,s})X^\mu $ & $\frac{1}{2}n_u n_d(n_d+1)$ &
			$\calO^{\tt AP}_{\chi u dd X}$ & 
			$\epsilon^{ijk}(\overline{\chi}\gamma_\mu\gamma_5 u_{i,p})(\overline{d^{\C}_{j,r} }i\gamma_5 d_{k,s}) X^\mu $ & $\frac{1}{2}n_un_d(n_d+1)$
			\\
			$\calO^{\tt T1V}_{\chi u dd X}$ & 
			$\epsilon^{ijk}(\overline{\chi}\sigma_{\mu\nu} u_{i,p})(\overline{d^{\C}_{j,r} }\gamma^\mu d_{k,s})X^\nu $ & $\frac{1}{2}n_un_d(n_d-1)$ &
			$\calO^{\tt VT1}_{\chi u dd X}$ & 
			$\epsilon^{ijk}(\overline{\chi}\gamma^\mu u_{i,p})(\overline{d^{\C}_{j,r} }\sigma_{\mu\nu} d_{k,s})X^\nu $ & $\frac{1}{2}n_un_d(n_d-1)$
			\\
			$\calO^{\tt T1A}_{\chi u dd X}$ & 
			$\epsilon^{ijk}(\overline{\chi}\sigma_{\mu\nu} u_{i,p})(\overline{d^{\C}_{j,r} }\gamma^\mu\gamma_5 d_{k,s})X^\nu $ & $\frac{1}{2}n_un_d(n_d+1)$ &
			$\calO^{\tt AT1}_{\chi u dd X}$ & 
			$\epsilon^{ijk}(\overline{\chi}\gamma^\mu\gamma_5 u_{i,p})(\overline{d^{\C}_{j,r} }\sigma_{\mu\nu} d_{k,s})X^\nu $ & $\frac{1}{2}n_un_d(n_d-1)$
			\\
			$\calO^{\tt T2V}_{\chi u dd X}$ & 
			$\epsilon^{ijk}(\overline{\chi} \sigma_{\mu\nu} i\gamma_5 u_{i,p})(\overline{d^{\C}_{j,r} }\gamma^\mu d_{k,s})X^\nu $ & $\frac{1}{2}n_un_d(n_d-1)$ &
			$\calO^{\tt VT2}_{\chi u dd X}$ & 
			$\epsilon^{ijk}(\overline{\chi}\gamma^\mu u_{i,p})(\overline{d^{\C}_{j,r} }\sigma_{\mu\nu} i\gamma_5 d_{k,s})X^\nu $ & $\frac{1}{2}n_un_d(n_d-1)$
			\\
			$\calO^{\tt T2A}_{\chi u dd X}$ & 
			$\epsilon^{ijk}(\overline{\chi}\sigma_{\mu\nu} i\gamma_5 u_{i,p})(\overline{d^{\C}_{j,r} }\gamma^\mu\gamma_5 d_{k,s})X^\nu $ & $\frac{1}{2}n_un_d(n_d+1)$  &
			$\calO^{\tt AT2}_{\chi u dd X}$ & 
			$\epsilon^{ijk}(\overline{\chi}\gamma^\mu\gamma_5 u_{i,p})(\overline{d^{\C}_{j,r} } \sigma_{\mu\nu} i\gamma_5 d_{k,s})X^\nu $ & $\frac{1}{2}n_un_d(n_d-1)$ 
			\\
			\hline
	\end{tabular}}
	\caption{Operator basis up to dim 7 for the fermion-vector-SM case. Here $\chi$ is a Majorana fermion field and $X_\mu$ a real vector field in dark sector. All operators in the table are non-hermitian, with $+\hc$ indicating the other half from the action of hermitian conjugation. The operators involving dark vector tensors alone are highlighted in green.}
	\label{tab:Xchi}
\end{table}

The operators in the fermion-vector-SM case can be constructed in a similar fashion as those operators in the scalar-fermion-SM case in Sec.\,\ref{sec:sca-fer-SM}. Denoting the dark fermion by $\chi$ and dark vector by $X_\mu$, Lorentz invariance forces the fermion $\chi$ to combine with a SM fermion to form a dim-3 current. Up to dim 6, only the neutrino field can appear in the current, $\overline{\nu_L}\Gamma \chi$; together with the vector field $X_\mu$, the dim-4 and dim-5 operators are determined uniquely by Lorentz invariance and EoM relations, i.e., $(\overline{\nu_{L}} \gamma^\mu\chi) X_\mu$ and $(\overline{\nu_{L}}\sigma^{\mu\nu}\chi) X_{\mu\nu}$, respectively. For the dim-6 case, since the dimension of $\overline{\nu_L}\Gamma \chi$ with an $X_\mu$ is already 4, the additional two dimensions can only be provided by either two derivatives or a photon field strength tensor. The former case with two additional derivatives leads to redundant operators by using the IBP and EoM relations. Thus, we are left with the only possibility from an additional photon strength tensor, and it is easy to find there are two such operators as shown in Table \ref{tab:Xchi}. 

Now we turn to the dim-7 part. These operators can be classified in terms of the number of fermion bilinears. The ones with a single bilinear can be constructed from the two dim-6 operators by attaching an additional derivative, i.e., the $(\overline{\nu_L}\chi) XFD$ type of operators. After taking into account the relations from IBP, EoMs, BIs, and GIs, they are selected to have the forms given in Table \ref{tab:Xchi}. The detail for their construction is provided in Appendix \ref{app2}. The operators with two fermion bilinears and an $X_\mu$ field has the lowest dimension 7. Thus, we only need to figure out bilinears with a free Lorentz index, which is to be contracted with the $X_\mu$, while their field content is totally identical with the scalar-fermion-SM case. Without much effort, one can obtain these non-redundant operators as shown in Table \ref{tab:Xchi}. Those operators not listed in the table can be transformed into the basis operators by using the FIs and Dirac gamma matrix identities Eq.\,\eqref{eq:GI}.
Note that all operators given in the table are non-hermitian, and their hermitian partners are indicated by the symbol $+\hc$. The operators formed by the dark vector field strength tensor is highlighted in green, among which the dim-5 operator ${\cal O}_{\nu\chi XD}^{\tt T}$ is the unique operator listed in the paper~\cite{Aebischer:2022wnl}. The BNV operators in the table can also induce the neutron dark decay, $n\to \chi X$. Their phenomenological implications will be discussed along with the decay $n\to \chi \phi$ in the scalar-fermion-SM case.

\section{Neutron lifetime anomaly from dark decay}
\label{neudark}

~~The DSEFT offers a phenomenologically rich landscape to explore, from terrestrial high intensity experiments to astrophysical and cosmological observables. 
One of its intriguing phenomenological aspects is inelastic DM, a concept proposed to explain the annual modulation signals claimed by DAMA while maintaining consistency with the limits from CDMS \cite{Tucker-Smith:2001myb}. 
The fermionic absorption DM is another example of the DSEFT description \cite{Dror:2019onn,Ge:2022ius}. The exotic interaction between two distinct dark vectors and a photon is discussed in Ref.\,\cite{Aebischer:2022wnl}, which can avoid many constraints. The DSEFT operators can also induce flavor changing neutral current processes with the dark particles mimicking the SM neutrinos, e.g., the recent Belle II anomaly for $B^+ \to K^+\nu\bar\nu$ \cite{belleIItalk2023,belleIIpara2023}, and thus are worth of study at flavor factories like the Belle II, Barbar, LHCb, BESIII, KOTO, and NA62~\cite{Gori:2020xvq, Krnjaic:2019rsv}. The exotic lepton flavor violating (LFV) processes such as $\ell_i \to \ell_j +\textrm{invisible}$ arise naturally as well in this framework, which are interesting to explore to enrich the study of LFV physics. Other experimental anomalies such as the muon $g-2$ can be explained by invoking some DSEFT operators \cite{Ge:2021cjz}. On the other hand, if dark sector particles are light enough, they can be directly generated in fixed-target experiments and at colliders, leading to general constraints on DSEFT operators \cite{Gori:2022vri}. Invisible dark sector particles can be probed as missing energy in fixed-target experiments \cite{Banerjee:2019pds,Berlin:2018bsc} or through mono-photon events at colliders \cite{Duerr:2019dmv}. Long-lived dark particles can be probed by displaced vertices \cite{Duerr:2019dmv,LHCb:2019vmc} or rare events in far detectors such as FASER \cite{Berlin:2018jbm}. Extremely light dark sector particles can also be probed by energy-loss arguments in astrophysical objects
such as supernovae \cite{Chang:2018rso} or white dwarfs \cite{Davidson:2000hf}. Furthermore, as previously mentioned, the BNV interactions from our DSEFT operators could potentially explain the neutron lifetime anomaly through dark decay while satisfying the constraint from DM-neutron annihilation, which we will thoroughly investigate in the following.

There is a longstanding discrepancy between the neutron lifetime measured 
in bottle-type and beam-type experiments \cite{Workman:2022ynf}. 
In the bottle-type experiment, the surviving neutrons are counted
to yield the total neutron decay width 
$\Gamma_n^{\tt tot} = 1/[(878.3\pm0.3)\,{\rm sec}]$ \cite{Mampe:1993an,Serebrov:2004zf,Pichlmaier:2010zz,Steyerl:2012zz,Arzumanov:2015tea}. While in the beam-type experiment, the number of protons from beta decay is counted, with the corresponding  decay width being
$\Gamma_n^{\beta} = 1/[(888\pm2)\,\rm sec] $ \cite{Byrne:1996zz,Yue:2013qrc}. 

A natural way to 
reconcile the tension between the 
two types of experiments
is to introduce neutron dark decay modes, 
with decaying products being dark particles which
slip away from the beam-type detection. 
The window for neutron dark decay 
width can be simply determined as the difference of the two observed values,
\begin{eqnarray}
	\Gamma_n^{\tt dark} = \Gamma_n^{\tt tot} - \Gamma_n^{\beta}
	= \tau_n^{-1} - \tau_{n,\beta}^{ -1} 
	= (8.184\pm 1.688)\times 10^{-30}\,{\rm GeV}.
\end{eqnarray}
For the neutron dark decay modes, there are many discussions in specific models, see \cite{Fornal:2018ubn,Elahi:2020urr,Khatibi:2023fwv} and references therein, but there are few from the EFT perspective \cite{Strumia:2021ybk,Alonso-Alvarez:2021oaj}. Here we find the DSEFT can nicely fit the gap. The dim-7 BNV DSEFT operators in Table \ref{tab:chiphi} and Table \ref{tab:Xchi} from the fermion-scalar-SM and fermion-vector-SM cases can naturally lead to neutron dark decay, if kinematics is allowed. In the following, we will investigate the implications of those  operators on the neutron lifetime anomaly as well as the DM-neutron annihilation.

\subsection{ Dark fermion-scalar case: $n \to \chi + \phi$ }

~~We first consider the $(\overline\chi u)(dd)\phi$-type operators in the fermion-scalar-SM case,
listed in the last sector of Table \ref{tab:chiphi}. 
For neutron dark decay, only the up and down quark flavors need to be picked up, and the flavor and color symmetries automatically result in four non-vanishing operators associated with down-quark vector and tensor currents,
\begin{subequations}
\label{eq:op:dark1}
	\begin{align}
		\calO_{\phi\chi }^{\tt VV} &= \epsilon^{ijk}(\overline{\chi}\gamma_\mu u_i)(\overline{d^\C_j}\gamma^\mu d_k)\phi, & \,
		\calO_{\phi\chi }^{\tt AV} &= \epsilon^{ijk}(\overline{\chi}\gamma_\mu \gamma_5 u_i)(\overline{d^\C_j}\gamma^\mu d_k)\phi,
		\\
		\calO_{\phi\chi }^{\tt T1} &= \epsilon^{ijk}(\overline{\chi}\sigma_{\mu\nu} u_i)(\overline{d^\C_j}\sigma^{\mu\nu} d_k)\phi,& \,
		\calO_{\phi\chi  }^{\tt T2} &= \epsilon^{ijk}(\overline{\chi}\sigma_{\mu\nu}\gamma_5 u_i)(\overline{d^\C_j}\sigma^{\mu\nu} d_k)\phi.
	\end{align}
\end{subequations}
Using Fierz transformation,
the four operators can be transformed into four other independent ones in terms of chiral quark fields,
\begin{subequations}
	\begin{align}
		\calO^{\tt LR}_{\phi\chi} &= \epsilon^{ijk}(\overline{\chi}P_L d_i)(\overline{d^\C_j}P_R u_k)\phi,&\,
		\calO^{\tt RL}_{\phi\chi} &= \epsilon^{ijk}(\overline{\chi}P_R d_i)(\overline{d^\C_j}P_L u_k)\phi,
		\\
		\calO^{\tt LL}_{\phi\chi} &= \epsilon^{ijk}(\overline{\chi}P_L d_i)(\overline{d^\C_j}P_L u_k)\phi,&\,
		\calO^{\tt RR}_{\phi\chi} &= \epsilon^{ijk}(\overline{\chi}P_R d_i)(\overline{d^\C_j}P_R u_k)\phi.
	\end{align}
\end{subequations}
The chiral basis will make their hadronic matrix elements  be easily extracted from the lattice QCD calculation.
The relationships of the Wilson coefficients for these two bases are 
\begin{subequations}
\begin{align}
	C^{\tt LR}_{\phi\chi} &=  4(C^{\tt VV}_{\phi\chi} + C^{\tt AV}_{\phi\chi}),&\,
	C^{\tt RL}_{\phi\chi} &=  4(C^{\tt VV}_{\phi\chi} - C^{\tt AV}_{\phi\chi}),
	\\
	C^{\tt LL}_{\phi\chi} &=  8(C^{\tt T1}_{\phi\chi} - C^{\tt T2}_{\phi\chi}),&\,
	C^{\tt RR}_{\phi\chi} & =  8(C^{\tt T1}_{\phi\chi} + C^{\tt T2}_{\phi\chi}).
\end{align}
\end{subequations}
To calculate the hadronic matrix elements, we denote the quark component of the effective operators 
as,
\begin{eqnarray}
\mathcal{N}^{\Gamma \Gamma^{\prime}} \equiv \epsilon^{ijk}
(P_{\Gamma}d_i) (\overline{d^\C_j} P_{\Gamma^{\prime}} u_k),
\end{eqnarray}
such that $\calO^{\Gamma \Gamma^{\prime}}_{\phi\chi} =  \bar{\chi} \mathcal{N}^{\Gamma \Gamma^{\prime}} \phi $
with $\Gamma, \Gamma^\prime \in \{ \cL,\cR \}$.

\textbf{Neutron dark decay}: 
Here we calculate the decay rate of the neutron dark decay, $n \to \chi +\phi$.
The neutron to vacuum transition activated by  $\mathcal{N}^{\Gamma \Gamma^{\prime}}$
can be parameterized as \cite{JLQCD:1999dld,Aoki:2017puj},
\begin{eqnarray}
\langle 0 | {\cal N}^{\Gamma \Gamma^\prime} | n \rangle 
= \alpha^{\Gamma\Gamma^\prime}_n P_\Gamma u_n,
\end{eqnarray}
where $\alpha^{\Gamma\Gamma^\prime}_n$ is a free parameter serving as a decay constant and $u_n$ is the neutron spinor wavefunction.  
The parity invariance of strong interaction implies 
$\alpha_n^{\tt LR} = -\alpha_n^{\tt RL}$ and 
$\alpha_n^{\tt LL} = -\alpha_n^{\tt RR}$.
In addition, the lattice QCD calculation yields
$\alpha_n \equiv \alpha_n^{\tt LR} 
=- \alpha_n^{\tt LL}  
= -0.0144(3)(21)\,{\rm GeV}^3$ \cite{Aoki:2017puj}.
Consequently,
when considering contributions from all operators,
there exists only one independent decay constant.
Therefore, the amplitude of $n \to \chi + \phi$ can be written as
\begin{eqnarray}
{\cal M}_{n\to \chi \phi} 
& = & \alpha_n 
\left[ 
(C^{\tt LR}_{\phi\chi}- C^{\tt LL}_{\phi\chi} )(\overline{u_\chi}P_L u_n)
- (C^{\tt RL}_{\phi\chi} - C^{\tt RR}_{\phi\chi}  )(\overline{u_\chi}P_R u_n)
\right]  \nonumber
\\ 
& = & 
4 \alpha_n 
\left[ 
(C^{\tt AV}_{\phi\chi} + 2 C^{\tt T2}_{\phi\chi} )(\overline{u_\chi}u_n) 
- (C^{\tt VV}_{\phi\chi} - 2 C^{\tt T1}_{\phi\chi})(\overline{u_\chi} \gamma_5 u_n)
\right]. 
\label{eq:amp_dec}
\end{eqnarray}
Due to different parity property of the two terms above, there is no interference between them for the spin-averaged matrix element square, we thus denote the two combinations of Wilson coefficients as,
\begin{align}
C^{a}_{\phi\chi}  & \equiv  C^{\tt VV}_{\phi\chi } - 2 C^{\tt T1}_{\phi\chi }, 
&
C^{b}_{\phi\chi}  & \equiv  C^{\tt AV}_{\phi\chi } + 2 C^{\tt T2}_{\phi\chi }, 
\end{align}
and introduce effective heavy scales via the relations, $\Lambda_a =| C^{a}_{\phi\chi}|^{-1/3}$ and  $\Lambda_b =| C^{b}_{\phi\chi}|^{-1/3}$.
Summing over the spin of the final state $\chi$ and averaging over the spin of the initial state neutron,
the decay width of $n \to \chi+ \phi$ takes the form, 
\begin{eqnarray}
	\Gamma_{n\to \chi \phi} =
	{\alpha_n^2 \lambda^{1\over 2}(m_n^2, m_\chi^2, m_\phi^2)  \over \pi m_n^3 } 
	\left\{ 
	\frac{1}{\Lambda_a^6} \left[(m_n-m_\chi)^2 - m_\phi^2 \right]
	+\frac{1}{\Lambda_b^6} \left[(m_n + m_\chi)^2 - m_\phi^2 \right] 
	\right\},
 \label{eq:n2chiphi}
\end{eqnarray}
where $\lambda(x,y,z)=x^2+y^2+z^2-2xy-2yz-2zx$ is the triangle function.

\begin{figure}
	\centering
	\includegraphics[width=6 cm]{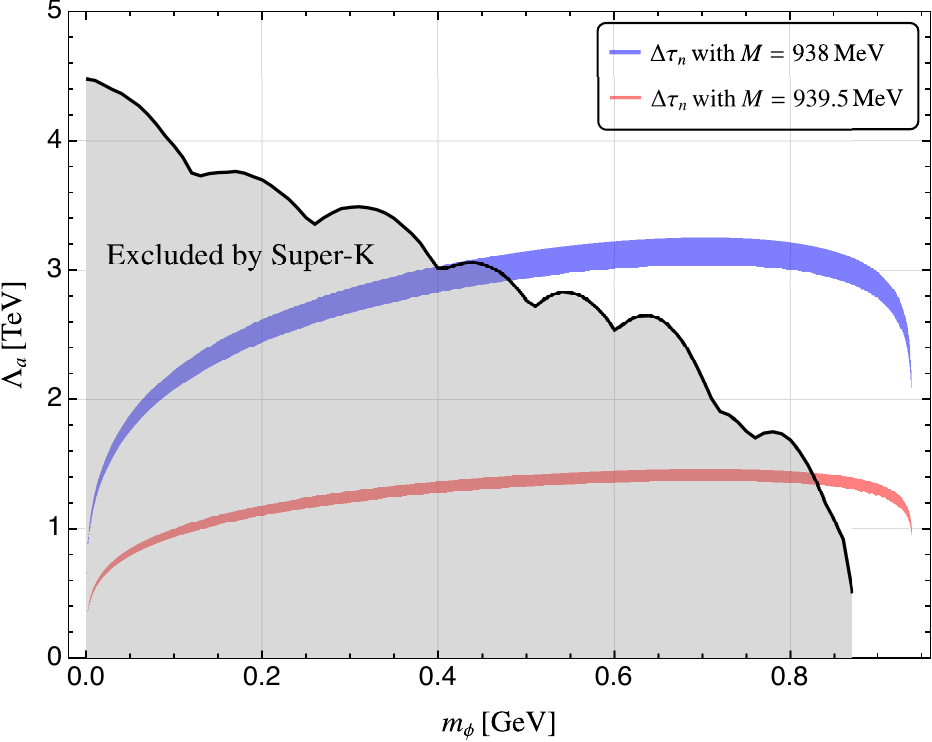} \quad\quad
	\includegraphics[width=6.1 cm]{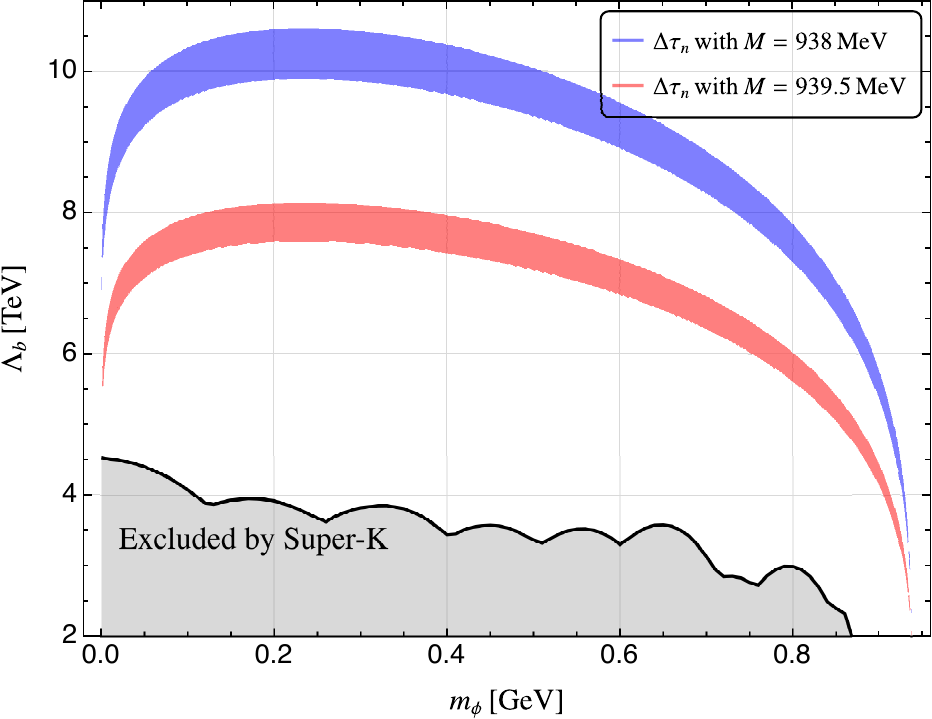}
	\caption{The allowed parameter space in the $m_{\phi}{\rm-}\Lambda$ plane for neutron dark decay with two benchmark values $M = 938 \,\rm MeV$ (blue) and $M = 939.5 \,\rm  MeV$ (red), where $\Lambda_b \to \infty$ ($\Lambda_a \to \infty$) is assumed in the left (right) panel. The gray region is excluded by Super-K.
	}
	\label{fig:scalar}
\end{figure}

The BNV operators mediating neutron dark decay could lead to the decay of stable bound nuclei such as beryllium and other stable ones, which would put strong constraints on the mass range of the dark particles.  
Ensuring the stability of nuclei in the presence of neutron dark decay is equivalent to requiring that the $Q$-value of nuclear decay is smaller than the separation energy for a neutron in the nucleus, denoted as $B_n$.
This
offers a mass window to 
enable neutron dark decay while avoiding the strong nuclear matter stability constraints.
Taking the lowest neutron separation energy from beryllium nucleus into consideration leads to \cite{Fornal:2018eol},
\begin{align}
	937.900\,{\rm MeV} < M\equiv m_\chi + m_\phi < 939.565\, {\rm MeV}.
\label{eq:mass_Window}
\end{align}
Without loss of generality, we consider two extreme cases with $M = 938$ or $ 939.5\,\rm MeV$ and show the allowed parameter space 
in the $m_\phi - \Lambda$ plane for neutron dark decay in 
Fig.\,\ref{fig:scalar},
in which the contributions from $\Lambda_a$ and $\Lambda_b$
are considered separately.
For case $a$,
with the fixed value $M = 938\, (939.5)\,\rm MeV$, 
the allowed parameter space expands a narrow band with $\Lambda_a$ increasing from 0.4 (1) TeV to 1.4 (3.2) TeV when $m_\phi$ runs from 0 to 0.8 GeV. The region between the two bands is also allowed once requiring the total dark mass $M$ is within the two extreme values. Similar situation happens for case $b$, but the behavior of the allowed region is reversed as $m_\phi$ increases. 

\textbf{DM-neutron annihilation constraint}: 
For the neutron dark decay window in Eq.\,\eqref{eq:mass_Window} and with the only available BNV interactions in Eq.\,\eqref{eq:op:dark1}, the dark particles cannot decay and are stable due to kinematic restrictions,
so that dark particles can serve as DM candidates.
Here we consider 
the dark fermion $\chi$ as the DM candidate,
and calculate the corresponding constraint from DM-neutron annihilation
process $\chi + n \to \phi + \pi^0$ at Super-Kamiokande (Super-K). 
The $\pi^0$ produced in the final state will decay into energetic diphoton, 
which can be easily detected by the water Cherekov detector.

Now we calculate the event rate of DM-neutron annihilation.
The transition form factor from a neutron state $n(k)$ to a meson state  $\pi^0 (p)$ with a momentum transfer $q=k-p$ can be parameterized as  \cite{Aoki:2017puj}
\begin{eqnarray}
\langle \pi^0|\mathcal{N}^{\Gamma \Gamma^{\prime}}| n \rangle=P_{\Gamma}\left[W_0^{\Gamma \Gamma^{\prime}}(t)-\frac{i \slashed{q}}{m_n} W_1^{\Gamma \Gamma^{\prime}}(t)\right] u_n(k),
\end{eqnarray}
where  $t=q^2$ and $m_n$
is the neutron mass.
The lattice calculation for $p \to \pi^0$ form factors is
given in Ref.\,\cite{Aoki:2017puj}
and $n \to \pi^0$ shares the same form factors due to isospin symmetry \cite{Aoki:2017puj}.
Parity invariance of strong interaction
implies
\begin{subequations}
\begin{align}
W_0^{\tt RL} (t) &= W_0^{\tt LR} (t), 
&
W_1^{\tt RL} (t) &=  W_1^{\tt LR} (t), 
\\
W_0^{\tt RR} (t) &=  W_0^{\tt LL} (t), 
&
W_1^{\tt RR} (t) &= W_1^{\tt LL} (t).  
\end{align}
\end{subequations}
Unlike the decay process, where parity transformation results in a negative sign, in this case, parity transformation yields a positive sign. This difference arises from the fact that the intrinsic parity of $\pi^0$ is odd.
In addition, the lattice calculation also numerically shows that
\begin{align}
W_0(t) & \equiv W_0^{\tt LR}(t) \simeq - W_0^{\tt LL}(t), 
&
W_1(t) & \equiv W_1^{\tt LR}(t) \simeq - W_1^{\tt LL}(t), 
\end{align}
which is similar to $\alpha_n^{\tt LR} = - \alpha_n^{\tt LL}$ for the decay constant.
Consequently, when considering contributions from all operators, there exist only two independent form factors. Therefore, the amplitude of $n (k) + \chi (k^\prime) \to \pi^0 (p) + \phi (p^\prime)$ can be expressed as follows,
\begin{eqnarray}
\mathcal{M}_{n \chi \to \pi^0 \phi}
&=& \langle \phi| \phi \bar{\chi}| \chi \rangle \langle \pi^0 \left| C^{\tt LR}_{\phi\chi} \mathcal{N}^{\tt LR} + C^{\tt RL}_{\phi\chi} \mathcal{N}^{\tt RL} + C^{\tt LL}_{\phi\chi} \mathcal{N}^{\tt LL} 
+C^{\tt RR}_{\phi\chi} \mathcal{N}^{\tt RR} \right| n \rangle \nonumber \\
& = & 4 \overline{v_\chi} (k^\prime) 
\left[  C^{a}_{\phi\chi}  - C^{b}_{\phi\chi} \gamma_5 \right]  \left[  W_0 (t) 
- \frac{i  \slashed{q} }{m_n} W_1 (t) \right] u_n(k).
\end{eqnarray}
Again, the contributions to the spin-averaged cross section from the two terms $C^{a}_{\phi\chi}$ and $C^{b}_{\phi\chi}$ will not interfere with each other due to parity. 
Taking the non-relativistic limit for the DM particle,
the spin-averaged cross section is,
\begin{eqnarray}
v_\chi \sigma_{\chi n \to \pi^0 \phi} 
& \simeq &
\frac{1}{4\pi m_n^2 (m_n+m_\chi)^4} 
\left\{
{ M_{n\chi\pi\phi}^{6} \over \Lambda_a^6} W_1^2 (t_0)
\right.
\nonumber
\\
&&+ 
\left. {M_{n\chi\pi\phi}^2 \over \Lambda_b^6} \left[ 4 m_n^2 (m_n + m_\chi)^2 W_0^2(t_0)  
+ (m_n^2+m_\phi^2-m_\pi^2-m_\chi^2)W_1^2(t_0) 
\right] 
\right\},
\end{eqnarray}
where $m_\chi$ ($m_\phi$, $m_\pi$) is the mass of 
$\chi$ ($\phi$, $\pi$), $v_\chi \simeq 10^{-3}$ is the local average velocity of DM
near the Earth, and 
\begin{subequations}
\begin{eqnarray}
t_0 &=& \frac{m_n m_\phi^2+m_\chi m_\pi^2}{m_n+m_\chi}-m_n m_\chi, \\
M_{n\chi\pi\phi} &=& \left[ (m_n+m_\chi)^2 - (m_\pi+m_\phi)^2 \right]
\left[ (m_n+m_\chi)^2 - (m_\pi-m_\phi)^2 \right]^{1/4}.
\end{eqnarray}
\end{subequations}
Hence, $\chi + n \to \pi^0 + \phi$ is an $s$-wave transition in our case and there is no $W_0$ contribution for the $\Lambda_a$ term.

The relation between the DM-nucleus cross section ($\sigma_{\chi N}$) and the DM-nucleon cross section ($\sigma_0$) approximately follows an $A^{2/3}$ scaling law, suggesting a predominantly surface process \cite{Astrua:2002zg}, namely
only the surface nucleons contribute to the interaction.
Consequently, this implies that $\sigma_{\chi N}= A^{2/3} \sigma_0$ \cite{Keung:2019wpw, Astrua:2002zg},
where $A$ is the mass number of the nucleus ($N$) with $Z$ protons, and  $\sigma_0 = \alpha \sigma_{\chi p}+(1-\alpha)\sigma_{\chi n}$ with $\alpha \equiv Z/A$. For our case, $\sigma_{\chi p}=0$.
With water as the target for Super-K, the interaction rate per second 
per gram of water is \cite{Keung:2019wpw}
\begin{eqnarray}
n_\chi v_\chi \left[(N_A \times 1/18)\times \sigma_{\chi O}
+ (N_A \times 2/18)\times \sigma_{\chi H}\right],
\end{eqnarray}
where  $n_\chi = \rho_\chi/m_\chi$ is the DM number density with $\rho_\chi = 0.3 ~\rm GeV/cm^3$ being the DM local mass density, $N_A = 6.022\times 10^{23}$ is the Avogadro number, and $N_A/18$ is the total number of $\rm H_2 O$ molecules
per gram of water. 

A search for nucleon decay via $n\to \bar{\nu}+\pi^0$ is carried out by Super-K Collaboration, using data from a combined 172.8 kton $\cdot$ years exposure of Super-K-I, -II, and -III \cite{Super-Kamiokande:2013rwg}. 
There is 22.5 kton of fiducial water in the Super-K Cherenkov detector. 
The signal efficiency for Super-K $n\to \bar{\nu} + \pi^0$ analysis
is 48.5\%, 44.0\%, and 48.5\% for SK-I (1489.2 live days), SK-II (798.6 live days), 
and SK-III (518.1 live days) runs, respectively.
Such a signal process is very similar to our signal process $\chi + n \to \phi + \pi^0$, in which only diphotons from $\pi^0$ decay are observed in the detector. 
Hence, we  proceed as Ref.\,\cite{Super-Kamiokande:2013rwg} to carry out our analysis.
We adopt the same signal efficiency
and smear the final state photons with the momentum resolution of $\sigma_\gamma = (0.6 + 2.6/\sqrt{p_\gamma}) \%$ \cite{Super-Kamiokande:2005mbp}, where $p_\gamma$ is the momentum of photons in the unit of GeV. Then the data is binned in the total momentum as Fig.\,(3) of Ref.\,\cite{Super-Kamiokande:2013rwg}.
We obtain constraints by requiring that the expected signal does not exceed the observed number of events by more than $2\sigma$ in any bin, namely, $N_s^i + N_b^i < N_o^i +2 \sigma_i$, where $N_s^i$, $N_b^i$, and  $N_o^i$ are the numbers of signal, background, and observed events of bin $i$, and $\sigma_i$ is the corresponding error bar of bin $i$.

The constraints from the null searching results by Super-K are shown in the gray region in Fig.\,\ref{fig:scalar}.
For case $a$ on the left panel, it can be seen that Super-K imposes very strong constraints on the $m_\phi$-$\Lambda_a$ plane. It excludes a large portion of parameter space, and only a small region with $ 0.4\,{\rm GeV} < m_\phi < m_n $ is allowed. While for case $b$ on the right panel, the constraints from Super-K have no impact on the desired region for the resolution of  neutron lifetime anomaly. This arises mainly because Super-K has a similar sensitivity in both cases while the neutron lifetime anomaly requires a heavier effective scale $\Lambda_b$ due to kinematic enhancement in Eq.\,\eqref{eq:n2chiphi}.  

\subsection{ Dark fermion-vector case: $n \to \chi + X$ }

Now we turn to consider the  $(\overline\chi u)(dd)X$-type operators
in the fermion-vector-SM case.
There are eight operators contributing to $n \to \chi + X$ with the vector and tensor down quark currents in Table \ref{tab:Xchi},
which are given by
\begin{subequations}
\label{ndarkdecay:vector1}
	\begin{align}
		\calO^{\tt SV }_{\chi X} &=
		\epsilon^{ijk}(\overline{\chi} u_{i})(\overline{d^{\C}_{j} }\gamma_\mu d_{k})X^\mu,& \,
		\calO^{\tt PV}_{\chi X} &=
		\epsilon^{ijk}(\overline{\chi} \gamma_5 u_{i})(\overline{d^{\C}_{j} }\gamma_\mu d_{k})X^\mu,
		\\
		\calO^{\tt T1V}_{\chi X} &=
		- i \epsilon^{ijk}(\overline{\chi}\sigma_{\mu\nu} u_{i})(\overline{d^{\C}_{j} }\gamma^\mu d_{k})X^\nu,& \,
		\calO^{\tt T2V}_{\chi X} &=
		- i  \epsilon^{ijk}(\overline{\chi} \sigma_{\mu\nu} \gamma_5 u_{i})(\overline{d^{\C}_{j} }\gamma^\mu d_{k})X^\nu,
		\\
		\calO^{\tt VT1 }_{\chi X} &=
		- i \epsilon^{ijk}(\overline{\chi}\gamma^\mu u_{i})(\overline{d^{\C}_{j} }\sigma_{\mu\nu} d_{k})X^\nu,& \,
		\calO^{\tt AT1 }_{\chi X} &=
		- i \epsilon^{ijk}(\overline{\chi}\gamma^\mu\gamma_5 u_{i})(\overline{d^{\C}_{j} }\sigma_{\mu\nu} d_{k})X^\nu,
		\\
		\calO^{\tt VT2 }_{\chi X} &=
		- i \epsilon^{ijk}(\overline{\chi}\gamma^\mu u_{i})(\overline{d^{\C}_{j} } \sigma_{\mu\nu}\gamma_5 d_{k})X^\nu, & \,
		\calO^{\tt AT2 }_{\chi X} &=
		- i \epsilon^{ijk}(\overline{\chi}\gamma^\mu\gamma_5 u_{i})(\overline{d^{\C}_{j} }\sigma_{\mu\nu}\gamma_5 d_{k})X^\nu,
	\end{align}
\end{subequations}
where the operators are relabelled 
and a factor $-i$ is attached for later convenience. 
In terms of chiral fields, the eight independent operators can be parameterized in the following way, 
\begin{subequations}
	\begin{eqnarray}
		\calO^{\tt LLL}_{\chi X} &=& \epsilon^{ijk}(\overline{\chi}\gamma_\mu P_L d_i)(\overline{d_j^{\C}} P_L u_k) X^\mu, 
		\\
		\calO^{\tt RRR}_{\chi X} &=& \epsilon^{ijk}(\overline{\chi}\gamma_\mu P_R d_i)(\overline{d^{\C}_j} P_R u_k) X^\mu,
		\\
		\calO^{\tt LRR}_{\chi X}&=& \epsilon^{ijk}(\overline{\chi}\gamma_\mu P_L d_i)(\overline{d^{\C}_j} P_R u_k) X^\mu  (\tt A), 
		\\
		\calO^{\tt RLL}_{\chi X} &=&   \epsilon^{ijk}(\overline{\chi}\gamma_\mu P_R d_i)(\overline{d^{\C}_j} P_L u_k) X^\mu  (\tt A),
		\\
		\calO^{\tt LLR}_{\chi X 1}  & = &  \epsilon^{ijk}(\overline{\chi} P_L d_i)(\overline{d^{\C}_j}\gamma_\mu P_R u_k) X^\mu, 
		\\ 
		\calO^{\tt RRL}_{\chi X 1}  & = & \epsilon^{ijk}(\overline{\chi} P_R d_i)(\overline{d^{\C}_j}\gamma_\mu P_L u_k) X^\mu, 
		\\
		 \calO^{\tt LLR}_{\chi X 2}  & = & \epsilon^{ijk}[ (\overline{\chi} P_L d_i)(\overline{u^{\C}_j}\gamma_\mu P_R d_k) 
			+ (\overline{\chi} P_L u_i)(\overline{d^{\C}_j}\gamma_\mu  P_R d_k) ]  X^\mu (\tt S), 
        \\
		\calO^{\tt RRL}_{\chi X 2} & = & \epsilon^{ijk} [(\overline{\chi} P_R d_i)(\overline{u^{\C}_j}\gamma_\mu P_L d_k) 
			+ (\overline{\chi} P_R u_i)(\overline{d^{\C}_j}\gamma_\mu  P_L d_k) ] X^\mu (\tt S), 
		\quad
	\end{eqnarray}
\end{subequations}
where a symbol ({\tt S}) or ({\tt A}) behind an operator
indicates its symmetry or antisymmetry under the exchange of 
the two quark fields with the same chirality.
 With such a reparametrization, 
from the perspective of chiral perturbation theory ($\chi$PT), we find they belong to some specific chiral irreducible representations of the chiral group $\rm SU(3)_\cL\otimes SU(3)_\cR$. Under $\rm SU(3)_\cL\otimes SU(3)_\cR$, 
the first two operators, $\calO^{\tt LLL}_{\chi X}$ and $\calO^{\tt RRR}_{\chi X}$,
belong to $\pmb{8}_\cL\otimes \pmb{1}_\cR$ and $\pmb{1}_\cL\otimes \pmb{8}_\cR$;
the third and fourth operators, $\calO^{\tt LRR}_{\chi X}$  and  $\calO^{\tt RLL}_{\chi X}$, belong to $\pmb{3}_\cL\otimes \bar{\pmb{3}}_\cR$ and $\bar{\pmb{3}}_\cL\otimes \pmb{3}_\cR$;
and the last four operators,
$\calO_{\chi X1}^{\cL\cL\cR}, \calO_{\chi X2}^{\cL\cL\cR}$ ($\calO_{\chi X1}^{\cR\cR\cL}, \calO_{\chi X2}^{\cR\cR\cL }$) belong to $\pmb{6}_\cL\otimes \pmb{3}_\cR$ ($\pmb{3}_\cL \otimes \pmb{6}_\cR$).
When matching onto $\chi$PT, the first four can be realized at leading order of chiral power counting, while the last four cannot.
Thus, we focus on the operators $\calO_{\chi X}^{\cL\cL\cL},\calO_{\chi X}^{\cL\cR\cR}$ and their parity partners, which can contribute to the neutron dark decay at leading order.

The relation for the Wilson coefficients of the first four operators are related to the operators in Eq.\,\eqref{ndarkdecay:vector1} as follows,
\begin{subequations}
\label{eq:Ope-vec}
	\begin{eqnarray}
		C_{\chi X}^{\cL\cL\cL} & = & 
		2 (C_{\chi X}^{\tt VT1} - C_{\chi  X}^{\tt AT1} - C_{\chi  X}^{\tt VT2} + C_{\chi  X}^{\tt AT2} ), 
		\\
		C_{\chi X}^{\cR\cR\cR} & = & 
		2 (C_{\chi X}^{\tt VT1} + C_{\chi  X}^{\tt AT1} + C_{\chi X}^{\tt VT2} + C_{\chi X}^{\tt AT2} ), 
		\\
		C_{\chi X}^{\cL\cR\cR} & = & 
		C_{\chi X}^{\tt SV} + C_{\chi X}^{\tt PV} - 3 C_{\chi X}^{\tt T1V} -3  C_{\chi X}^{\tt T2V}, 
		\\
		C_{\chi X}^{\cR\cL\cL} & = & 
		C_{\chi X}^{\tt SV} - C_{\chi X}^{\tt PV} - 3 C_{\chi  X}^{\tt T1V} +3  C_{\chi X}^{\tt T2V}.
	\end{eqnarray}
\end{subequations}
Since these four operators share the same quark factors,
i.e., $\mathcal{N}^{\Gamma \Gamma^{\prime}}$, as in the dark fermion-scalar case, 
the same decay constant ($\alpha_n$)
and form factors ($W_0(t)$ and $W_1(t)$) apply here.

\textbf{Neutron dark decay:}
Taking into account the contribution from all four operators in Eq.\,\eqref{eq:Ope-vec},
the amplitude of $n \to \chi + X$ for the vector case
is given by
\begin{eqnarray}
{\cal M}_{n \to \chi X}  =  
\alpha_n \overline{u_\chi}\gamma_\mu [ C_{\chi X}^a - C_{\chi X}^b \gamma_5 ]u_n  \epsilon_X^{*\mu}, 
\end{eqnarray}
where 
\begin{eqnarray}
C_{\chi X}^a & \equiv & {1\over 2}(C_{\chi X}^{\cL\cR\cR} -C_{\chi X}^{\cL\cL\cL} 
- C_{\chi X}^{\cR\cL\cL} + C_{\chi X}^{\cR\cR\cR} )  
=  C_{\chi X}^{\tt PV} -3  C_{\chi  X}^{\tt T2V} 
+ 2 C_{\chi X}^{\tt AT1} + 2 C_{\chi X}^{\tt VT2}, 
\\
C_{\chi X}^b & \equiv & {1\over 2}(C_{\chi X}^{\cL\cR\cR} -C_{\chi X}^{\cL\cL\cL} 
+ C_{\chi X}^{\cR\cL\cL} - C_{\chi X}^{\cR\cR\cR} ) 
= C_{\chi X}^{\tt SV} - 3 C_{\chi X}^{\tt T1V}
- 2 C_{\chi X}^{\tt VT1} - 2C_{\chi X}^{\tt AT2}. 
\end{eqnarray}
Summing over the final state spin and averaging the initial state spin,
the decay width takes the form
\begin{eqnarray}
	\Gamma_{n\to \chi X} =
	{\alpha_n^2 \lambda^{1\over 2}(m_n^2, m_\chi^2, m_X^2)  \over 16\pi m_n^3 m_X^2} 
	\left(|C_{\chi X}^a |^2 F_a
	+ |C_{\chi X}^b|^2 F_b \right),
\end{eqnarray}
where $m_X$ is the mass of dark vector $X$ and
\begin{eqnarray}
F_{ a,b} =  [(m_n \pm m_\chi)^2 + 2m_X^2] [(m_n \mp m_\chi)^2 -m_X^2]. 
\end{eqnarray}
From the decay width formula, it is obvious the result is singular when $m_X \to 0$ as stemming from the longitudinal polarization component, a well-known problem for a massive vector field. To bypass the problem, we follow Ref.\,\cite{He:2022ljo} and assume the corresponding Wilson coefficient is proportional to the vector mass due to underlying dynamics, and reparameterize them as  $C_{\chi X}^a \equiv m_X / \Lambda_a^4$ and $C_{\chi X}^b \equiv m_X / \Lambda_b^4$ respectively,
which leads to 
\begin{eqnarray}
	\Gamma_{n\to \chi X} = 
	{\alpha_n^2 \lambda^{1\over 2}(m_n^2, m_\chi^2, m_X^2)  \over 16\pi m_n^3} 
	\left( { F_{a} \over \Lambda_a^8 } + { F_{b} \over \Lambda_{b}^8}
	\right). 
\end{eqnarray}

\begin{figure}
\centering
\includegraphics[width=5.7 cm]{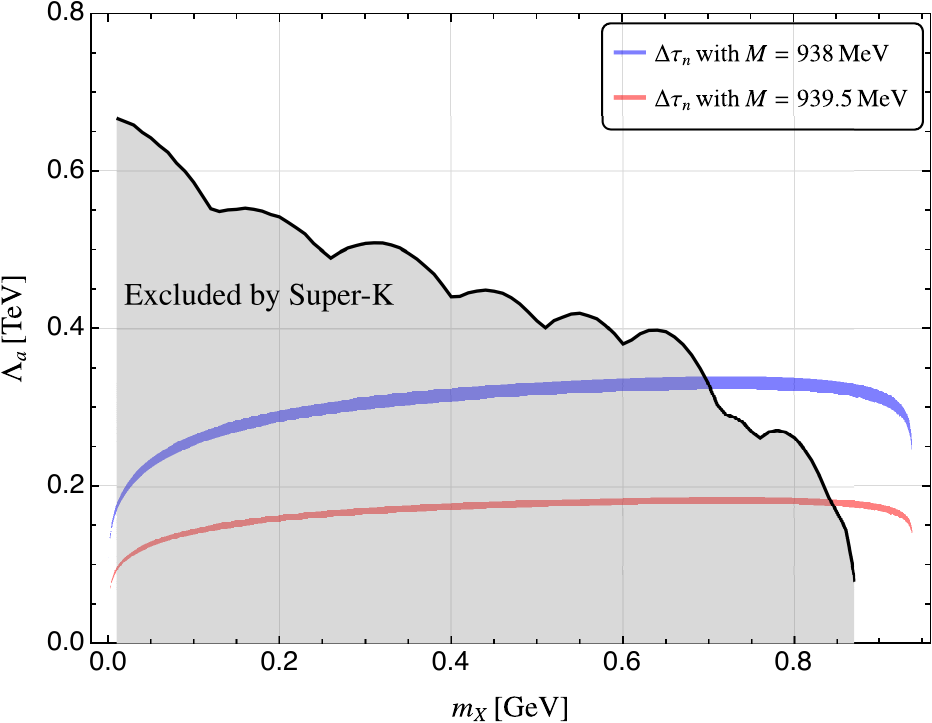} \quad\quad
\includegraphics[width=5.7 cm]{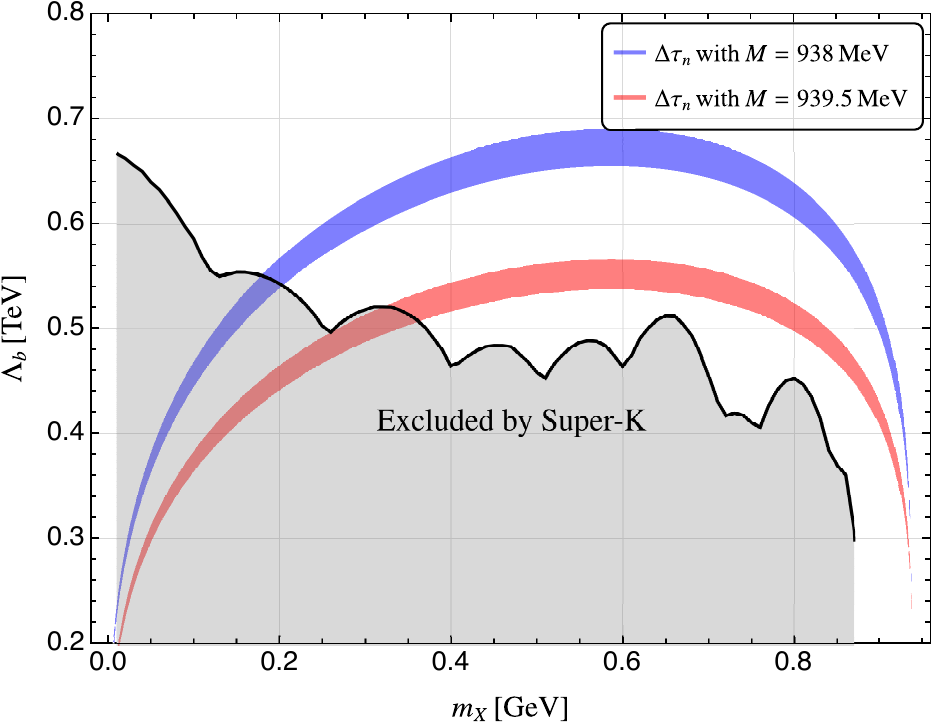}
\caption{The allowed parameter space in the $m_{X}{\rm-}\Lambda$ plane for neutron dark decay with two benchmark values $M = 938 ~\rm MeV$ (blue) and $M = 939.5 ~\rm  MeV$ (red), where $\Lambda_b \to \infty$ ($\Lambda_a \to \infty$) is assumed in the left (right) panel. The gray region is excluded by Super-K.}
	\label{fig:vector}
\end{figure}

The allowed parameter space for specific values of 
$M \equiv m_\chi + m_X$ in 
the $m_X - \Lambda$ plane to accommodate neutron dark decay
is shown in Fig.\,\ref{fig:vector},
in which the contributions from $\Lambda_a$ and $\Lambda_b$ terms
are considered separately.
Similar to $n \to \chi + \phi$ from the scalar-fermion case, the allowed parameter space is within the two bands once the dark mass sum $M$ is between the two extreme values. 
A weaker constraint on $\Lambda_{a,b}$ than that in the scalar case arises because the decay rate is now suppressed by two more powers of $\Lambda_{a,b}$.
In addition, the parameter space from case $a$ is almost a flat band, while for the case $b$ it is  a round arch shape.

\textbf{DM-neutron annihilation constraint:}
For the vector case, with the contribution from all four operators in Eq.\,\eqref{eq:Ope-vec},
the amplitude for DM-neutron annihilation,
$n (k) + \chi (k^\prime) \to \pi^0 (p) + X (p^\prime)$, 
is given by
\begin{eqnarray}
 \mathcal{M}_{n \chi \to \pi^0  X }
 =   \overline{v_\chi} (k^\prime) \slashed{\epsilon}^*_X
\left[  C^b_{\chi X} - C^a_{\chi X}  \gamma_5 \right]  \left[  W_0 (t) 
- \frac{i  \slashed{q} }{m_n} W_1 (t) \right] u_n(k),
\end{eqnarray}
where $\epsilon_X$ is the polarization vector of dark vector $X$.
Following the same procedure as in the scalar case, we can derive the Super-K constraint for the vector case,
as shown in the gray region in Fig.\,\ref{fig:vector}.
However, we omit the detailed form of the cross section here due to its complexity.
From Fig.\,\ref{fig:vector}, for case $a$,
the stringent constraint from Super-K excludes the majority of the parameter space for neutron dark decay, leaving a small viable window at $m_\chi \gtrsim 0.7$ GeV.
For case $b$, there exists a sizable region that can explain the neutron lifetime anomaly, with only the lighter mass region ($m_X \lesssim 0.2\, \rm GeV$) being excluded. 

Therefore, we observe that the allowed parameter space for $\Lambda_a$ in dark decay is subject to strong constraints imposed by Super-K, regardless of whether it is in the scalar or vector case. In contrast, the parameter space associated with $\Lambda_b$ exhibit significantly larger surviving regions.
The observation of neutron stars with masses exceeding one solar mass can also impose stringent constraints on neutron dark decay, as discussed in Ref.\,\cite{Grinstein:2018ptl,McKeen:2018xwc,Baym:2018ljz} and references therein. However, it is worth noting that repulsive self-interactions among dark sector particles, as explored in Ref.\ \cite{McKeen:2018xwc,Grinstein:2018ptl}, have the potential to alleviate the neutron star constraint.

\section{Conclusion}
\label{sec:conc}

~~In this work we have established a dark sector effective field theory to examine effective interactions involving two distinct light dark sector particles and SM particles up to dim 7. We have considered various possibilities with the spin of dark particles up to 1, including real scalars ($\phi$ and $S$), Majorana fermions ($\chi$ and $\psi$), and real vectors ($X_\mu$ and $V_\mu$). Our investigation spans the scenarios in which these dark particles interact with at least one SM particle, categorized into scalar-scalar-SM ($\phi S$-SM), fermion-fermion-SM ($\chi\psi$-SM), vector-vector-SM ($X V$-SM), scalar-fermion-SM ($\phi \chi$-SM), scalar-vector-SM ($\phi X$-SM), and fermion-vector-SM ($\chi X$-SM) cases. These scenarios shed light on the diverse ways in which the dark sector may connect with the visible SM sector. As a phenomenological application we have explored the implications of baryon number violating interactions in the scalar-fermion-SM and fermion-vector-SM cases, particularly in the context of neutron dark decay. We also considered the constraints from DM-neutron annihilation using data from Super-Kamiokande. We found neutron dark decay in each scenario can accommodate the anomaly, at the same time, without contradicting with the Super-Kamiokande
limit. 

\section*{Acknowledgement}
\addcontentsline{toc}{section}{\numberline{}Acknowledgements}

This work was supported in part by the Guangdong Major Project of Basic and Applied Basic Research No. 2020B0301030008, and by the Grants 
No.~NSFC-12035008, 
No.~NSFC-11975130, 
No.~NSFC-12247151,
and No.~NSFC-12305110.

\appendix

\section{Construction of the $(\overline\Psi\Psi)XVD^2$-type operators}
\label{app1}

In this Appendix, we make a detailed analysis on the construction of dim-7 operators in the class $(\overline\Psi\Psi)XVD^2$ with two dark vectors ($X_\mu$ and $V_\mu$) and two derivatives. We realize that the form $(\overline\Psi\Psi)XVD^2$ can give rise to operators consisting of dark field tensors $X^{\mu\nu}$ and (or) $V^{\mu\nu}$. Based on Lorentz invariance, it is easy to figure out the operators with two dark tensors,\footnote{Without loss of generality, here we work with the flavor diagonal case for simplicity.}
\begin{subequations}
    \label{eq:type1}
    \begin{align}
        \tilde\calO_{\Psi XV D^2 1}^{\tt S}&=(\overline{\Psi }\Psi) X_{\mu\nu} V^{\mu\nu},& \, \tilde\calO_{\Psi XV D^2 2}^{\tt S}&=(\overline{\Psi}\Psi )\tilde X_{\mu\nu}  V^{\mu\nu},
		\\
		\tilde\calO_{\Psi XV D^2 1}^{\tt P}&=(\overline{\Psi } i\gamma_5 \Psi ) X_{\mu\nu}  V^{\mu\nu},&\,  \tilde\calO_{\Psi XV D^2 2}^{\tt P}&=(\overline{\Psi } i\gamma_5 \Psi) \tilde X_{\mu\nu}  V^{\mu\nu},
		\\
		\tilde\calO_{\Psi XV D^2}^{\tt T1}&=(\overline{\Psi } \sigma^{\mu\nu} \Psi) X_{\mu\rho}  V_{\nu }^{~\rho},&\,\tilde\calO_{\Psi XV D^2}^{\tt T2}&=(\overline{\Psi} \sigma^{\mu\nu}i \gamma_5\Psi ) X_{\mu\rho}V_{\nu}^{~\rho}.
    \end{align} 
\end{subequations}
For the operators with one dark tensor, which can be either $X^{\mu\nu}$ or $V^{\mu\nu}$, the other derivative can act on the fermion bilinear or the remaining vector field as a symmetric Lorentz tensor once the IBP relation is considered. For the former case, we can form the following operators, 
\begin{subequations}
    \label{eq:type2}
    \begin{align}
        \calO_{\Psi XVD^2 1}^{\tt S}&=(\overline{\Psi} i \overleftrightarrow{D_\mu} \Psi) X_\nu V^{\mu\nu},&\, \calO_{\Psi XVD^2 2}^{\tt S}&=(\overline{\Psi} i\overleftrightarrow{D_\mu} \Psi) X_\nu \tilde V^{\mu\nu},
        \\
		\calO_{\Psi XVD^2 3}^{\tt S}&=(\overline{\Psi} i\overleftrightarrow{D_\mu} \Psi) X^{\mu\nu} V_\nu, &\, \calO_{\Psi XVD^2 4}^{\tt S}&=(\overline{\Psi} i\overleftrightarrow{D_\mu} \Psi)\tilde X^{\mu\nu} V_\nu,
		\\
		\calO_{\Psi XVD^2 1}^{\tt P}&=(\overline{\Psi} \gamma_5 \overleftrightarrow{D_\mu} \Psi) X_\nu V^{\mu\nu}, &\,
		\calO_{\Psi XVD^2 2}^{\tt P}&=(\overline{\Psi} \gamma_5 \overleftrightarrow{D_\mu} \Psi) X_\nu \tilde V^{\mu\nu},
        \\
		\calO_{\Psi XVD^2 3}^{\tt P}&=(\overline{\Psi} \gamma_5 \overleftrightarrow{D_\mu} \Psi) X^{\mu\nu} V_\nu,&\,
		\calO_{\Psi XVD^2 4}^{\tt P}&=(\overline{\Psi} \gamma_5 \overleftrightarrow{D_\mu} \Psi) \tilde X^{\mu\nu} V_\nu,
	    \\
		\calO_{\Psi XVD^2 1}^{\tt T1}&=(\overline{\Psi} \sigma_{\mu\nu} i\overleftrightarrow{D_\rho} \Psi)X^\mu V^{\nu\rho},&\,
		\calO_{\Psi XVD^2 1}^{\tt T2}&=(\overline{\Psi} \sigma_{\mu\nu}\gamma_5 \overleftrightarrow{D_\rho} \Psi) X^\mu V^{\nu\rho},
		\\
		\calO_{\Psi XVD^2 2}^{\tt T1}&=(\overline{\Psi_p} \sigma_{\mu\nu} i\overleftrightarrow{D_\rho} \Psi_r) X^{\mu\rho} V^\nu,&\,
		\calO_{\Psi XVD^2 2}^{\tt T2}&=(\overline{\Psi} \sigma_{\mu\nu}\gamma_5 \overleftrightarrow{D_\rho} \Psi) X^{\mu\rho} V^\nu.
    \end{align}
\end{subequations}
For the latter case, those operators formed from the symmetric tensor $\partial_\mu X_\nu + \partial_\nu X_\mu$ or $\partial_\mu V_\nu + \partial_\nu V_\mu$ can be reduced into a linear combination of operators
in Eqs.\,\eqref{eq:type1} and \eqref{eq:type2} up to EoM and total derivative terms. For instance,
\begin{eqnarray}
(\overline{\Psi_p} \sigma_{\mu\nu} \Psi_r) (\partial_\rho X^\mu + \partial^\mu X_\rho) V^{\nu\rho}
= \calO_{\Psi XVD^2 1}^{\tt S} - \calO_{\Psi XVD^2 2}^{\tt P}
+\fbox{EoM}+\fbox{T}.
\end{eqnarray}
Similar relations can be established for the operators containing a $\gamma_5$ or with $X\leftrightarrow V$.

Complexity arises when we inspect the rest of operators in the $(\overline\Psi\Psi)XVD^2$ class that cannot form any dark field tensor. To start, we arrange the fermion field $\overline{\Psi}$ to be derivative free from IBP relation and classify these operators into four categories according to their Lorentz structure:\\
\noindent 
{\bf Type A}: Arranging $D^2$ to $(\overline\Psi\Psi)X^\nu V_\nu$ and $(\overline\Psi\gamma_5\Psi)X^\nu V_\nu$ to form the following operators:
\begin{subequations}
	\label{eq:typea}
	\begin{align}
		&\calO_{a1}\equiv(\overline\Psi D_\mu\Psi) (\partial^\mu X^\nu) V_\nu,\quad  &\calO^\prime_{a1}&\equiv(\overline\Psi \gamma_5 D_\mu\Psi) (\partial^\mu X^\nu)V_\nu,
		\\
		&\calO_{a2}\equiv(\overline\Psi D_\mu\Psi) X^\nu  (\partial^\mu V_\nu),\quad  &\calO^\prime_{a2}&\equiv(\overline\Psi \gamma_5 D_\mu\Psi)X^\nu  (\partial^\mu V_\nu),
		\\
		&\calO_{a3}\equiv(\overline\Psi \Psi) (\partial^\mu X^\nu)  (\partial_\mu V_\nu),\quad  &\calO^\prime_{a3}&\equiv(\overline\Psi \gamma_5 D_\mu\Psi)(\partial^\mu X^\nu)  (\partial_\mu V_\nu).
	\end{align}
\end{subequations}
\noindent 
{\bf Type B}: Arranging $D_\mu D_\nu$ to $(\overline\Psi\Psi)X^\mu V^\nu$ and $(\overline\Psi\gamma_5\Psi)X^\mu V^\nu$ to form the following operators:
\begin{subequations}
	\label{eq:typeb}
	\begin{align}
		&\calO_{b1}\equiv(\overline\Psi D_\mu\Psi) (\partial_\nu X^\mu)  V^\nu,\quad  &\calO^\prime_{b1}&\equiv(\overline\Psi \gamma_5 D_\mu\Psi) (\partial_\nu X^\mu)  V^\nu,
		\\
		&\calO_{b2}\equiv(\overline\Psi D_\nu\Psi) X_\mu (\partial^\mu V^\nu),\quad  &\calO^\prime_{b2}&\equiv(\overline\Psi \gamma_5 D_\nu\Psi) X_\mu (\partial^\mu V^\nu),
		\\
		&\calO_{b3}\equiv(\overline\Psi \Psi) (\partial_\nu X_\mu) (\partial^\mu V^\nu),\quad  &\calO^\prime_{b3}&\equiv(\overline\Psi\gamma_5 \Psi) (\partial_\nu X_\mu) (\partial^\mu V^\nu),
		\\
		&\calO_{b4}\equiv(\overline\Psi D_\mu D_\nu\Psi) X^\mu V^\nu,\quad  &\calO^\prime_{b4}&\equiv(\overline\Psi \gamma_5 D_\mu D_\nu\Psi) X^\mu V^\nu.
	\end{align}
\end{subequations}
\noindent 
{\bf Type C}: Arranging $D^2$ to $(\overline\Psi\sigma_{\mu\nu}\Psi)X^\mu V^\nu$ and $(\overline\Psi\sigma_{\mu\nu}\gamma_5\Psi)X^\mu V^\nu$ to form the following operators:
\begin{subequations}
	\label{eq:typec}
	\begin{align}
		&\calO_{c1}\equiv(\overline\Psi \sigma_{\mu\nu}D_\rho\Psi) (\partial^\rho X^\mu) V^\nu,\quad  &\calO^\prime_{c1}&\equiv(\overline\Psi \sigma_{\mu\nu}\gamma_5 D_\rho\Psi) (\partial^\rho X^\mu) V^\nu,
		\\
		&\calO_{c2}\equiv(\overline\Psi \sigma_{\mu\nu}D_\rho\Psi) X^\mu (\partial^\rho V^\nu),\quad  &\calO^\prime_{c2}&\equiv(\overline\Psi \sigma_{\mu\nu}\gamma_5 D_\rho\Psi) X^\mu (\partial^\rho V^\nu),
		\\
		&\calO_{c3}\equiv(\overline\Psi \sigma_{\mu\nu}\Psi) (\partial^\rho X^\mu) (\partial_\rho V^\nu),\quad  &\calO^\prime_{c3}&\equiv(\overline\Psi \sigma_{\mu\nu}\gamma_5\Psi) (\partial^\rho X^\mu) (\partial_\rho V^\nu).
	\end{align}
\end{subequations}
\noindent 
{\bf Type D}: Arranging $D^\nu D_\rho$ to $(\overline\Psi\sigma_{\mu\nu}\Psi)X^\mu V^\rho$ and $(\overline\Psi\sigma_{\mu\nu}\gamma_5\Psi)X^\mu V^\rho$ to form the following operators:
\begin{subequations}
	\label{eq:typed}
	\begin{align}
		&\calO_{d1}\equiv(\overline\Psi \sigma_{\mu\nu}D_\rho\Psi) (\partial^\nu X^\mu) V^\rho,\quad  &\calO^\prime_{d1}&\equiv(\overline\Psi \sigma_{\mu\nu}\gamma_5 D_\rho\Psi) (\partial^\nu X^\mu) V^\rho,
		\\
		&\calO_{d2}\equiv(\overline\Psi \sigma_{\mu\nu}D_\rho\Psi) X^\mu (\partial^\nu V^\rho),\quad  &\calO^\prime_{d2}&\equiv(\overline\Psi \sigma_{\mu\nu}\gamma_5 D_\rho\Psi) X^\mu (\partial^\nu V^\rho),
		\\
		&\calO_{d3}\equiv(\overline\Psi \sigma_{\mu\nu}\Psi) (\partial_\rho X^\mu) (\partial_\nu V^\rho),\quad  &\calO^\prime_{d3}&\equiv(\overline\Psi \sigma_{\mu\nu}\gamma_5\Psi) (\partial_\rho X^\mu) (\partial_\nu V^\rho).
	\end{align}
\end{subequations}
For simplicity, we have omitted those operators in Type D with $X\leftrightarrow V$. In the following discussion, we will demonstrate that all of these operators in Type C and D are reducible and can be expressed in terms of operators in other categories. 

Firstly, all operators in Type C can be reduced into linear combinations of operators in Type A, B, D, and EoM terms. This can be easily realized by utilizing the following relations,
\begin{subequations}
  \begin{eqnarray}
\label{eq:q1}
    \sigma_{\mu\nu}D_\rho&=&\frac{1}{2}\left(\sigma_{\mu\rho} D_\nu-\sigma_{\nu\rho} D_\mu+\sigma_{\mu\nu} D_\rho \right)-{1 \over2}\left(\epsilon_{\mu\nu\rho\alpha}\gamma_5 D^\alpha
+\epsilon_{\mu\nu\rho\alpha}\gamma^\alpha\gamma_5\slashed{D}\right),
\\
\label{eq:parten}
    \overleftarrow{D_\rho}\sigma_{\mu\nu}&=&{1\over 2}\left(\overleftarrow{D_{\nu}}\sigma_{\mu\rho}-\overleftarrow{D_\mu}\sigma_{\nu\rho}+\overleftarrow{D_\rho}\sigma_{\mu\nu}\right)+\frac{1}{2}\left(\epsilon_{\mu\nu\rho\alpha}\overleftarrow{\slashed{D}}\gamma_5\gamma^\alpha+\epsilon_{\mu\nu\rho\alpha}\gamma_5\overleftarrow{D^\alpha}\right).
\end{eqnarray}  
\end{subequations}
We now examine the operators in Type D.
It is obvious that both $\calO_{d1}$ and $\calO^\prime_{d1}$ actually belong to operators in Eq.\,\eqref{eq:type2}. By using the IBP, EoM, and other gamma matrix identities, $\calO_{d3}$ and $\calO^\prime_{d3}$ can be transformed into the operators in Type B and those with a dark field tensor in Eq.\,\eqref{eq:type2}. 
In the case of $\calO_{d2}$ and $\calO^\prime_{d2}$, they can be converted into operators that involve two derivatives acting on a fermion tensor bilinear. The resultant operators can be further reduced into the ones 
in Type B as well as those involving a SM field tensor $F^{\mu\nu}$ or $G^{A,\mu\nu}$, along with various EoM operators. In summary, 
all operators in Type C and D with a tensor bilinear structure ($\sigma_{\mu\nu}$) can be reduced into opertors with a scalar bilinear as well as operators involving a dark field tensor. 

Next, we examine operators in Type A and B. Since the operators with a pesudoscalar bilinear share the same structure as those with a scalar bilinear up to an additional $\gamma_5$, we thus only need to focus on the reduction of the latter ones. Three combinations of the seven operators with a scalar bilinear in Type A and B categories can be reduced into operators in Eq.\,\eqref{eq:type1} and Eq.\,\eqref{eq:type2} up to EoM terms,
\begin{subequations}
	\begin{eqnarray}
        \label{eq:opXV:1}
		2(\calO_{a3}-\calO_{b3})&=&\tilde\calO_{\Psi XV D^2 1}^{\tt S},
		\\
        \label{eq:opXV:2}
		2(\calO_{a1}-\calO_{b1})&=&-i\calO_{\Psi XVD^2 3}^{\tt S}-\frac{1}{2}\tilde\calO_{\Psi XV D^2 1}^{\tt S}+\fbox{EoM},
		\\
        \label{eq:opXV:3}
		2(\calO_{a2}-\calO_{b2})&=&-i\calO_{\Psi XVD^2 1}^{\tt S}-\frac{1}{2}\tilde\calO_{\Psi XV D^2 1}^{\tt S}+\fbox{EoM}.
	\end{eqnarray}
\end{subequations}
Therefore, we are left with four possible operators needing to be checked as independent basis operators. We choose the following four new combinations out of the seven operators, 
\begin{subequations}
    \label{eq:op:four}
	\begin{eqnarray}
        \label{eq:op:four:1}
		\calO_{\Psi XV D^2 5}^{\tt S}&\equiv&(\overline{\Psi}\Psi) \partial_{(\mu} X_{\nu)} \partial^{(\mu} V^{\nu)}=2(\calO_{a3}+\calO_{b3}),
		\\
        \label{eq:op:four:2}
		\calO_{\Psi XV D^2 6}^{\tt S}&\equiv& (\overline{\Psi} i\overleftrightarrow{D_\mu} \Psi) \partial^{(\mu} X^{\nu)} V_\nu=2\left(\calO_{a1}+\calO_{b1}\right)+\frac{i}{2} \tilde\calO_{\Psi XV D^2 1}^{\tt S},
		\\
		\calO_{\Psi XV D^2 7}^{\tt S}&\equiv& (\overline{D_{\mu} \Psi} D_{\nu} \Psi) X^{(\mu} V^{\nu)} =-2 \calO_{b4}-\left(\calO_{b1}+\calO_{b2}\right)+\fbox{F,G},
		\\
        \label{eq:opXV2}
		\calO_{\Psi XV D^2}^{\tt R}&\equiv& (\overline{\Psi} \overleftrightarrow{D_\mu} \Psi) X_{\nu} \partial^{(\mu} V^{\nu)} =2 i \left(\calO_{a2}+\calO_{b2}\right)+\frac{1}{2}\tilde\calO_{\Psi XV D^2 1}^{\tt S},
	\end{eqnarray}
\end{subequations}
where $\fbox{F,G}$ stands for operators with the two covariant derivatives being replaced by photon or gluon field strength tensor. 
It can be shown that the last operator $\calO_{\Psi XV D^2}^{\tt R}$ is still reducible. Therefore, except for operators in Eq.\,\eqref{eq:type1} and Eq.\,\eqref{eq:type2}, only three new operators need to be included in the basis, we have chosen them to be $\calO_{\Psi XV D^2 5,6,7}^{\tt S}$, as shown in Table \ref{tab:XV:3}. The similar argument applies to the operators  with a pesudoscalar bilinear, the independent operators are listed in Table \ref{tab:XV:3} as well.

\section{Construction of the $(\overline{\nu_L}\chi) XFD$-type operators}
\label{app2}

In this Appendix, 
we consider the construction of the dim-7 operator class 
$(\overline{\nu_L}\chi) XFD$ in the case with one fermion ($\chi$) and one vector ($X_\mu$) dark fields. Color and charge conservation selects the SM
fermion to be the neutrino field $\nu_L$ and the field strength tensor to be the photon 
field $F_{\mu\nu}$. 
First, the neutrino field can be arranged to be derivative free by IBP, 
then we form the following most general possibilities, 
\begin{subequations}
	\label{eq:a1}
	\begin{align}
		&\calO_{1}\equiv(\overline{{\nu}_{L}}\partial_\mu\chi)X_\nu F^{\mu\nu},\quad  &\calO_{2}&\equiv(\overline{{\nu}_{L}}\partial_\mu\chi)X_\nu \tilde{F}^{\mu\nu},
		\\
		&\calO_{3}\equiv(\overline{{\nu}_{L}}\chi)\partial_\mu X_\nu F^{\mu\nu},\quad  &\calO_{4}&\equiv(\overline{{\nu}_{L}}\chi)\partial_\mu X_\nu \tilde{F}^{\mu\nu},
		\\
		&\calO_{5}\equiv(\overline{{\nu}_{L}}\sigma_{\mu\nu}\partial_\rho\chi) X^\rho F^{\mu\nu},\quad &\calO_{6}&\equiv(\overline{{\nu}_{L}}\sigma_{\mu\nu}\partial_\rho\chi) X^\mu F^{\nu\rho},
		\\
		&\calO_{7}\equiv(\overline{{\nu}_{L}}\sigma_{\mu\nu}\chi) \partial^\mu X^\rho F^{\nu}_{~\rho},\quad &\calO_{8}&\equiv(\overline{{\nu}_{L}}\sigma_{\mu\nu}\chi) \partial_\rho X^\mu F^{\nu\rho},
		\\
		&\calO_{9}\equiv(\overline{{\nu}_{L}}\sigma_{\mu\nu}\chi) X_\rho \partial^\rho F^{\mu\nu},\quad &\calO_{10}&\equiv(\overline{{\nu}_{L}}\sigma_{\mu\nu}\chi) X_\rho \partial^\mu F^{\rho\nu}.
	\end{align}
\end{subequations}
Among them, there are four redundant operators. It can be shown that the two operators $\calO_9$ and $\calO_{10}$ with a derivative on the field strength tensor are related. Through the use of BI, the operator $\calO_{9}$ transforms as,
\begin{eqnarray}
\calO_{9}&\overset{\text{BI}}{=}& -(\overline{{\nu}_{L}}\sigma_{\mu\nu}\chi)X_\rho \partial^\nu F^{\rho\mu}-(\overline{{\nu}_{L}}\sigma_{\mu\nu}\chi)X_\rho \partial^\mu F^{\nu\rho}
= 2\calO_{10},
\end{eqnarray}
where we have used the antisymmetric property of the Lorentz indices inside $\sigma_{\mu\nu}$. Furthermore, the operator $\calO_{10}$ can be reduced to a combination of $\calO_1$, $\calO_3$, and $\calO_7$ by exploiting the IBP, GI, and EoM relations,
\begin{eqnarray}
	\nonumber
	\calO_{10}
  &\overset{\tt IBP}{=}&-\partial^\mu(\overline{{\nu}_{L}}\sigma_{\mu\nu}\chi) X_\rho F^{\rho\nu}-(\overline{{\nu}_{L}}\sigma_{\mu\nu}\chi)\partial^\mu X_\rho F^{\rho\nu}+\fbox{T}
	\\
	\nonumber
	&\overset{\tt GI}{\underset{\tt EoM}{=}}& (\overline{\nu_{L}}i\overleftarrow{\partial_{\nu}}\chi-\overline{{\nu}_{L}} i\partial_{\nu}\chi) X_\rho F^{\rho\nu}-(\overline{{\nu}_{L}}\sigma_{\mu\nu}\chi)\partial^\mu X_\rho F^{\rho\nu}+\fbox{EoM}
	\\
	\nonumber
	&\overset{\tt IBP}{\underset{\tt EoM}{=}}& -2(\overline{{\nu}_{L}} i\partial_{\nu}\chi) X_\rho F^{\rho\nu}-(\overline{{\nu}_{L}}\chi)i\partial_\nu X_\rho F^{\rho\nu}-(\overline{{\nu}_{L}}\sigma_{\mu\nu}\chi)\partial^\mu X_\rho F^{\rho\nu}+\fbox{EoM}
	\\
	&\Rightarrow&2i\calO_1+i\calO_3 +\calO_{7}.
\end{eqnarray}
The operators $\calO_5$ and $\calO_6$ are related to each other. 
From Eq.\,\eqref{eq:q1} with $D_\rho\rightarrow \partial_\rho$,
the operator $\calO_5$ can be reduced into a linear combination of ${\cal O}_2, {\cal O}_6$, together with EoM operators. 
The similar relationship can be established for $\calO_7$ and $\calO_8$. Applying IBP relation for $\calO_{8}$, it transforms into,
\begin{eqnarray}
	\label{eq:opchiXd}
	\calO_{8}
	&\overset{\tt IBP}{=}& -\calO_{6}-(\overline{{\nu}_{L}}\overleftarrow{\partial_\rho}\sigma_{\mu\nu}\chi)X^\mu F^{\nu\rho}+\fbox{T}.
\end{eqnarray}
In consideration of Eq.\,\eqref{eq:parten} with $D_\rho\rightarrow \partial_\rho$, the second term $(\overline{{\nu}_{L}}\overleftarrow{\partial_\rho}\sigma_{\mu\nu}\chi)X^\mu F^{\nu\rho}$ on the right-hand side is reduced as follows,
\begin{eqnarray}
    \nonumber
    (\overline{{\nu}_{L}}\overleftarrow{\partial_\rho}\sigma_{\mu\nu}\chi)X^\mu F^{\nu\rho}&\Rightarrow&(\overline{{\nu}_{L}}\overleftarrow{\partial_\nu}\sigma_{\mu\rho}\chi)X^\mu F^{\nu\rho}-(\overline{{\nu}_{L}}\overleftarrow{\partial_\mu}\sigma_{\nu\rho}\chi)X^\mu F^{\nu\rho}
    \\
    &\overset{\tt IBP}{\underset{\tt EoM}{\Rightarrow}}&4i\calO_1+2i\calO_3+\calO_5+\calO_6+2\calO_7+\calO_8,
\end{eqnarray}
which implies $\calO_7\Rightarrow -\calO_8$ up to other kept operators.

To sum up, we conclude that there are six independent operators for the $(\overline{\nu_L}\chi) XFD$ class without considering hermitian conjugation, and we choose them as shown in Table \ref{tab:Xchi}. Their relationships with the operators in Eqs.\,\eqref{eq:a1} are,
\begin{subequations}
\begin{align}
\calO_{\nu\chi X FD1}^{\tt S} & =  2\calO_3, 
&
\calO_{\nu\chi X FD 2}^{\tt S} & = 2\calO_4,
\\
\calO_{\nu\chi X FD 3}^{\tt S} & =-i\left(2\calO_{1}+\calO_{3}\right)+\fbox{EoM},
&
\calO_{\nu\chi X FD 4}^{\tt S} & =-i\left(2\calO_{2}+\calO_{4}\right),
		\\
\calO_{\nu\chi X FD}^{\tt T1} & =\calO_7-\calO_8, 
& 
\calO_{\nu\chi X FD}^{\tt T2} & =i\left(2\calO_{6}+\calO_{8}\right).
\end{align}
\end{subequations}

\bibliography{references.bib}{}
\bibliographystyle{utphys28mod}

\end{document}